\documentclass[longauth]{aa}

\usepackage{graphicx}
\usepackage{txfonts}
\usepackage{gensymb}
\usepackage[colorlinks=true, linkcolor=blue, citecolor=blue, draft=False]{hyperref}
\usepackage{placeins}

\begin{document} 

\title{GOODS-ALMA 2.0: Starbursts in the main sequence reveal compact star formation regulating galaxy evolution prequenching}

\author{
C. G\'omez-Guijarro
\inst{1}
\and
D. Elbaz
\inst{1}
\and
M. Xiao
\inst{1,2}
\and
V. I. Kokorev
\inst{3,4}
\and
G. E. Magdis
\inst{3,5,4}
\and
B. Magnelli
\inst{1}
\and
E. Daddi
\inst{1}
\and
F. Valentino
\inst{3,4}
\and
M. T. Sargent
\inst{6,7}
\and
M. Dickinson
\inst{8}
\and
M. B\'ethermin
\inst{9}
\and
M. Franco
\inst{10}
\and
A. Pope
\inst{11}
\and
B. S. Kalita
\inst{1}
\and
L. Ciesla
\inst{9}
\and
R. Demarco
\inst{12}
\and
H. Inami
\inst{13}
\and
W. Rujopakarn
\inst{14,15,16}
\and
X. Shu
\inst{17}
\and
T. Wang
\inst{2}
\and
L. Zhou
\inst{2,18}
\and
D. M. Alexander
\inst{19}
\and
F. Bournaud
\inst{1}
\and
R. Chary
\inst{20}
\and
H. C. Ferguson
\inst{21}
\and
S. L. Finkelstein
\inst{22}
\and
M. Giavalisco
\inst{11}
\and
D. Iono
\inst{23,24}
\and
S. Juneau
\inst{8}
\and
J. S. Kartaltepe
\inst{25}
\and
G. Lagache
\inst{9}
\and
E. Le Floc'h
\inst{1}
\and
R. Leiton
\inst{12}
\and
L. Leroy
\inst{1}
\and
L. Lin
\inst{26}
\and
K. Motohara
\inst{27}
\and
J. Mullaney
\inst{28}
\and
K. Okumura
\inst{1}
\and
M. Pannella
\inst{29,30}
\and
C. Papovich
\inst{31,32}
\and
E. Treister
\inst{33}
}

\institute{
AIM, CEA, CNRS, Universit\'e Paris-Saclay, Universit\'e Paris Diderot, Sorbonne Paris Cit\'e, F-91191 Gif-sur-Yvette, France
\email{carlos.gomezguijarro@cea.fr}
\and
School of Astronomy and Space Science, Nanjing University, Nanjing 210093, PR China
\and
Cosmic Dawn Center (DAWN), Denmark
\and
Niels Bohr Institute, University of Copenhagen, Jagtvej 128, DK-2200 Copenhagen N, Denmark
\and
DTU-Space, Technical University of Denmark, Elektrovej 327, DK-2800 Kgs. Lyngby, Denmark
\and
Astronomy Centre, Department of Physics and Astronomy, University of Sussex, Brighton BN1 9QH, UK
\and
International Space Science Institute (ISSI), Hallerstrasse 6, CH-3012 Bern, Switzerland
\and
Community Science and Data Center/NSF’s NOIRLab, 950 N. Cherry Ave., Tucson, AZ 85719, USA
\and
Aix Marseille Universit\'e, CNRS, LAM, Laboratoire d'Astrophysique de Marseille, Marseille, France
\and
Centre for Astrophysics Research, University of Hertfordshire, Hatfield AL10 9AB, UK
\and
Astronomy Department, University of Massachusetts, Amherst, MA 01003, USA
\and
Departamento de Astronom\'ia, Facultad de Ciencias F\'sicas y Matem\'aticas, Universidad de Concepci\'on, Concepci\'on, Chile
\and
Hiroshima Astrophysical Science Center, Hiroshima University, 1-3-1 Kagamiyama, Higashi-Hiroshima, Hiroshima 739-8526, Japan
\and
Department of Physics, Faculty of Science, Chulalongkorn University, 254 Phayathai Road, Pathumwan, Bangkok 10330, Thailand
\and
National Astronomical Research Institute of Thailand (Public Organization), Don Kaeo, Mae Rim, Chiang Mai 50180, Thailand
\and
Kavli IPMU (WPI), UTIAS, The University of Tokyo, Kashiwa, Chiba 277-8583, Japan
\and
Department of Physics, Anhui Normal University, Wuhu, Anhui 241000, PR China
\and
Key Laboratory of Modern Astronomy and Astrophysics (Nanjing University), Ministry of Education, Nanjing 210093, PR China
\and
Centre for Extragalactic Astronomy, Department of Physics, Durham University, Durham DH1 3LE, UK
\and
Infrared Processing and Analysis Center, MS314-6, California Institute of Technology, Pasadena, CA 91125, USA
\and
Space Telescope Science Institute, 3700 San Martin Drive, Baltimore, MD 21218, USA
\and
Department of Astronomy, The University of Texas at Austin, Austin, TX 78712, USA
\and
National Astronomical Observatory of Japan, National Institutes of Natural Sciences, 2-21-1 Osawa, Mitaka, Tokyo 181-8588, Japan
\and
SOKENDAI (The Graduate University for Advanced Studies), 2-21-1 Osawa, Mitaka, Tokyo 181-8588, Japan
\and
School of Physics and Astronomy, Rochester Institute of Technology, 84 Lomb Memorial Drive, Rochester, NY 14623, USA
\and
Institute of Astronomy \& Astrophysics, Academia Sinica, Taipei 10617, Taiwan
\and
Institute of Astronomy, Graduate School of Science, The University of Tokyo, 2-21-1 Osawa, Mitaka, Tokyo 181-0015, Japan
\and
Department of Physics and Astronomy, University of Sheffield, Sheffield S3 7RH, UK
\and
Astronomy Unit, Department of Physics, University of Trieste, via Tiepolo 11, I-34131 Trieste, Italy
\and
Fakult\"at f\"ur Physik der Ludwig-Maximilians-Universit\"at, D-81679 M\"unchen, Germany
\and
Department of Physics and Astronomy, Texas A\&M University, College Station, TX, 77843-4242, USA
\and
George P. and Cynthia Woods Mitchell Institute for Fundamental Physics and Astronomy, Texas A\&M University, College Station, TX, 77843-4242, SA
\and
Instituto de Astrof\'isica, Facultad de F\'isica, Pontificia Universidad Cat\'olica de Chile, Casilla 306, Santiago 22, Chile
}

\date{}

\abstract
{Compact star formation appears to be generally common in dusty star-forming galaxies (SFGs). However, its role in the framework set by the scaling relations in galaxy evolution remains to be understood. In this work we follow up on the galaxy sample from the GOODS-ALMA 2.0 survey, an ALMA blind survey at 1.1\,mm covering a continuous area of 72.42\,arcmin$^2$ using two array configurations. We derived physical properties, such as star formation rates, gas fractions, depletion timescales, and dust temperatures for the galaxy sample built from the survey. There exists a subset of galaxies that exhibit starburst-like short depletion timescales, but they are located within the scatter of the so-called main sequence of SFGs. These are dubbed starbursts in the main sequence and display the most compact star formation and they are characterized by the shortest depletion timescales, lowest gas fractions, and highest dust temperatures of the galaxy sample, compared to typical SFGs at the same stellar mass and redshift. They are also very massive, accounting for $\sim 60\%$ of the most massive galaxies in the sample ($\log (M_{\rm{*}}/M_{\odot}) > 11.0$). We find trends between the areas of the ongoing star formation regions and the derived physical properties for the sample, unveiling the role of compact star formation as a physical driver of these properties. Starbursts in the main sequence appear to be the extreme cases of these trends. We discuss possible scenarios of galaxy evolution to explain the results drawn from our galaxy sample. Our findings suggest that the star formation rate is sustained in SFGs by gas and star formation compression, keeping them within the main sequence even when their gas fractions are low and they are presumably on the way to quiescence.}

\keywords{galaxies: evolution -- galaxies: fundamental parameters -- galaxies: high-redshift -- galaxies: star formation -- galaxies: structure -- submillimeter: galaxies}

\titlerunning{GOODS-ALMA 2.0: Starbursts in the MS reveal compact star formation regulating galaxy evolution}
\authorrunning{C. G\'omez-Guijarro et al.}

\maketitle

\section{Introduction} \label{sec:intro}

The cosmic history of stellar mass assembly in galaxy evolution is directly linked with the star formation history. The cosmic star formation rate (SFR) density manifests a gradual growth from the epochs of formation of the first galaxies until it peaks at $z \sim 2$, followed by a factor 10 decline until $z \sim 0$ \citep[see][for a review]{madau14}.

Numerous studies revealed the existence of a tight correlation between the SFR and the stellar mass of star-forming galaxies (SFGs), the so-called main sequence (MS) of SFGs \citep[e.g.,][]{brinchmann04,daddi07,elbaz07,noeske07,whitaker12}. The MS shows signs of a small scatter at least up to $z \sim 4$ \citep[$\sim 0.3$\,dex; e.g.,][]{elbaz07,noeske07,whitaker12,speagle14,schreiber15}, which suggests that secular evolution is the dominant mode of stellar assembly, involving a steady equilibrium between gas inflows, outflows, and consumption in star formation \citep[e.g.,][]{daddi10,tacconi10,genzel10,dekel13,feldmann15}. The average SFR at a fixed stellar mass is higher with increasing redshift, which is reflected in the rise of the MS normalization with increasing redshift \citep[e.g.,][]{noeske07,whitaker12,speagle14,schreiber15,leslie20}. Outliers to the MS exist at all redshifts characterized by their elevated SFR compared to MS galaxies at a fixed stellar mass and redshift. Galaxies of this kind are commonly referred to as starbursts galaxies (SBs). In addition to the latter, various other definitions of SBs are widespread in the literature \citep[see][for a compilation]{heckman05}. Particularly important for this work are the definitions of SBs of an anomalously rapid gas consumption reflected in short depletion timescales compared to normal SFGs at a fixed stellar mass and redshift, as well as the definition based on the elevated SFR compared to MS galaxies at a fixed stellar mass and redshift.

Understanding the evolution of the SFR density and the secular mode of stellar mass assembly implied by the MS requires one to study the cold gas reservoirs that fuel star formation. Investigation of the evolution of the cold gas density has uncovered that it closely follows the evolution of the SFR density \citep[e.g.,][]{riechers19,decarli19,magnelli20}. In the last decade, numerous studies have progressively revealed the increase in the gas content and decrease in the depletion timescales with increasing redshift \citep[e.g.,][]{daddi08,daddi10,tacconi10,genzel10,magdis12,bethermin15,dessaugeszavadsky20,donevski20,kokorev21}. More recently, large statistical galaxy samples have established the gas scaling relations for the gas fractions ($f_{\rm{gas}} = M_{\rm{gas}}/(M_{\rm{gas}} + M_{\rm{*}})$) and depletion timescales ($\tau_{\rm{dep}} = M_{\rm{gas}}/\rm{SFR}$), with their dependencies on a specific star formation rate ($\rm{sSFR} = \rm{SFR} / M_{\rm{*}}$), stellar mass, as well as lookback time \citep[e.g.,][]{scoville17,tacconi18,liu19}. The overall larger amount of gas looks to be the main responsible factor for the average increase in the SFR with increasing redshift, while at a fixed stellar mass and redshift, both a mix of higher gas fractions and lower depletion timescales are invoked to explain the existence of SBs.

Gas appears to be the main regulator of galaxy evolution and galaxy growth seems to be a relatively smooth process over long timescales, as shown by the gas evolution dominating the behavior of the cosmic SFR density and the existence of the MS. Self-regulation seems to be typical among galaxies, as only a minor fraction of SFGs deviate from the equilibrium, normally associated with disruptive dynamical events, such as galaxy mergers \citep[e.g.,][]{joseph85,kartaltepe12,cibinel19}. An always increasing number of galaxies fall off the MS when the star formation shuts off or drastically decreases \citep[e.g.,][]{brammer11,straatman15,davidzon17}, which is the so-called quenching of star formation that leads to the formation of quiescent galaxies (QGs).

However, the detailed mechanisms governing the overall smooth and self-regulated galaxy evolution set by the gas scaling relations and the MS framework remain to be understood. The study of the properties of SFGs, especially those within the scatter of the MS, offers the means to explore these mechanisms further. In the last years, the Atacama Large Millimeter/submillimeter array (ALMA) has opened new frontiers \citep[see][for a review]{hodge20}. A number of studies have indicated that star formation traced by submm/mm dust continuum emission takes place in compact areas smaller than the stellar sizes \citep[e.g.,][]{simpson15,ikarashi15,hodge16,fujimoto17,fujimoto18,gomezguijarro18,gomezguijarro21,elbaz18,lang19,rujopakarn19,gullberg19,franco20b}. Similar results have been also found in radio emission \citep[e.g.,][]{murphy17,jimenezandrade19,jimenezandrade21}. Furthermore, other recent works have discovered the existence of a population of SBs within the scatter of the MS, owing to their short depletion timescales and enhanced star formation surface densities, exhibiting compact star formation traced by submm/mm dust continuum emission \citep{elbaz18} or, similarly, compact radio emission \citep{jimenezandrade19}. Besides, a population of galaxies within the scatter of the MS displayed compact star formation as traced by CO lines \citep{puglisi19}.

Nevertheless, the role of compact star formation and of the population of SBs in the MS in the more general smooth and self-regulated galaxy growth is yet to be comprehended. Several studies have advocated for compaction events predicted by galaxy formation models in which extended SFGs in the MS can secularly evolve into compact SFGs in the MS by funneling gas to their central regions yielding the build up of their stellar cores \citep[e.g.,][]{dekel13,zolotov15,tacchella16}. Other studies have advocated that SBs dominated by a violent episode of star formation typical of gas-rich mergers that move well above the scatter of the MS are also capable of funneling gas to the center of collision and quickly build up compact stellar cores \citep[e.g.,][]{mihos96,hopkins06,toft14,gomezguijarro18,puglisi21}. In the latter scenario compact SFGs in the MS would be fading SBs where most of the gas has already been consumed, located within the scatter of the MS, but on their way to quiescence \citep[e.g.,][]{elbaz18,gomezguijarro19,puglisi21}.

In this work, we aim at studying these unknowns about the role of compact star formation and the origin, nature, and role of SBs in the MS in galaxy evolution. We follow up on the galaxy sample presented in \citet{gomezguijarro21} from the GOODS-ALMA 2.0 survey, an ALMA blind survey at 1.1\,mm covering a continuous area of 72.42\,arcmin$^2$ using two array configurations at a similar and homogeneous depth over the whole field. The combined mosaic with both configurations reaches an average sensitivity of 68.4$\,\mu$Jy beam$^{-1}$ at an average angular resolution of 0\farcs447\,$\times$\,0\farcs418. The layout of the paper is as follows. An overview of the GOODS-ALMA survey, galaxy sample, and multiwavelength ancillary data involved in this work is given in Sect.~\ref{sec:sample_data}. In Sect.~\ref{sec:properties} we present the dust and stellar-based properties of the galaxy sample. Sect.~\ref{sec:scl_rel} is dedicated to study the physical properties of the galaxy sample in the framework of the MS and the scaling relations for depletion timescales, gas fractions, and dust temperatures. We investigate the role of compact star formation, traced by the dust continuum emission, as a physical driver of the galaxy sample behavior in relation with the scaling relations in Sect.~\ref{sec:compact}. In Sect.~\ref{sec:discussion} we discuss and interpret our results in the broader cosmological context along with literature studies. We summarize the main findings and conclusions in Sect.~\ref{sec:summary}.

Throughout this work we adopt a concordance cosmology $[\Omega_\Lambda,\Omega_M,h]=[0.7,0.3,0.7]$ and a Salpeter initial mass function (IMF) \citep{salpeter55}. When magnitudes are quoted they are in the AB system \citep{oke74}.

\section{Sample and data} \label{sec:sample_data}

\subsection{ALMA 1.1\,mm galaxy survey in GOODS-South} \label{subsec:ags}

GOODS-ALMA is a 1.1\,mm galaxy survey in the Great Observatories Origins Deep Survey South field \citep[GOODS-South;][]{dickinson03,giavalisco04} carried out with ALMA Band 6. GOODS-ALMA extends over a continuous area of 72.42\,arcmin$^2$ (primary beam response level $\geq 20\%$) centered at $\alpha = $3$^{\rm{h}}$32$^{\rm{m}}$30$^{\rm{s}}$, $\delta = -27$\degree48\arcmin00 with a homogeneous average sensitivity employing two different array configurations: Cycle 3 observations (program 2015.1.00543.S; PI: D. Elbaz) provided the high-resolution small spatial scales using a more extended array configuration (high resolution dataset), while Cycle 5 observations (program 2017.1.00755.S; PI: D. Elbaz) provided the low-resolution large spatial scales employing a more compact array configuration (low resolution dataset). The high resolution dataset was presented in \citet{franco18} (GOODS-ALMA 1.0). The low resolution dataset and its combination with the high resolution (combined dataset) was presented in \citet{gomezguijarro21} (GOODS-ALMA 2.0), improving the $uv$ coverage and sensitivity. On average, the high resolution and low resolution mosaics have similar sensitivities of 89.0 and 95.2\,$\mu$Jy beam$^{-1}$ at an angular resolution of 0\farcs251\,$\times$\,0\farcs232 and 1\farcs33\,$\times$\,0\farcs935, respectively (synthesized beam FWHM along the major\,$\times$\,minor axis). The combined mosaic reaches an average sensitivity of 68.4$\,\mu$Jy beam$^{-1}$ at an average angular resolution of 0\farcs447\,$\times$\,0\farcs418.

In this work, we used the GOODS-ALMA 2.0 source catalog presented in \citet{gomezguijarro21}, a sample composed of 88 sources, 50\% detected at $\rm{S/N_{peak}} \geq 5$ (sources detected with a purity of 100\% associated with the absence of negative detections at the same significance) and 50\% detected at $3.5 \leq \rm{S/N_{peak}} \leq 5$ aided by priors. We refer the reader to \citet{gomezguijarro21} for further details about the observations, data processing, source catalog, and 1.1\,mm flux density and size measurements.

\subsection{Multiwavelength ancillary data} \label{subsec:ancil_data}

Ultraviolet (UV) to near-infrared (near-IR) data were taken from the $K_s$-band-selected ZFOURGE catalog by \citet{straatman16}. ZFOURGE (PI: I. Labb\'e) uses five near-IR medium bands ($J_1$, $J_2$, $J_3$, $H_s$, and $H_l$) and the $K_s$-band. It combines dedicated FourStar/$K_s$-band observations with prexisting $K$-band imaging to create super-deep detection images: in the CDFS field (containing GOODS-South) VLT/HAWK-I/$K$ from HUGS \citep{fontana14}, VLT/ISAAC/$K$ from GOODS, ultra deep data in the HUDF region \citep{retzlaff10}, CFHST/WIRCAM/$K$ from TENIS \citep{hsieh12}, and Magellan/PANIC/$K$ in HUDF (PI: I. Labb\'e). The ancillary CDFS UV to near-IR filters include VLT/VIMOS/$U,R$-imaging \citep{nonino09}, \textit{HST}/ACS/$B,V,I,Z$-imaging \citep{giavalisco04,wuyts08}, ESO/MPG/WFI/$U_{38},V,R_c$-imaging \citep{erben05,hildebrandt06}, \textit{HST}/WFC3/$F098M,F105W$,$F125W,F140W,F160W$ and \textit{HST}/ACS$F606W,F814W$-imaging \citep{grogin11,koekemoer11,windhorst11,brammer12}, and 11 Subaru/{Suprime-Cam} optical medium bands \citep{cardamone10}. \textit{Spitzer}/IRAC/3.6 and 4.5\,$\mu$m images are the ultradeep mosaics from the IUDF \citep{labbe15}, using data from the IUDF (PI: I. Labb\'e) and IGOODS (PI: P. Oesch) programs, combined with GOODS (PI: M. Dickinson), ERS (PI: G. Fazio), S-CANDELS (PI: G. Fazio), SEDS (PI: G. Fazio), and UDF2 (PI: R. Bouwens). Mid-IR \textit{Spitzer}/IRAC/5.8 and 8.0\,$\mu$m images are from GOODS (PI: M. Dickinson).

Additionally, we used mid-IR to submm data including \textit{Spitzer}/MIPS/24\,$\mu$m images from GOODS (PI: M. Dickinson), \textit{Herschel}/PACS/70, 100, 160\,$\mu$m from GOODS-\textit{Herschel} \citep[][PI: D. Elbaz]{elbaz11} and PEP \citep{lutz11} combined \citep{magnelli13}, and \textit{Herschel}/SPIRE/250, 350, 500\,$\mu$m from HerMES \citep{oliver12}. In radio, VLA observations in GOODS-South (PI: W. Rujopakarn) were taken at 3\,GHz (10\,cm) covering the entire GOODS-ALMA field (Rujopakarn et al., in prep) and at 6\,GHz (5\,cm) around the HUDF with partial coverage of GOODS-ALMA \citep{rujopakarn16}. For the \textit{Herschel} data, the PACS PEP/GOODS-\textit{Herschel} reach 3$\sigma$ depths of 0.9, 0.6, and 1.3\,mJy at 70, 100, 160\,$\mu$m, respectively \citep{magnelli13}. The SPIRE HerMES reach 5$\sigma$ depths of 4.3, 3.6, and 5.2\,mJy at 250, 350, 500\,$\mu$m, respectively \citep{oliver12}. In the case of SPIRE, its large beam yields a high confusion limit, requiring tailored de-blending methodologies to provide fluxes from the highly confused low-resolution data to optical counterparts. We used a catalog (T. Wang, private communication) built using a state-of-the-art de-blending methodology similar to that applied in the "super-deblended" catalogs in the GOODS-North \citep{liu18} and COSMOS \citep{jin18} fields. Briefly, source extraction was run on the SPIRE images starting with all the MIPS 24\,$\mu$m priors down to 20\,$\mu$Jy. This flux level yielded a prior density of $\sim 1$\,source/beam, which maximized the efficiency of source extraction. A first PSF-fitting run determined flux densities and uncertainties. After that, a second run was performed by freezing faint sources to their predicted fluxes if they had a brighter neighbor within 1.5 pixels. Flux measurements were crossed checked with those predicted from spectral energy distribution (SED) fitting to the mid-IR to submm data. Overall, this methodology helps to break blending degeneracies in crowded regions, yielding $\times 2$--10 lower RMS levels and more accurate flux density measurements for the majority of nonfreezed sources and cleaner residual images.

\section{Dust and stellar-based properties} \label{sec:properties}

\subsection{Redshifts and stellar masses} \label{subsec:zmstar}

The redshifts and stellar masses for the 88 ALMA sources that compose our galaxy sample were presented in \citet{gomezguijarro21}. These values are recapitulated in Tables~\ref{tab:sample_prop_100pur} and \ref{tab:sample_prop_prior}. In summary, we searched for stellar counterparts in the ZFOURGE catalog by \citet{straatman16}, which provides photometric redshift and stellar mass estimations. Spectroscopic redshifts were obtained through a recent compilation in GOODS-South (N. Hathi, private communication), along with additional recent surveys in the field: VANDELS \citep{garilli21}, the MUSE-Wide survey \citep{herenz17,urrutia19}, and ASPECS LP \citep{decarli19,gonzalezlopez19,boogaard19} \citep[see][for the spectroscopic redshift references therein]{gomezguijarro21}. When the redshifts were updated by new spectroscopic values compared to the photometric redshifts from ZFOURGE, the stellar mass estimations were updated employing the same methodology as ZFOURGE: photometry was fit using \texttt{FAST++}\footnote{https://github.com/cschreib/fastpp}, an updated version of the SED fitting code \texttt{FAST} \citep{kriek09} employed in ZFOURGE. The stellar population models were from \citet{bruzual03}, with exponentially declining star formation histories, \citet{calzetti00} dust attenuation law, and fixed solar metallicity. The stellar masses were multiplied by a factor of 1.7 to scale them from a Chabrier \citep{chabrier03} to a Salpeter \citep{salpeter55} IMF \citep[e.g.,][]{reddy06,santini12,elbaz18}. In addition, the redshifts and stellar masses of the six optically dark galaxies in \citet{zhou20}, namely AGS4, AGS11, AGS15, AGS17, AGS24, and AGS25 (A2GS2, A2GS15, A2GS10, A2GS7, A2GS29, and A2GS17, respectively), were substituted by the values in that work, where they were studied in detail. Tailored photometry was obtained for four galaxies that were affected by blending (A2GS28, A2SGS30, A2GS33, and A2GS60) and for an extra galaxy with no counterpart in the ZFOURGE catalog (A2GS38), employing updated \textit{HST} data from the HLF-GOODS-S v$2.0$ mosaics \citep{illingworth16,whitaker19} homogenized to the WFC3/$F160W$-band PSF, along with the ground-based bands in ZFOURGE homogenized to a common Moffat PSF profile \citep[0\farcs9 FWHM; see][]{straatman16}, and the \textit{Spitzer}/IRAC data. The photometry was carried out following the methodology described in \citet{gomezguijarro18} for crowded and blended sources: all the bands affected by blending were fit using \texttt{GALFIT} \citep{peng02}, with priors set to the number of sources in the $F160W$-band image. For the bands unaffected by blending, aperture photometry was performed using aperture diameters of 0\farcs7 for \textit{HST} \citep[as in][]{whitaker19} and 1\farcs2 for the ground-based data \citep[as in][]{straatman16} with aperture corrections to account for the flux losses outside the aperture. PSF photometry with \texttt{GALFIT} was used for the \textit{Spitzer}/IRAC bands. Uncertainties were derived from empty aperture measurements.

\subsection{Infrared SED fitting} \label{subsec:fir_sed}

SEDs in the mid-IR to mm for the galaxy sample can be divided in two groups depending on whether they have a \textit{Herschel} counterpart or not. Therefore, we performed two types of SED fitting to the mid-IR to mm photometry.

For the galaxies with a \textit{Herschel} counterpart (69/88) we employed the panchromatic SED fitting tool \texttt{Stardust}\footnote{https://github.com/VasilyKokorev/stardust} developed by \citet{kokorev21}. In summary, this code performs a multi-component fit that combines linearly stellar libraries, AGN torus templates, and IR models of dust emission from star formation. All three components are fit simultaneously yet independently from each other, without assuming an energy balance between UV emission absorbed and re-emitted by dust at far-IR and mm wavelengths. An energy-balanced solution implies cospatial stellar and dust continuum emission \citep[e.g.,][]{dacunha08}, an assumption that is applicable and physically motivated in a wide variety of SFGs, but perhaps not valid for all dusty SFGs (DSFGs). Stars and dust are not always physically connected, with samples of DSFGs exhibiting differences in their stellar and dust continuum distributions sometimes spatially offset from each other \citep[e.g.,][]{simpson15,hodge16,gomezguijarro18,elbaz18,franco18}. The code finds the best linear combination of a set of basic templates (similar to eigenvectors in principal component analysis), instead of using a vast library of templates. The models used to create the basic templates are: 1) a stellar library (12 templates) composed of an updated version of the single stellar population synthesis models in \citet{brammer08}. The stellar component helps to constrain the AGN contribution in the near/mid-IR; 2) an AGN library consisting in empirically-derived templates from \citet{mullaney11} that describe the AGN emission from 6 to 100\,$\mu$m rest-frame, using both high and low-luminosity AGN templates; 3) an IR library composed of 4\,862 \citet{draineli07} dust emission templates with the correction from \citet{draine14}. These models combine two components: a cold dust component coming from the diffuse interstellar medium (ISM) heated by a minimum radiation field, $U_{\rm{min}}$, and a warm dust component associated with photodissociation regions (PDRs) heated by a power law distribution of starlight with intensities ranging from $U_{\rm{min}}$ to $U_{\rm{max}}$. The relative contribution between the two components is quantified by the parameter $\gamma$, giving the fraction of dust mass associated with the warm dust component in PDRs. The properties of the dust grains in the models are parametrized by the polycyclic aromatic hydrocarbon (PAH) index $q_{\rm{PAH}}$, yielding the fraction of dust mass in the form of PAH grains; 4) radio continuum data are not fit by the code. The radio model is a power law $S_\nu \propto \nu^{\alpha}$ with a fixed spectral index of $\alpha = -0.75$. The output radio luminosity at 1.4\,GHz rest-frame is such that satisfies the IR-radio correlation by \citet{delvecchio21}, which depends on the stellar mass of a given galaxy.

As described by \citet{draineli07}, the infinitesimal portion of the dust mass $dM_{\rm{dust}}$ exposed to a radiation field with intensities between $U$ and $U+dU$ can be expressed as a combination of a $\delta$-function, accounting for the cold dust component heated by $U_{\rm{min}}$, and a power law distribution of starlight from $U_{\rm{min}}$ to $U_{\rm{max}}$:

\begin{equation}
\label{eq:mdust}
\frac{dM_{\rm{dust}}}{dU} = (1 - \gamma) M_{\rm{dust}} \delta(U - U_{\rm{min}}) + \gamma M_{\rm{dust}} \frac{\alpha - 1}{U_{\rm{min}}^{1-\alpha} - U_{\rm{max}}^{1-\alpha}} U^{-\alpha}
\end{equation}
\noindent with $U_{\rm{min}} \leq U_{\rm{max}}$ and $\alpha \neq 1$. The values ($q_{\rm PAH}$, $U_{\rm min}$, $\gamma$) are obtained from the result of the fit, considering a fixed $U_{\rm{max}} = 10^6$ and $\alpha = 2$  following \citet{draine07}. The total dust mass ($M_{\rm{dust}}$) is then calculated from the normalization and the total IR luminosity by integrating the fit in the range 8--1000\,$\mu$m rest-frame. Considering the fraction of the total IR luminosity that is due to AGN contribution ($f_{\rm{AGN}}$) from the fitting procedure, it is ensured that the IR luminosity ($L_{\rm{IR}}$) used hereafter accounts for star formation only:

\begin{equation}
\label{eq:lir}
L_{\rm{IR}} = (1 - f_{\rm{AGN}}) \int_{8\mu m}^{1000\mu m} L_{\nu}(\lambda) \times \frac{c}{\lambda^2} d\lambda\,.
\end{equation}

For galaxies without a \textit{Herschel} counterpart (19/88), we employed an iterative approach using the IR template library of \citet{schreiber18}. This library considers an evolution of the dust temperature ($T_{\rm{dust}}$) with redshift and distance to the MS ($\Delta \rm{MS}$), defined as the ratio of the SFR to the SFR of the MS at the same stellar mass and redshift ($\Delta \rm{MS} = \rm{SFR/SFR_{MS}}$), as well as the dependency of the mid-to-total infrared color ($\rm{IR8} = L_{\rm{IR}}/L_8$) with $\Delta \rm{MS}$. Therefore, it is possible to define MS ($\Delta \rm{MS} = 1$) and SB ($\Delta \rm{MS} = 5$) templates. The templates are normalized to $M_{\rm{dust}} = 1$\,$M_{\odot}$. The total $M_{\rm{dust}}$ is then calculated by re-normalizing to the 1.1\,mm flux density and the total $L_{\rm{IR}}$ by integrating the SED in the range 8--1000\,$\mu$m rest-frame.

For each galaxy, the output $L_{\rm{IR}}$ along with the galaxy redshift and stellar mass (see Sect.~\ref{subsec:zmstar}) was used to calculate its $\Delta \rm{MS}$. If the resulting $\Delta \rm{MS}$ after normalizing the MS template was within the 1$\sigma$ scatter of the MS, which corresponds to $\Delta \rm{MS} < 2$ \citep[$\sim 0.3$\,dex; e.g.,][]{noeske07,whitaker12,schreiber15} and, simultaneously, $\Delta \rm{MS}$ after normalizing the SB template went in the direction of the MS reaching at least within the 2$\sigma$ scatter of the MS ($\Delta \rm{MS} < 4$), then a MS template was considered as the appropriate one. The opposite was true, if the resulting $\Delta \rm{MS}$ after normalizing with the SB template was outside the 2$\sigma$ scatter of the MS ($\Delta \rm{MS} > 4$) and, simultaneously, $\Delta \rm{MS}$ after normalizing with the MS template went in the direction of the SBs reaching further than the 2$\sigma$ scatter of the MS, then a SB template was considered as the appropriate one. For all the galaxies without a \textit{Herschel} counterpart in our sample, the above conditions were satisfied and we chose either a MS or a SB template for each galaxy. Although not relevant for the analysis in this work, it is interesting to note that the latter was even true for the galaxies with a \textit{Herschel} counterpart, except just in two cases (namely A2GS53 and A2GS62) where $\Delta \rm{MS} < 2$ with the MS template, but $\Delta \rm{MS} > 4$ with the SB template. From the estimates as obtained by \texttt{Stardust} (the two have a \textit{Herschel} counterpart) both are in the MS within a factor 3 ($0.33 < \Delta \rm{MS} < 3$, 0.5\,dex).

When comparing $M_{\rm{dust}}$ estimates it is important to take into account possible physical differences in the dust models. As explained by \citet{schreiber18} the dust models in their IR template library assume dust grain species with different emissivities that lead to systematically lower $M_{\rm{dust}}$ values compared to those from \citet{draineli07} dust models. Therefore, in order for the $M_{\rm{dust}}$ estimates to be comparable in galaxies with and without a \textit{Herschel} counterpart, we scaled the $M_{\rm{dust}}$ estimates obtained using the dust SED libraries by \citet{schreiber18} to account for the physical difference in the dust models following \citet{schreiber18}.

As a sanity test, we compared $M_{\rm{dust}}$ and $L_{\rm{IR}}$ estimates for the galaxies with a \textit{Herschel} counterpart as obtained by both \texttt{Stardust} and the iterative approach using the dust SED libraries by \citet{schreiber18}. The results are consistent on average with a relatively large dispersion with a median relative difference $(M_{\rm{dust}}^{\rm{Sch18}} - M_{\rm{dust}}^{\rm{Stardust}}) / M_{\rm{dust}}^{\rm{Stardust}} = -0.01 \pm 0.45$ and $(L_{\rm{IR}}^{\rm{Sch18}} - L_{\rm{IR}}^{\rm{Stardust}}) / L_{\rm{IR}}^{\rm{Stardust}} = -0.07 \pm 0.71$ (where the uncertainty is the median absolute deviation). A large dispersion was expected since for the iterative approach using the dust SED libraries by \citet{schreiber18} only the 1.1\,mm flux densities were considered.

In Tables~\ref{tab:sample_prop_100pur} and \ref{tab:sample_prop_prior} we present $M_{\rm{dust}}$ and $L_{\rm{IR}}$ estimates as obtained by \texttt{Stardust} for the galaxies with a \textit{Herschel} counterpart and the iterative approach using the dust SED libraries by \citet{schreiber18} for the galaxies without a \textit{Herschel} counterpart. The uncertainties were calculated performing 10\,000 Monte-Carlo simulations perturbing the photometry randomly within the uncertainties. Appendix~\ref{sec:appendix_a} collects the different SED fits to the mid-IR to mm photometry for the galaxy sample. Inspecting the results, there are two galaxies with \textit{Herschel} counterparts for which the mid-IR to mm photometry is not satisfactory as it is still affected by blending (namely A2GS7 and A2GS12). The de-blending methodology for SPIRE was optimized for the entire \textit{Herschel} mosaic. Therefore, there could still exist some specific single source cases reflecting some level of contamination by blending (as it is the case for A2GS7 and A2GS12). We treated them as if they were galaxies without a \textit{Herschel} counterpart instead, with their $M_{\rm{dust}}$ and $L_{\rm{IR}}$ obtained following the iterative approach with the dust SED libraries by \citet{schreiber18}. Note also that, having reassessed the \textit{Herschel} counterpart assignation in this work for the sources in common with \citet{franco20b}, it resulted in some additional galaxies to have \textit{Herschel} detections respect to \citet{franco20b}.

\subsection{Star formation rates} \label{subsec:sfr}

The total SFR accounts for the contribution of the obscured star formation probed in the IR ($\rm{SFR_{IR}}$) and the unobscured star formation probed in the UV ($\rm{SFR_{UV}}$):

\begin{equation}
\label{eq:sfr}
\rm{SFR} = \rm{SFR_{IR}} + \rm{SFR_{UV}}\,.
\end{equation}

We calculated the obscured star formation through $L_{\rm{IR}}$ as derived in Sect.~\ref{subsec:fir_sed} using the \citet{kennicutt98} conversion (we note that we considered the fraction of the total IR luminosity that is due to AGN contribution and $L_{\rm{IR}}$ accounts for star formation only):

\begin{equation}
\label{eq:sfr_ir}
\rm{SFR_{IR}} (M_{\odot} yr^{-1}) = 1.72 \times 10^{-10} \textit{L}_{\rm{IR}} (L_{\odot})\,.
\end{equation}

The unobscured star formation can be obtained through the UV luminosity ($L_{\rm{UV}}$) using the conversion from \citet{daddi04}:

\begin{equation}
\label{eq:sfr_uv}
\rm{SFR_{UV}} (M_{\odot} yr^{-1}) = 2.16 \times 10^{-10} \textit{L}_{\rm{UV}} (L_{\odot})\,,
\end{equation}

\noindent where $L_{\rm{UV}}$ is given by the rest-frame 1500\,\AA~luminosity:

\begin{equation}
\label{eq:luv}
L_{\rm{UV}} (L_{\odot}) = 4 \pi d_L^2 \frac{\nu_{1500}}{(1 + z)} \frac{10^{-0.4(48.6 + m_{1500})}}{3.826 \times 10^{33}}\,,
\end{equation}

\noindent with $d_L$ the luminosity distance (cm), $\nu_{\rm{1500}}$ the frequency (Hz) that corresponds to $\lambda = 1500$\,\AA, and $m_{\rm{1500}}$ the rest-frame 1500\,\AA~(AB) magnitude that we obtained from the ZFOURGE catalog for the stellar counterpart of each ALMA source. For the galaxies which we updated the default ZFOURGE redshifts and photometry for as explained in Sect.~\ref{subsec:zmstar}, we derived rest-frame 1500\,\AA~magnitudes employing the same methodology as ZFOURGE: a top hat filter centered at 1500\,\AA~with a 350\,\AA~width.

In comparison with the estimates in \citet{franco20b} for the sources in common with this work, SFR estimates are consistent on average with a small dispersion of $20\%$.

\subsection{Gas masses} \label{subsec:mgas}

The gas mass reservoir ($M_{\rm{gas}}$) in galaxies can be determined through the dust emission. We employed the gas-to-dust mass ratio ($\delta_{\rm{GDR}}$) with a metallicity dependency \citep[e.g.,][]{magdis11,magdis12,berta16}:

\begin{equation}
\label{eq:mgas}
M_{\rm{gas}} = \delta_{\rm{GDR}}(Z) M_{\rm{dust}}\,.
\end{equation}

This technique converts $M_{\rm{dust}}$ to $M_{\rm{gas}}$ through a well-established relation between $\delta_{\rm{GDR}}$ and the gas-phase metallicity of galaxies ($\delta_{\rm{GDR}}$--$Z$), seen both in the local universe and at high redshift \citep[e.g.,][]{leroy11,magdis12,remyruyer14,genzel15,coogan19}. Here we used the $\delta_{\rm{GDR}}$--$Z$ relation of \citet{magdis12}:

\begin{equation}
\label{eq:gdr_z}
\log \delta_{\rm{GDR}} = (10.54 \pm 1.00) - (0.99 \pm 0.12) (12 + \log \rm{(O/H)})\,.
\end{equation}

Direct metallicity measurements are unknown for our galaxy sample and, in general, elusive for DSFGs. We employed a mass-metallicity relation \citep[MZR; e.g.,][]{erb06} to calculate the metallicities for the galaxy sample. In particular, we used the redshift-dependent MZR expression of \citet{genzel15}:

\begin{equation}
\label{eq:mzr}
12 + \log \rm{(O/H)} = \textit{a} - 0.087 (\log (\textit{M}_{*}/1.7) - \textit{b})^2\,,
\end{equation}

\noindent where $a = 8.74$ and $b = 10.4 + 4.46 \log (1 + z) - 1.78 \log (1 + z)^2$. We included a 1.7 factor to convert our stellar masses and SFRs from a Salpeter \citep{salpeter55} to a Chabrier \citep{chabrier03} IMF \citep[e.g.,][]{reddy06,santini12,elbaz18}, as the latter is the IMF used in the MZR by \citet{genzel15}. The resulting metallicities are calibrated in the \citet{pettini04} (PP04 N2) scale, the calibration used in Eq.~\ref{eq:gdr_z} by \citet{magdis12}. We adopted an uncertainty of 0.2\,dex in the metallicities \citep{magdis12}.

Having calculated the gas-phase metallicities, we used Eq.~\ref{eq:gdr_z} to derive $\delta_{\rm{GDR}}$ for each galaxy and converted $M_{\rm{dust}}$ into $M_{\rm{gas}}$ using Eq.~\ref{eq:mgas}. We note that these $M_{\rm{gas}}$ estimates yield the total gas budget of the galaxies, including the molecular ($M_{\rm{H_2}}$) and the atomic phase ($M_{\rm{HI}}$): $M_{\rm{gas}} = M_{\rm{H_2}} + M_{\rm{HI}}$. A variety of studies have suggested that for the redshifts and relatively massive galaxies probed in our work, the molecular gas dominates over the atomic gas within the physical scales probed by the dust continuum observations at 1.1\,mm \citep[e.g.,][]{obreschkow09,tacconi10,daddi10}.

In Tables~\ref{tab:sample_prop_100pur} and \ref{tab:sample_prop_prior} we present $M_{\rm{gas}}$ estimates for the galaxies in our sample along with gas fractions ($f_{\rm{gas}} = M_{\rm{gas}}/(M_{\rm{gas}} + M_{\rm{*}})$) and depletion timescales ($\tau_{\rm{dep}} = M_{\rm{gas}}/\rm{SFR}$), the latter being by definition the inverse of the star formation efficiency ($\rm{SFE} = 1 / \tau_{\rm{dep}}$).

However, another description of metallicities is the so-called fundamental metallicity relation \citep[FMR;][]{mannucci10}. The FMR incorporates an additional dependency of metallicity on the SFR. We obtained alternative $M_{\rm{gas}}$ estimates using the FMR expression of \citet{mannucci10}:

\begin{multline}
\label{eq:fmr}
12 + \log \rm{(O/H)} = 8.90 + 0.37m - 0.14s - 0.19m^2 \\+ 0.12ms - 0.054s^2\,,
\end{multline}

\noindent where $m = \log (M_{*}/1.7) - 10$ and $s = \log \rm{SFR}/1.7$ (solar units). Once again, a 1.7 factor is included to convert our stellar masses and SFRs from a Salpeter \citep{salpeter55} to a Chabrier \citep{chabrier03} IMF, as the latter is the IMF used in the FMR by \citet{mannucci10}. The resulting metallicities are calibrated for the \citet{kewley02} (KD02) photoionization models. We converted them to the PP04 N2 scale, the calibration used in Eq.~\ref{eq:gdr_z} by \citet{magdis12}, following the conversion recipes by \citet{kewley08}.

The median relative difference between MZR and FMR $M_{\rm{gas}}$ estimates is $(M_{\rm{gas}}^{\rm{FMR}} - M_{\rm{gas}}^{\rm{MZR}}) / M_{\rm{gas}}^{\rm{MZR}} = 0.37 \pm 0.22$. While the dispersion is relatively small, the MZR $M_{\rm{gas}}$ estimates are systematically lower than those from the FMR. This is expected as the FMR dependency on SFR results in galaxies with a higher SFR to have lower metallicities at a fixed stellar mass and, thus, higher $\delta_{\rm{GDR}}$ and $M_{\rm{gas}}$.

In any case, metallicities are poorly constrained in general for DSFGs, SBs and at $z > 3$ \citep[e.g.,][]{tan13,kewley13,steidel14}. The extent to which the MZR or FMR are preferable for DSFGs is an open question and beyond the scope of this paper. In this work the estimates reported and used hereafter are those from MZR. We opted for this choice as the scaling relations for $\tau_{\rm{dep}}$ and $f_{\rm{gas}}$ from \citet{tacconi18} employed in the following Sects.~\ref{sec:scl_rel} and \ref{sec:compact} used the MZR and, particularly, that of \citet{genzel15}, allowing for a direct comparison. The MZR in \citet{genzel15} accounts for a redshift evolution and it is applied in \citet{tacconi18} for galaxies at $0 < z < 4$ and stellar masses ranging $\log (M_{\rm{*}}/M_{\odot}) = 9.0$--11.8, which span almost the totality of our galaxy sample. In addition, MZR was also the choice for the follow up on the GOODS-ALMA 1.0 galaxy sample presented in \citet{franco20b}, permitting a consistent analysis and direct comparison. Regardless, the results presented in the following Sects.~\ref{sec:scl_rel} and \ref{sec:compact} using the MZR are reproduced for the FMR and a fixed solar metallicity in Appendix~\ref{sec:appendix_c}.

Another possibility to derive $M_{\rm{gas}}$ via the dust emission is by employing a single-band measurement of the dust continuum flux in the Rayleigh-Jeans (RJ) side of the IR SED \citep[e.g.,][]{scoville14,groves15,schinnerer16}. For this technique we followed \citet{scoville16}, resulting in $M_{\rm{gas}}$ estimates consistent on average within $25\%$ and a relatively large dispersion compared to those obtained from the $\delta_{\rm{GDR}}$--$Z$ technique, both with the MZR and the FMR. The median relative differences are $(M_{\rm{gas}}^{\rm{RJ}} - M_{\rm{gas}}^{\rm{GDR}-Z(MZR)}) / M_{\rm{gas}}^{\rm{GDR}-Z(MZR)} = 0.23 \pm 0.84$ and $(M_{\rm{gas}}^{\rm{RJ}} - M_{\rm{gas}}^{\rm{GDR}-Z(FMR)}) / M_{\rm{gas}}^{\rm{GDR}-Z(FMR)} = -0.15 \pm 0.87$. We note that the RJ technique also accounts for the total gas budget of the galaxies, including $M_{\rm{H_2}}$ and $M_{\rm{HI}}$. While the RJ technique is a very powerful tool to derive $M_{\rm{gas}}$ estimates for large galaxy samples at a relatively cheap observational cost, it has limitations as it does not take into account the evolution of the gas-phase metallicity or the mass-weighted $T_{\rm{dust}}$ of galaxies with redshift at a fixed stellar mass \citep[e.g.,][]{genzel15,berta16,schinnerer16,magdis17,harrington21}. Therefore, we adopted $M_{\rm{gas}}$ estimates from the $\delta_{\rm{GDR}}$--$Z$ technique instead.

In comparison with the estimates in \citet{franco20b} for the sources in common with this work, $M_{\rm{gas}}$ estimates are on average $10\%$ systematically higher with a relatively small dispersion of $30\%$. This systematic difference is in perfect agreement with the systematically $10\%$ higher 1.1\,mm flux densities reported in \citet{gomezguijarro21} compared to those in \citet{franco18,franco20a}

\subsection{Dust temperatures} \label{subsec:tdust}

Complementary to the IR SED fits in Sect.~\ref{subsec:fir_sed}, we also fit the mid-IR to mm photometry of the galaxies with a \textit{Herschel} counterpart employing a single-temperature optically-thin modified blackbody (MBB) with the purpose of deriving luminosity-weighted $T_{\rm{dust}}$ estimates \citep[e.g.,][]{hwang10,magdis11,magdis12,magnelli12,magnelli14}. The MBB is given by:

\begin{equation}
\label{eq:tdust}
S_{\nu} \propto \frac{\nu^{3 + \beta}}{e^{\frac{h \nu}{k T_{\rm{dust}}}} - 1}
\end{equation}

\noindent where $T_{\rm{dust}}$ is the effective dust temperature and $\beta$ is the effective dust emissivity index.

Following the criteria of \citet{hwang10}, also applied by \citet{franco20b}: 1) we considered observed datapoints at $\lambda_{\rm{rf}} \geq 0.55$\,$\lambda_{\rm{peak}}$ to avoid emission from small grains that dominate the dust emission at shorter wavelengths, with $\lambda_{\rm{peak}}$ given by the \texttt{Stardust} fit; 2) we required at least one datapoint at $0.55$\,$\lambda_{\rm{peak}} \leq \lambda_{\rm{rf}} \leq \lambda_{\rm{peak}}$ (met by all galaxies except two, namely A2GS20 and A2GS53); and 3) we required at least one datapoint at $\lambda_{\rm{rf}} \geq \lambda_{\rm{peak}}$ (excluding the radio data dominated by synchrotron emission). We assumed a fixed value of $\beta = 1.5$ typical for DSFGs \citep[e.g.,][]{hildebrand83,kovacs06,gordon10}.

In Tables~\ref{tab:sample_prop_100pur} and \ref{tab:sample_prop_prior} we present $T_{\rm{dust}}$ estimates for the galaxies with a \textit{Herschel} counterpart and the result of the MBB fit is depicted in Appendix~\ref{sec:appendix_a}. The uncertainties were calculated performing 10\,000 Monte-Carlo simulations perturbing the photometry randomly within the uncertainties. There are some galaxies lacking a $T_{\rm{dust}}$ estimate, in summary: 1) galaxies without a \textit{Herschel} counterpart for which there was not enough information to attempt a $T_{\rm{dust}}$ estimation (A2GS8, A2GS10, A2GS17, A2GS18, A2GS27, A2GS29, A2GS33, A2GS37, A2GS38, and A2GS40); 2) galaxies with a \textit{Herschel} counterpart for which the mid-IR to mm photometry was not satisfactory (A2GS7 and A2GS12); 3) galaxies that did not meet the MBB fitting criteria indicated above (A2GS20 and A2GS53); 4) galaxies for which the MBB fit did not converge (A2GS25, A2GS26, A2GS35, A2GS44, A2GS48, A2GS50, A2GS51, A2GS64, and A2GS81), typically galaxies with only SPIRE 250\,$\mu$m and ALMA 1.1\,mm data points.

In comparison with the estimates in \citet{franco20b} for the sources in common with this work, $T_{\rm{dust}}$ estimates are perfectly consistent on average with a very low dispersion of $10\%$.

\subsection{AGN} \label{subsec:agn}

Tracking the presence of AGN within our galaxy sample is interesting to unveil potential signs of coevolution between AGN and star formation activity, but it is also important to check potential biases in the derived dust and stellar-based properties in the galaxies identified as AGN. We identified AGN based on X-ray, IR, or radio excesses that can not be explained based solely by star formation. First, we searched for X-ray-bright AGN by cross matching our galaxy sample with the \textit{Chandra} Deep Field-South 7\,Ms source catalog \citep{luo17}. Galaxies identified as AGN were those with a X-ray counterpart with an intrinsic X-ray luminosity $\log (L_{\rm{X,int}}/L_{\odot}) > 42.5$ (absorption-corrected soft and hard X-ray luminosity, typically applied as a threshold for X-ray bright AGN \citep[e.g.,][]{luo17}). We note that for the galaxies for which the redshifts differ from those in the X-ray catalog, we corrected the intrinsic X-ray luminosity assuming an intrinsic X-ray photon index $\Gamma = 1.8$ \citep{luo17}. 24/88 galaxies were identified as AGN based on their X-ray excess. However, it is important to note that X-ray emission from massive X-ray binaries in galaxies with high SFR such as those studied in this work could be comparable to the level expected for X-ray-bright AGN \citep[e.g.,][]{mineo12}. Therefore, it is plausible that in some of these galaxies identified as AGN based on the classical widespread definition for X-ray excess, the latter could be explained based solely by star formation or at least a significant fraction of it. Second, we also checked for IR-bright AGN defined as those galaxies showing a significant AGN contribution to the total IR luminosity $f_{\rm{AGN}} > 20\%$ (as derived from the IR SED fitting in Sect.~\ref{subsec:fir_sed}, typically applied as a threshold for IR-bright AGN, above which their effects are significant in the determination of $L_{\rm{IR}}$ and SFR \citep[e.g.,][]{valentino21}). While there is a small level of AGN contribution to the IR SED in some galaxies (see Figs.~\ref{fig:fir_seds_1} and \ref{fig:fir_seds_2}), for none of them this level is large enough to be classified as AGN. Finally, we also checked for radio-bright AGN applying a radio-excess criterion \citep[e.g.,][]{donley05,delmoro13,delvecchio17}. Galaxies identified as AGN were those having a significantly low IR/radio ratio ($q_{\rm{IR}}$) located $> 3\sigma$ below the scatter of the infrared-radio correlation \citep[IRRC; e.g.,][]{dejong85,helou85,condon92}, following the IRRC definition of \citet{delvecchio21} for a given stellar mass and redshift. While some galaxies display a small radio excess compared to the predictions of the IRRC for their stellar mass and redshift (see Figs.~\ref{fig:fir_seds_1} and \ref{fig:fir_seds_2}), for none of them this level is large enough to be classified as AGN. Although not classified as radio-bright AGN, A2GS34 exhibits an inverted radio spectra, which could be a sign of a compact radio core \citep[e.g.,][]{kellermann69}. Nevertheless, A2GS34 was already classified as a X-ray-bright AGN. In summary, we find evidence for AGN in 27\% of the galaxies in our sample.

\subsection{Completeness and selection limits} \label{subsec:compl}

\begin{figure}
\begin{center}
\includegraphics[width=\columnwidth]{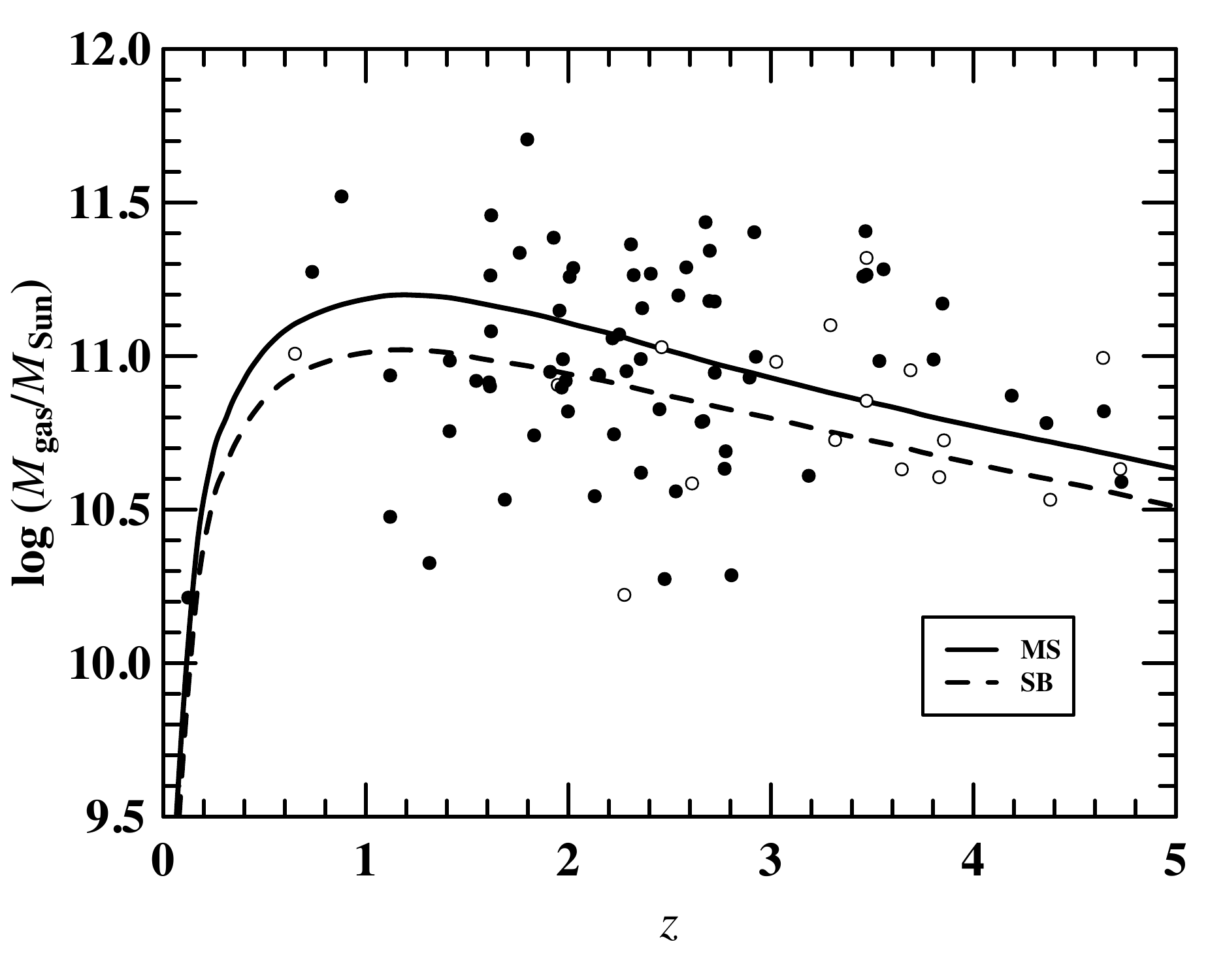}
\caption{$M_{\rm{gas}}$ versus redshift for our GOODS-ALMA 2.0 galaxy sample. Filled symbols represent galaxies with a \textit{Herschel} counterpart and open symbols galaxies without a \textit{Herschel} counterpart. The black line indicates the $M_{\rm{gas}}$ completeness limit as a function of redshift (solid and dashed for MS and SB templates, respectively). This limit correspond to the GOODS-ALMA 2.0 $\sim 100\%$ completeness, associated with flux densities $S_{1.1\rm{mm}} > 1$\,mJy for sources with dust continuum sizes up to 1\arcsec~FWHM \citep[see][]{gomezguijarro21} for the median metallicity of our galaxy sample.}
\label{fig:mgas_compl}
\end{center}
\end{figure}

The GOODS-ALMA 2.0 survey reaches a $\sim 100\%$ completeness at flux densities $S_{1.1\rm{mm}} > 1$\,mJy for sources with dust continuum sizes up to 1\arcsec~FWHM \citep[see][]{gomezguijarro21}. In Fig.~\ref{fig:mgas_compl} we show the $M_{\rm{gas}}$ limit as a function of redshift that corresponds to this flux density completeness threshold. This $M_{\rm{gas}}$ limit was calculated from the $M_{\rm{dust}}$ limit by using both a MS and SB template from the dust SED libraries by \citet{schreiber18} normalized at $S_{1.1\rm{mm}} = 1$\,mJy as a function of redshift. Then, we employed the $\delta_{\rm{GDR}}$--$Z$ technique for the median metallicity of our galaxy sample as calculated from the MZR. Lower (higher) metallicities would lead to higher (lower) $\delta_{\rm{GDR}}$ and $M_{\rm{gas}}$ (if the FMR was assumed, metallicities would be also systematically lower at a fixed stellar mass and, thus, leading to higher $\delta_{\rm{GDR}}$ and $M_{\rm{gas}}$ as well).

As a sanity test, we checked whether the galaxies in our sample were expected to be detected by \textit{Herschel} or not. In order to do this, we employed the dust SED libraries by \citet{schreiber18} to predict flux ratios between ALMA 1.1\,mm and the six \textit{Herschel} bands of PACS (70, 100, 160\,$\mu$m) and SPIRE (250, 350, 500\,$\mu$m). These flux ratio predictions were calculated for both MS ($\Delta \rm{MS} = 1$) and SB ($\Delta \rm{MS} = 5$) templates as a function of redshift. In Appendix~\ref{sec:appendix_b} we show them in Fig.~\ref{fig:fluxratio_pred} and they are tabulated in steps of 0.1 in redshift in Table~\ref{tab:fluxratio_pred}. We compared the flux predictions with the \textit{Herschel} flux limits given by \citet{leiton15}, where the authors reported 2.2\,mJy for PACS 160\,$\mu$m and 2.5\,mJy for SPIRE 250\,$\mu$m at 68\% completeness in GOODS, and values three times above these flux limits resulted in a completeness of $\sim 90\%$. There were only four galaxies for which the SPIRE 250\,$\mu$m flux predictions were above three times these flux limits and, thus, the galaxies should have been detected by \textit{Herschel}. In these cases the PACS 160\,$\mu$m flux predictions were also above three times the flux limits. However, in these four cases, the chosen template between MS and SB was always a SB template, with the fluxes being consistent with the limits for three of them if the chosen template was the MS one instead (namely A2GS10, A2GS40, and A2GS86). It is also important to consider that trends in $T_{\rm{dust}}$ with redshift and $\Delta \rm{MS}$ come from statistical samples of the general population of SFGs \citep[e.g.,][]{magnelli14,bethermin15,schreiber18}, but there are exceptions of newly discovered galaxy populations that might not follow these general trends. Notably, there exists a population of SBs within the scatter of the MS, owing to their higher SFR surface densities compared to normal SFGs in the MS \citep{elbaz18}. These three galaxies could be closer to the MS, but having warmer $T_{\rm{dust}}$ than predicted by the general trends, which could be characteristic of SBs within the MS. At the moment, with only a single datapoint at 1.1\,mm it is not possible to measure $T_{\rm{dust}}$ in these galaxies and confirm their nature and most appropriate template, so we opted for maintaining the general approach and applied the SB template to them.

\begin{table*}
\scriptsize
\caption{Dust and stellar-based properties}
\label{tab:sample_prop_100pur}
\centering
\begin{tabular}{lcccccccccc}
\hline\hline
ID & z & $\log (M_{\rm{*}}/M_{\odot})$ & $\log (L_{\rm{IR}}/L_{\odot})$ & SFR & $\Delta \rm{MS}$ & $\log (M_{\rm{dust}}/M_{\odot})$ & $T_{\rm{dust}}$ & $\log (M_{\rm{gas}}/M_{\odot})$ & $f_{\rm{gas}}$ & $\tau_{\rm{dep}}$ \\  &  &  &  & ($M_{\odot}$ yr$^{-1}$) &  &  & (K) &  &  & (Myr) \\ (1) & (2) & (3) & (4) & (5) & (6) & (7) & (8) & (9) & (10) & (11) \\
\hline
A2GS1   & 2.309 & 11.06 & 12.80 $\pm$ 0.01 & 1093 $\pm$  10 &  5.59 $\pm$ 0.05 & 9.03 $\pm$ 0.01 & 42.4$_{-0.2}^{+0.2}$ & 11.09 $\pm$ 0.20 & 0.52 $\pm$ 0.11 &  112 $\pm$    51 \\
A2GS2   & 3.556 & 11.09 & 12.92 $\pm$ 0.01 & 1428 $\pm$  31 &  3.79 $\pm$ 0.08 & 8.93 $\pm$ 0.01 & 45.0$_{-0.5}^{+1.0}$ & 11.07 $\pm$ 0.20 & 0.49 $\pm$ 0.11 &   82 $\pm$    37 \\
A2GS3$\dagger$   & 2.582 & 11.44 & 12.86 $\pm$ 0.01 & 1251 $\pm$  13 &  2.96 $\pm$ 0.03 & 9.03 $\pm$ 0.01 & 44.1$_{-0.3}^{+0.3}$ & 11.03 $\pm$ 0.20 & 0.28 $\pm$ 0.09 &   85 $\pm$    39 \\
A2GS4   & 2.918 & 10.76 & 12.92 $\pm$ 0.01 & 1416 $\pm$  24 & 10.03 $\pm$ 0.17 & 8.97 $\pm$ 0.01 & 46.0$_{-0.3}^{+0.4}$ & 11.17 $\pm$ 0.20 & 0.72 $\pm$ 0.09 &  103 $\pm$    47 \\
A2GS5   & 1.797 & 11.09 & 12.55 $\pm$ 0.01 &  614 $\pm$  11 &  4.37 $\pm$ 0.08 & 9.44 $\pm$ 0.01 & 32.8$_{-0.2}^{+0.2}$ & 11.45 $\pm$ 0.20 & 0.70 $\pm$ 0.10 &  460 $\pm$   210 \\
A2GS6   & 3.46  & 11.14 & 12.88 $\pm$ 0.01 & 1312 $\pm$  29 &  3.25 $\pm$ 0.07 & 8.93 $\pm$ 0.01 & 45.6$_{-0.5}^{+1.2}$ & 11.05 $\pm$ 0.20 & 0.45 $\pm$ 0.11 &   85 $\pm$    39 \\
A2GS7   & 3.467 & 10.52 & 13.00 $\pm$ 0.03 & 1750 $\pm$ 100 & 17.7  $\pm$ 1.0  & 8.95 $\pm$ 0.03 & (...)                & 11.28 $\pm$ 0.20 & 0.85 $\pm$ 0.06 &  108 $\pm$    50 \\
A2GS8(*)   & 3.29  & 11.44 & 12.59 $\pm$ 0.04 &  665 $\pm$  57 &  1.02 $\pm$ 0.09 & 8.91 $\pm$ 0.04 & (...)                & 10.95 $\pm$ 0.20 & 0.24 $\pm$ 0.09 &  133 $\pm$    63 \\
A2GS9   & 2.696 & 10.90 & 12.30 $\pm$ 0.03 &  345 $\pm$  22 &  1.95 $\pm$ 0.12 & 8.92 $\pm$ 0.03 & 31.3$_{-2.4}^{+4.6}$ & 11.06 $\pm$ 0.20 & 0.59 $\pm$ 0.11 &  330 $\pm$   150 \\
A2GS10(*)  & 3.47 & 10.56 & 12.91 $\pm$ 0.03 & 1415 $\pm$  85 & 13.03 $\pm$ 0.78 & 8.86 $\pm$ 0.03 & (...)                & 11.18 $\pm$ 0.20 & 0.80 $\pm$ 0.07 &  106 $\pm$    49 \\
A2GS11$\dagger$  & 2.41  & 11.33 & 12.61 $\pm$ 0.01 &  694 $\pm$  13 &  2.16 $\pm$ 0.04 & 9.05 $\pm$ 0.01 & 39.7$_{-0.3}^{+0.3}$ & 11.06 $\pm$ 0.20 & 0.35 $\pm$ 0.10 &  165 $\pm$    76 \\
A2GS12  & 3.847 & 10.87 & 12.62 $\pm$ 0.03 &  724 $\pm$  52 &  2.89 $\pm$ 0.21 & 8.83 $\pm$ 0.03 & (...)                & 11.05 $\pm$ 0.20 & 0.60 $\pm$ 0.11 &  156 $\pm$    73 \\
A2GS13  & 1.619 & 10.96 & 12.48 $\pm$ 0.01 &  521 $\pm$   3 &  5.27 $\pm$ 0.03 & 9.18 $\pm$ 0.01 & 34.4$_{-0.2}^{+0.2}$ & 11.19 $\pm$ 0.20 & 0.63 $\pm$ 0.11 &  300 $\pm$   140 \\
A2GS14  & 1.956 & 11.19 & 12.32 $\pm$ 0.01 &  361 $\pm$   4 &  1.95 $\pm$ 0.02 & 8.97 $\pm$ 0.01 & 36.3$_{-0.3}^{+0.1}$ & 10.97 $\pm$ 0.20 & 0.38 $\pm$ 0.11 &  260 $\pm$   120 \\
A2GS15  & 3.47 & 10.24 & 12.36 $\pm$ 0.02 &  391 $\pm$  22 &  7.53 $\pm$ 0.42 & 8.81 $\pm$ 0.02 & 39.2$_{-1.7}^{+1.3}$ & 11.25 $\pm$ 0.20 & 0.91 $\pm$ 0.04 &  460 $\pm$   210 \\
A2GS16$\dagger$  & 2.45  & 11.39 & 12.33 $\pm$ 0.03 &  364 $\pm$  21 &  1.01 $\pm$ 0.06 & 8.69 $\pm$ 0.03 & 40.7$_{-0.9}^{+1.1}$ & 10.69 $\pm$ 0.20 & 0.17 $\pm$ 0.06 &  130 $\pm$    62 \\
A2GS17(*)  & 4.64  & 10.39 & 12.84 $\pm$ 0.04 & 1200 $\pm$ 110 & 11.5  $\pm$ 1.1  & 8.54 $\pm$ 0.04 & (...)                & 11.00 $\pm$ 0.20 & 0.80 $\pm$ 0.07 &   84 $\pm$    40 \\
A2GS18(*)  & 3.689 & 10.56 & 12.63 $\pm$ 0.05 &  738 $\pm$  79 &  6.33 $\pm$ 0.68 & 8.52 $\pm$ 0.05 & (...)                & 10.85 $\pm$ 0.20 & 0.66 $\pm$ 0.10 &   96 $\pm$    46 \\
A2GS19$\dagger$  & 2.225 & 11.43 & 12.39 $\pm$ 0.01 &  427 $\pm$  10 &  1.35 $\pm$ 0.03 & 8.60 $\pm$ 0.01 & 43.4$_{-0.7}^{+0.6}$ & 10.57 $\pm$ 0.20 & 0.12 $\pm$ 0.05 &   85 $\pm$    39 \\
A2GS20  & 2.68  & 10.72 & 11.96 $\pm$ 0.04 &  157 $\pm$  13 &  1.34 $\pm$ 0.11 & 9.20 $\pm$ 0.04 & (...)                & 11.39 $\pm$ 0.20 & 0.82 $\pm$ 0.07 & 1540 $\pm$   730 \\
A2GS21  & 2.698 & 10.30 & 12.57 $\pm$ 0.01 &  639 $\pm$  19 & 14.24 $\pm$ 0.43 & 8.89 $\pm$ 0.01 & 37.1$_{-0.7}^{+1.6}$ & 11.22 $\pm$ 0.20 & 0.90 $\pm$ 0.04 &  260 $\pm$   120 \\
A2GS22  & 2.72  & 10.77 & 12.22 $\pm$ 0.02 &  286 $\pm$  14 &  2.14 $\pm$ 0.11 & 8.90 $\pm$ 0.02 & 35.4$_{-1.8}^{+1.0}$ & 11.07 $\pm$ 0.20 & 0.67 $\pm$ 0.10 &  410 $\pm$   190 \\
A2GS23  & 2.00  & 10.76 & 12.25 $\pm$ 0.01 &  306 $\pm$   8 &  3.33 $\pm$ 0.09 & 8.97 $\pm$ 0.01 & 35.3$_{-1.4}^{+2.9}$ & 11.08 $\pm$ 0.20 & 0.67 $\pm$ 0.10 &  390 $\pm$   180 \\
A2GS24  & 2.32  & 11.38 & 12.33 $\pm$ 0.01 &  364 $\pm$  12 &  1.13 $\pm$ 0.04 & 9.13 $\pm$ 0.01 & 36.0$_{-0.4}^{+0.9}$ & 11.12 $\pm$ 0.20 & 0.35 $\pm$ 0.10 &  360 $\pm$   160 \\
A2GS25  & 2.543 & 10.13 & 12.81 $\pm$ 0.01 & 1099 $\pm$  11 & 38.65 $\pm$ 0.39 & 8.82 $\pm$ 0.01 & 48.9$_{-0.5}^{+0.5}$ & 11.21 $\pm$ 0.20 & 0.92 $\pm$ 0.03 &  145 $\pm$    66 \\
A2GS26  & 1.927 & 10.64 & 11.55 $\pm$ 0.04 &   61 $\pm$   6 &  0.89 $\pm$ 0.09 & 9.22 $\pm$ 0.04 & (...)                & 11.36 $\pm$ 0.20 & 0.84 $\pm$ 0.06 & 3700 $\pm$  1800 \\
A2GS27(*)  & 4.72  & 11.01 & 12.39 $\pm$ 0.08 &  426 $\pm$  77 &  0.96 $\pm$ 0.17 & 8.39 $\pm$ 0.08 & (...)                & 10.61 $\pm$ 0.21 & 0.29 $\pm$ 0.10 &   96 $\pm$    50 \\
A2GS28  & 1.967 & 10.71 & 12.22 $\pm$ 0.02 &  283 $\pm$  10 &  3.48 $\pm$ 0.12 & 8.60 $\pm$ 0.02 & 38.1$_{-0.8}^{+0.9}$ & 10.72 $\pm$ 0.20 & 0.50 $\pm$ 0.11 &  183 $\pm$    84 \\
A2GS29(*)  & 3.47 & 11.32 & 12.40 $\pm$ 0.05 &  436 $\pm$  54 &  0.75 $\pm$ 0.09 & 8.69 $\pm$ 0.05 & (...)                & 10.76 $\pm$ 0.21 & 0.22 $\pm$ 0.08 &  131 $\pm$    64 \\
A2GS30$\dagger$  & 3.80  & 11.46 & 12.81 $\pm$ 0.01 & 1099 $\pm$  37 &  1.28 $\pm$ 0.04 & 8.75 $\pm$ 0.01 & 59.1$_{-1.6}^{+1.5}$ & 10.81 $\pm$ 0.20 & 0.18 $\pm$ 0.07 &   58 $\pm$    27 \\
A2GS31$\dagger$  & 2.15  & 11.18 & 12.27 $\pm$ 0.01 &  322 $\pm$   8 &  1.51 $\pm$ 0.04 & 8.77 $\pm$ 0.01 & 38.1$_{-0.8}^{+0.9}$ & 10.79 $\pm$ 0.20 & 0.29 $\pm$ 0.09 &  188 $\pm$    86 \\
A2GS32$\dagger$  & 2.251 & 11.55 & 12.39 $\pm$ 0.02 &  426 $\pm$  20 &  1.15 $\pm$ 0.05 & 8.95 $\pm$ 0.02 & 36.8$_{-0.5}^{+0.9}$ & 10.90 $\pm$ 0.20 & 0.18 $\pm$ 0.07 &  188 $\pm$    86 \\
A2GS33(*)  & (...) & (...) &            (...) &          (...) &            (...) &           (...) & (...)                &            (...) &           (...) &            (...) \\
A2GS34$\dagger$  & 1.613 & 11.43 & 12.00 $\pm$ 0.01 &  170 $\pm$   4 &  0.98 $\pm$ 0.02 & 8.84 $\pm$ 0.01 & 33.0$_{-0.4}^{+0.7}$ & 10.78 $\pm$ 0.20 & 0.18 $\pm$ 0.07 &  350 $\pm$   160 \\
A2GS35$\dagger$  & 4.73  & 10.94 & 12.77 $\pm$ 0.02 & 1001 $\pm$  41 &  2.65 $\pm$ 0.11 & 8.23 $\pm$ 0.02 & (...)                & 10.48 $\pm$ 0.20 & 0.26 $\pm$ 0.09 &   30 $\pm$    14 \\
A2GS36$\dagger$  & 2.36  & 11.27 & 12.06 $\pm$ 0.02 &  196 $\pm$  10 &  0.69 $\pm$ 0.03 & 8.52 $\pm$ 0.02 & 34.5$_{-1.0}^{+3.3}$ & 10.54 $\pm$ 0.20 & 0.16 $\pm$ 0.06 &  175 $\pm$    81 \\
A2GS37(*)  & 3.85  & 11.16 & 12.33 $\pm$ 0.07 &  372 $\pm$  63 &  0.76 $\pm$ 0.13 & 8.54 $\pm$ 0.07 & (...)                & 10.67 $\pm$ 0.21 & 0.25 $\pm$ 0.09 &  126 $\pm$    65 \\
A2GS38(*)  & (...) & (...) &            (...) &          (...) &            (...) &           (...) & (...)                &            (...) &           (...) &            (...) \\
A2GS39  & 2.36  & 10.61 & 12.44 $\pm$ 0.01 &  478 $\pm$   7 &  6.02 $\pm$ 0.09 & 8.77 $\pm$ 0.01 & 40.9$_{-0.4}^{+0.4}$ & 10.96 $\pm$ 0.20 & 0.69 $\pm$ 0.10 &  192 $\pm$    88 \\
A2GS40(*)  & 2.46  & 10.21 & 12.40 $\pm$ 0.09 &  432 $\pm$  86 & 13.1 $\pm$ 2.6 & 8.57 $\pm$ 0.09 & (...)               & 10.92 $\pm$ 0.22 & 0.84 $\pm$ 0.07 &  192 $\pm$   100 \\
A2GS41  & 1.759 & 10.54 & 11.88 $\pm$ 0.01 &  129 $\pm$   4 &  2.58 $\pm$ 0.08 & 9.06 $\pm$ 0.01 & 28.3$_{-0.8}^{+1.0}$ & 11.20 $\pm$ 0.20 & 0.82 $\pm$ 0.07 & 1220 $\pm$   560 \\
A2GS42$\dagger$  & 2.29  & 11.12 & 12.13 $\pm$ 0.02 &  233 $\pm$   9 &  1.09 $\pm$ 0.04 & 8.80 $\pm$ 0.02 & 34.9$_{-1.0}^{+2.2}$ & 10.84 $\pm$ 0.20 & 0.35 $\pm$ 0.10 &  300 $\pm$   140 \\
A2GS43  & 3.54  & 10.54 & 12.88 $\pm$ 0.02 & 1319 $\pm$  52 & 12.45 $\pm$ 0.49 & 8.53 $\pm$ 0.02 & 65.8$_{-1.4}^{+3.0}$ & 10.85 $\pm$ 0.20 & 0.67 $\pm$ 0.10 &   54 $\pm$    25 \\
A2GS44$\dagger$  & 4.19  & 11.05 & 12.71 $\pm$ 0.02 &  888 $\pm$  34 &  2.12 $\pm$ 0.08 & 8.56 $\pm$ 0.02 & (...)                & 10.74 $\pm$ 0.20 & 0.33 $\pm$ 0.10 &   62 $\pm$    28 \\
\hline
\end{tabular}
\tablefoot{A2GS1 to A2GS44 correspond to the 100\% pure source catalog in \citet{gomezguijarro21} (1) ALMA source ID with galaxies without a \textit{Herschel} counterpart followed by *; (2) redshift with spectroscopic redshifts shown with three decimal digits \citep[see][]{gomezguijarro21} for the spectroscopic redshift references therein); (3) stellar mass; (4) IR luminosity (8--1000\,$\mu$m rest-frame) as obtained from \texttt{Stardust} for the galaxies with a \textit{Herschel} counterpart accounting for star formation only (without AGN contribution) or from the iterative approach using the dust SED libraries by \citet{schreiber18} for the galaxies without a \textit{Herschel} counterpart (see Sect.~\ref{subsec:fir_sed}); (5) total SFR accounting for the contribution of the obscured star formation probed in the IR ($\rm{SFR_{IR}}$) and the unobscured star formation probed in the UV ($\rm{SFR_{UV}}$); (6) distance to the MS (where the MS is from \citet{schreiber15}) defined as the ratio of the SFR to the SFR of the MS at the same stellar mass and redshift ($\Delta \rm{MS} = \rm{SFR/SFR_{MS}}$); (7) dust mass as obtained from \texttt{Stardust} for the galaxies with a \textit{Herschel} counterpart or from the iterative approach using the dust SED libraries by \citet{schreiber18} for the galaxies without a \textit{Herschel} counterpart (see Sect.~\ref{subsec:fir_sed}); (8) dust temperature obtained from a MBB fit for the galaxies with a \textit{Herschel} counterpart (see Sect.~\ref{subsec:tdust}); (9) gas mass as obtained from the metallicity-dependent gas-to-dust mass ratio technique (see Sect.~\ref{subsec:mgas}); (10) gas fraction ($f_{\rm{gas}} = M_{\rm{gas}}/(M_{\rm{gas}} + M_{\rm{*}})$); (11) depletion timescale ($\tau_{\rm{dep}} = M_{\rm{gas}}/\rm{SFR}$) (see Sect~\ref{subsec:mgas}). A2GS33 and A2GS38 lack any estimate as these are $K_s$ dropouts (A2GS33 a candidate \textit{Spitzer}/IRAC 3.6\,$\mu$m dropout) and, thus, they lack of robust redshift and stellar mass estimates to attempt an IR SED fitting and to derive dust and stellar-based properties \citep[see][]{gomezguijarro21}. We note that among the galaxies with a \textit{Herschel} counterpart there are two galaxies for which the mid-IR to mm photometry is not satisfactory and we treated them following the iterative approach with the dust SED libraries by \citet{schreiber18} (namely A2GS7 and A2GS12). Besides, some of the galaxies lack a $T_{\rm{dust}}$ estimate as they did not meet the MBB fitting criteria or the mid-IR to millimeter photometry was scarce and the fit did not converge (see Sect~\ref{subsec:tdust}). Galaxies classified as SBs in the MS following the definition in Sect.~\ref{sec:scl_rel} are highlighted with $\dagger$.}
\end{table*}

\begin{table*}
\scriptsize
\caption{Dust and stellar-based properties}
\label{tab:sample_prop_prior}
\centering
\begin{tabular}{lcccccccccc}
\hline\hline
ID & z & $\log (M_{\rm{*}}/M_{\odot})$ & $\log (L_{\rm{IR}}/L_{\odot})$ & SFR & $\Delta \rm{MS}$ & $\log (M_{\rm{dust}}/M_{\odot})$ & $T_{\rm{dust}}$ & $\log (M_{\rm{gas}}/M_{\odot})$ & $f_{\rm{gas}}$ & $\tau_{\rm{dep}}$ \\  &  &  &  & ($M_{\odot}$ yr$^{-1}$) &  &  & (K) &  &  & (Myr) \\ (1) & (2) & (3) & (4) & (5) & (6) & (7) & (8) & (9) & (10) & (11) \\
\hline
A2GS45$\dagger$  & 2.77  & 11.06 & 11.99 $\pm$ 0.03 &  168 $\pm$  13 &  0.66 $\pm$  0.05 & 8.50 $\pm$ 0.03 & 37.8$_{-4.8}^{+2.8}$ & 10.59 $\pm$ 0.20 & 0.26 $\pm$ 0.09 &   230 $\pm$   110 \\
A2GS46  & 1.910 & 10.96 & 12.59 $\pm$ 0.01 &  673 $\pm$   5 &  5.37 $\pm$  0.04 & 8.64 $\pm$ 0.01 & 41.6$_{-0.5}^{+0.4}$ & 10.68 $\pm$ 0.20 & 0.35 $\pm$ 0.10 &    72 $\pm$    33 \\
A2GS47(*)  & 3.83  & 10.03 & 12.30 $\pm$ 0.13 &  340 $\pm$ 100 &  9.51 $\pm$  2.86 & 8.16 $\pm$ 0.13 & (...)                & 10.73 $\pm$ 0.24 & 0.83 $\pm$ 0.08 &   156 $\pm$    97 \\
A2GS48  & 2.926 & 10.07 & 12.09 $\pm$ 0.03 &  211 $\pm$  14 &  7.29 $\pm$  0.48 & 8.54 $\pm$ 0.03 & (...)                & 11.00 $\pm$ 0.20 & 0.90 $\pm$ 0.04 &   470 $\pm$   220 \\
A2GS49  & 1.973 & 10.31 & 11.83 $\pm$ 0.02 &  116 $\pm$   6 &  3.50 $\pm$  0.19 & 8.63 $\pm$ 0.02 & 34.8$_{-2.3}^{+5.0}$ & 10.87 $\pm$ 0.20 & 0.78 $\pm$ 0.08 &   580 $\pm$   270 \\
A2GS50  & 2.89  & 10.31 & 11.97 $\pm$ 0.04 &  159 $\pm$  13 &  3.21 $\pm$  0.27 & 8.55 $\pm$ 0.04 & (...)                & 10.91 $\pm$ 0.20 & 0.80 $\pm$ 0.07 &   500 $\pm$   240 \\
A2GS51  & 2.36  & 10.83 & 11.81 $\pm$ 0.04 &  110 $\pm$  11 &  0.85 $\pm$  0.09 & 8.83 $\pm$ 0.04 & (...)                & 10.96 $\pm$ 0.20 & 0.57 $\pm$ 0.11 &   790 $\pm$   380 \\
A2GS52  & 0.88  &  9.54 & 10.83 $\pm$ 0.06 &   12 $\pm$   2 &  4.14 $\pm$  0.60 & 9.08 $\pm$ 0.06 & 20.4$_{-2.0}^{+3.7}$ & 11.38 $\pm$ 0.21 & 0.99 $\pm$ 0.01 & 20000 $\pm$ 10000 \\
A2GS53  & 2.024 & 10.53 & 11.39 $\pm$ 0.06 &   43 $\pm$   6 &  0.75 $\pm$  0.11 & 9.12 $\pm$ 0.06 & (...)                & 11.29 $\pm$ 0.21 & 0.85 $\pm$ 0.06 &  4400 $\pm$  2200 \\
A2GS54  & 0.735 & 10.06 & 11.30 $\pm$ 0.01 &   34 $\pm$   1 &  4.14 $\pm$  0.10 & 8.94 $\pm$ 0.01 & 23.2$_{-0.3}^{+0.4}$ & 11.03 $\pm$ 0.20 & 0.90 $\pm$ 0.04 &  3100 $\pm$  1400 \\
A2GS55  & 1.545 & 10.76 & 11.84 $\pm$ 0.01 &  118 $\pm$   3 &  1.77 $\pm$  0.04 & 8.73 $\pm$ 0.01 & 32.4$_{-0.6}^{+0.5}$ & 10.79 $\pm$ 0.20 & 0.51 $\pm$ 0.11 &   520 $\pm$   240 \\
A2GS56$\dagger$  & 2.78  & 11.04 & 12.40 $\pm$ 0.02 &  437 $\pm$  22 &  1.78 $\pm$  0.09 & 8.45 $\pm$ 0.02 & 48.0$_{-1.3}^{+3.6}$ & 10.55 $\pm$ 0.20 & 0.24 $\pm$ 0.08 &    81 $\pm$    38 \\
A2GS57$\dagger$  & 4.64  & 11.15 & 12.86 $\pm$ 0.01 & 1233 $\pm$  38 &  2.06 $\pm$  0.06 & 8.50 $\pm$ 0.01 & 54.6$_{-1.0}^{+1.1}$ & 10.67 $\pm$ 0.20 & 0.25 $\pm$ 0.09 &    38 $\pm$    17 \\
A2GS58$\dagger$  & 1.83  & 11.03 & 12.14 $\pm$ 0.01 &  239 $\pm$   8 &  1.81 $\pm$  0.06 & 8.57 $\pm$ 0.01 & 38.3$_{-0.6}^{+1.4}$ & 10.59 $\pm$ 0.20 & 0.27 $\pm$ 0.09 &   161 $\pm$    74 \\
A2GS59$\dagger$  & 2.475 & 11.40 & 12.06 $\pm$ 0.03 &  199 $\pm$  12 &  0.54 $\pm$  0.03 & 8.20 $\pm$ 0.03 & 48.3$_{-2.5}^{+3.3}$ & 10.19 $\pm$ 0.20 & 0.06 $\pm$ 0.03 &    78 $\pm$    36 \\
A2GS60  & 1.120 & 10.53 & 11.30 $\pm$ 0.01 &   34 $\pm$   1 &  1.13 $\pm$  0.01 & 8.33 $\pm$ 0.01 & 28.0$_{-0.3}^{+0.2}$ & 10.38 $\pm$ 0.20 & 0.41 $\pm$ 0.11 &   700 $\pm$   320 \\
A2GS61  & 1.615 & 11.40 & 12.16 $\pm$ 0.01 &  249 $\pm$   5 &  1.47 $\pm$  0.03 & 9.17 $\pm$ 0.01 & 31.1$_{-0.6}^{+0.8}$ & 11.11 $\pm$ 0.20 & 0.34 $\pm$ 0.10 &   510 $\pm$   230 \\
A2GS62  & 1.120 & 10.82 & 11.76 $\pm$ 0.01 &   99 $\pm$   1 &  2.04 $\pm$  0.02 & 8.79 $\pm$ 0.01 & 28.1$_{-0.2}^{+0.3}$ & 10.78 $\pm$ 0.20 & 0.47 $\pm$ 0.11 &   600 $\pm$   280 \\
A2GS63  & 3.19  & 10.54 & 12.37 $\pm$ 0.01 &  400 $\pm$  13 &  4.26 $\pm$  0.14 & 8.22 $\pm$ 0.01 & 52.9$_{-1.6}^{+1.1}$ & 10.52 $\pm$ 0.20 & 0.49 $\pm$ 0.11 &    82 $\pm$    38 \\
A2GS64  & 2.67  & 10.67 & 12.07 $\pm$ 0.06 &  204 $\pm$  27 &  1.96 $\pm$  0.26 & 8.51 $\pm$ 0.06 & (...)                & 10.71 $\pm$ 0.21 & 0.52 $\pm$ 0.12 &   250 $\pm$   120 \\
A2GS65$\dagger$  & 2.22  & 11.15 & 12.36 $\pm$ 0.01 &  395 $\pm$  10 &  1.85 $\pm$  0.05 & 8.86 $\pm$ 0.01 & 40.2$_{-0.4}^{+1.0}$ & 10.89 $\pm$ 0.20 & 0.35 $\pm$ 0.10 &   194 $\pm$    89 \\
A2GS66$\dagger$  & 1.686 & 10.81 & 12.03 $\pm$ 0.01 &  185 $\pm$   3 &  2.28 $\pm$  0.04 & 8.31 $\pm$ 0.01 & 37.2$_{-1.2}^{+0.6}$ & 10.37 $\pm$ 0.20 & 0.27 $\pm$ 0.09 &   127 $\pm$    58 \\
A2GS67(*)  & 0.650 & 10.30 & 11.92 $\pm$ 0.08 &  142 $\pm$  26 & 11.81 $\pm$  2.20 & 8.63 $\pm$ 0.08 & (...)                & 10.65 $\pm$ 0.21 & 0.69 $\pm$ 0.11 &   310 $\pm$   170 \\
A2GS68  & 1.413 & 10.10 & 11.60 $\pm$ 0.01 &   69 $\pm$   2 &  4.63 $\pm$  0.13 & 8.37 $\pm$ 0.01 & 35.2$_{-0.8}^{+1.0}$ & 10.59 $\pm$ 0.20 & 0.76 $\pm$ 0.08 &   540 $\pm$   250 \\
A2GS69  & 1.414 & 10.89 & 11.93 $\pm$ 0.01 &  147 $\pm$   2 &  2.00 $\pm$  0.03 & 8.82 $\pm$ 0.01 & 31.4$_{-0.4}^{+0.3}$ & 10.83 $\pm$ 0.20 & 0.46 $\pm$ 0.11 &   460 $\pm$   210 \\
A2GS70(*)  & 2.61  &  9.61 & 12.28 $\pm$ 0.11 &  331 $\pm$  81 & 37.46 $\pm$  9.20 & 8.41 $\pm$ 0.11 & (...)                & 11.05 $\pm$ 0.23 & 0.97 $\pm$ 0.02 &   340 $\pm$   200 \\
A2GS71(*)  & 3.026 & 10.26 & 12.48 $\pm$ 0.07 &  553 $\pm$  82 & 11.90 $\pm$  1.76 & 8.52 $\pm$ 0.07 & (...)                & 10.91 $\pm$ 0.21 & 0.82 $\pm$ 0.07 &   150 $\pm$    75 \\
A2GS72(*)  & 2.28  &  9.37 & 11.99 $\pm$ 0.21 &  171 $\pm$  80 & 38.85 $\pm$ 18.30 & 8.20 $\pm$ 0.21 & (...)                & 10.91 $\pm$ 0.29 & 0.97 $\pm$ 0.02 &   480 $\pm$   390 \\
A2GS73  & 1.987 & 10.15 & 12.25 $\pm$ 0.01 &  309 $\pm$   5 & 13.40 $\pm$  0.22 & 8.46 $\pm$ 0.01 & 42.1$_{-0.7}^{+0.6}$ & 10.77 $\pm$ 0.20 & 0.81 $\pm$ 0.07 &   185 $\pm$    84 \\
A2GS74  & 1.61  & 11.31 & 11.71 $\pm$ 0.02 &   89 $\pm$   4 &  0.58 $\pm$  0.02 & 8.89 $\pm$ 0.02 & 29.3$_{-0.7}^{+2.3}$ & 10.84 $\pm$ 0.20 & 0.25 $\pm$ 0.09 &   780 $\pm$   360 \\
A2GS75$\dagger$  & 1.618 & 11.25 & 12.35 $\pm$ 0.01 &  386 $\pm$   4 &  2.65 $\pm$  0.02 & 8.91 $\pm$ 0.01 & 36.9$_{-0.5}^{+0.3}$ & 10.87 $\pm$ 0.20 & 0.30 $\pm$ 0.10 &   194 $\pm$    88 \\
A2GS76  & 2.53  & 11.24 & 11.76 $\pm$ 0.05 &   98 $\pm$  12 &  0.32 $\pm$  0.04 & 8.52 $\pm$ 0.05 & 34.5$_{-3.0}^{+4.0}$ & 10.55 $\pm$ 0.21 & 0.17 $\pm$ 0.07 &   360 $\pm$   170 \\
A2GS77$\dagger$  & 2.805 & 10.55 & 12.02 $\pm$ 0.03 &  179 $\pm$  13 &  2.14 $\pm$  0.15 & 7.98 $\pm$ 0.03 & 56.0$_{-4.1}^{+7.4}$ & 10.23 $\pm$ 0.20 & 0.32 $\pm$ 0.10 &    91 $\pm$    43 \\
A2GS78(*)  & 3.65  & 10.15 & 12.28 $\pm$ 0.13 &  335 $\pm$  96 &  7.48 $\pm$  2.15 & 8.17 $\pm$ 0.13 & (...)                & 10.67 $\pm$ 0.24 & 0.77 $\pm$ 0.10 &   141 $\pm$    87 \\
A2GS79  & 1.998 & 10.94 & 12.42 $\pm$ 0.01 &  454 $\pm$   5 &  3.52 $\pm$  0.04 & 8.54 $\pm$ 0.01 & 43.7$_{-0.5}^{+0.3}$ & 10.60 $\pm$ 0.20 & 0.32 $\pm$ 0.10 &    87 $\pm$    40 \\
A2GS80$\dagger$  & 1.314 & 11.06 & 11.55 $\pm$ 0.01 &   61 $\pm$   2 &  0.74 $\pm$  0.02 & 8.29 $\pm$ 0.01 & 36.0$_{-1.6}^{+2.0}$ & 10.25 $\pm$ 0.20 & 0.13 $\pm$ 0.05 &   290 $\pm$   130 \\
A2GS81  & 2.66  & 10.63 & 11.75 $\pm$ 0.05 &   97 $\pm$  11 &  1.02 $\pm$  0.11 & 8.57 $\pm$ 0.05 & (...)                & 10.79 $\pm$ 0.20 & 0.59 $\pm$ 0.11 &   630 $\pm$   300 \\
A2GS82(*)  & 4.38  & 10.61 & 12.15 $\pm$ 0.13 &  242 $\pm$  74 &  1.51 $\pm$  0.46 & 8.22 $\pm$ 0.13 & (...)                & 10.57 $\pm$ 0.24 & 0.48 $\pm$ 0.14 &   153 $\pm$    97 \\
A2GS83$\dagger$  & 2.130 & 11.00 & 12.08 $\pm$ 0.01 &  207 $\pm$   5 &  1.32 $\pm$  0.03 & 8.37 $\pm$ 0.01 & 42.1$_{-0.9}^{+0.7}$ & 10.43 $\pm$ 0.20 & 0.21 $\pm$ 0.08 &   123 $\pm$    56 \\
A2GS84  & 4.36  & 10.74 & 12.89 $\pm$ 0.03 & 1322 $\pm$  76 &  6.13 $\pm$  0.35 & 8.35 $\pm$ 0.03 & 66.8$_{-1.0}^{+1.7}$ & 10.66 $\pm$ 0.20 & 0.45 $\pm$ 0.11 &    34 $\pm$    16 \\
A2GS85  & 2.72  & 10.77 & 12.07 $\pm$ 0.03 &  201 $\pm$  12 &  1.51 $\pm$  0.09 & 8.71 $\pm$ 0.03 & 38.7$_{-1.9}^{+2.4}$ & 10.88 $\pm$ 0.20 & 0.56 $\pm$ 0.11 &   380 $\pm$   170 \\
A2GS86(*)  & 1.95  & 10.12 & 12.16 $\pm$ 0.12 &  251 $\pm$  71 & 11.91 $\pm$  3.38 & 8.45 $\pm$ 0.12 & (...)                & 10.76 $\pm$ 0.23 & 0.81 $\pm$ 0.08 &   230 $\pm$   140 \\
A2GS87(*)  & 3.32  & 10.56 & 12.08 $\pm$ 0.13 &  207 $\pm$  60 &  2.01 $\pm$  0.59 & 8.41 $\pm$ 0.13 & (...)                & 10.71 $\pm$ 0.24 & 0.59 $\pm$ 0.13 &   250 $\pm$   150 \\
A2GS88  & 0.123 & 10.21 & 10.35 $\pm$ 0.01 &    4 $\pm$   1 &  0.90 $\pm$  0.01 & 8.13 $\pm$ 0.01 & 22.8$_{-0.1}^{+0.1}$ & 10.05 $\pm$ 0.20 & 0.41 $\pm$ 0.11 &  2900 $\pm$  1300 \\
\hline
\end{tabular}
\tablefoot{A2GS45 to A2GS88 correspond to the prior-based source catalog in \citet{gomezguijarro21}.}
\end{table*}

\section{The main sequence and scaling relations framework in galaxy evolution} \label{sec:scl_rel}

In this section we study the properties derived in Sect.~\ref{sec:properties} in the framework of scaling relations in galaxy evolution. We place them in relation with the MS and the scaling relations for depletion timescales, gas fractions, and dust temperatures.

\begin{figure*}
\begin{center}
\includegraphics[width=\columnwidth]{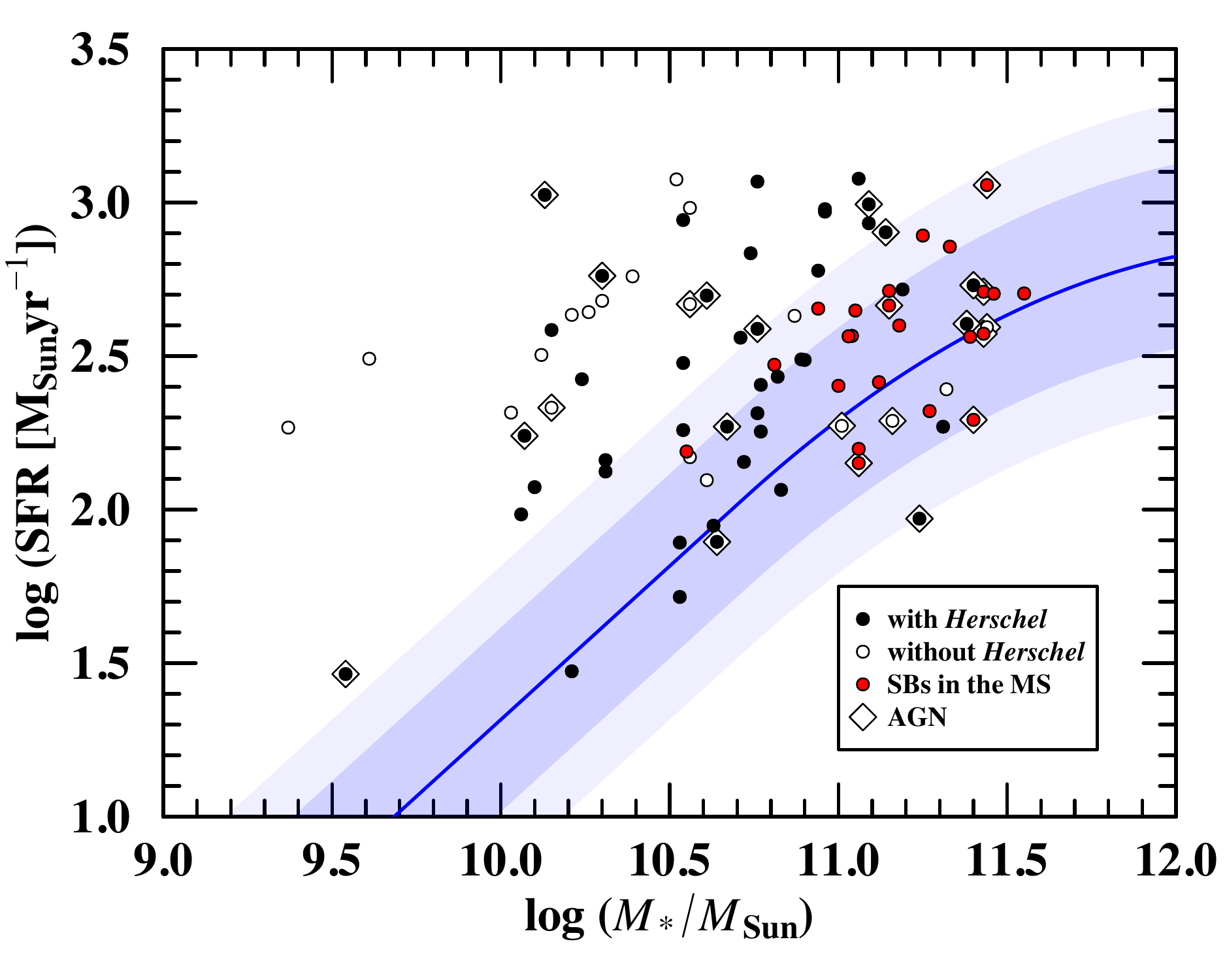}
\includegraphics[width=\columnwidth]{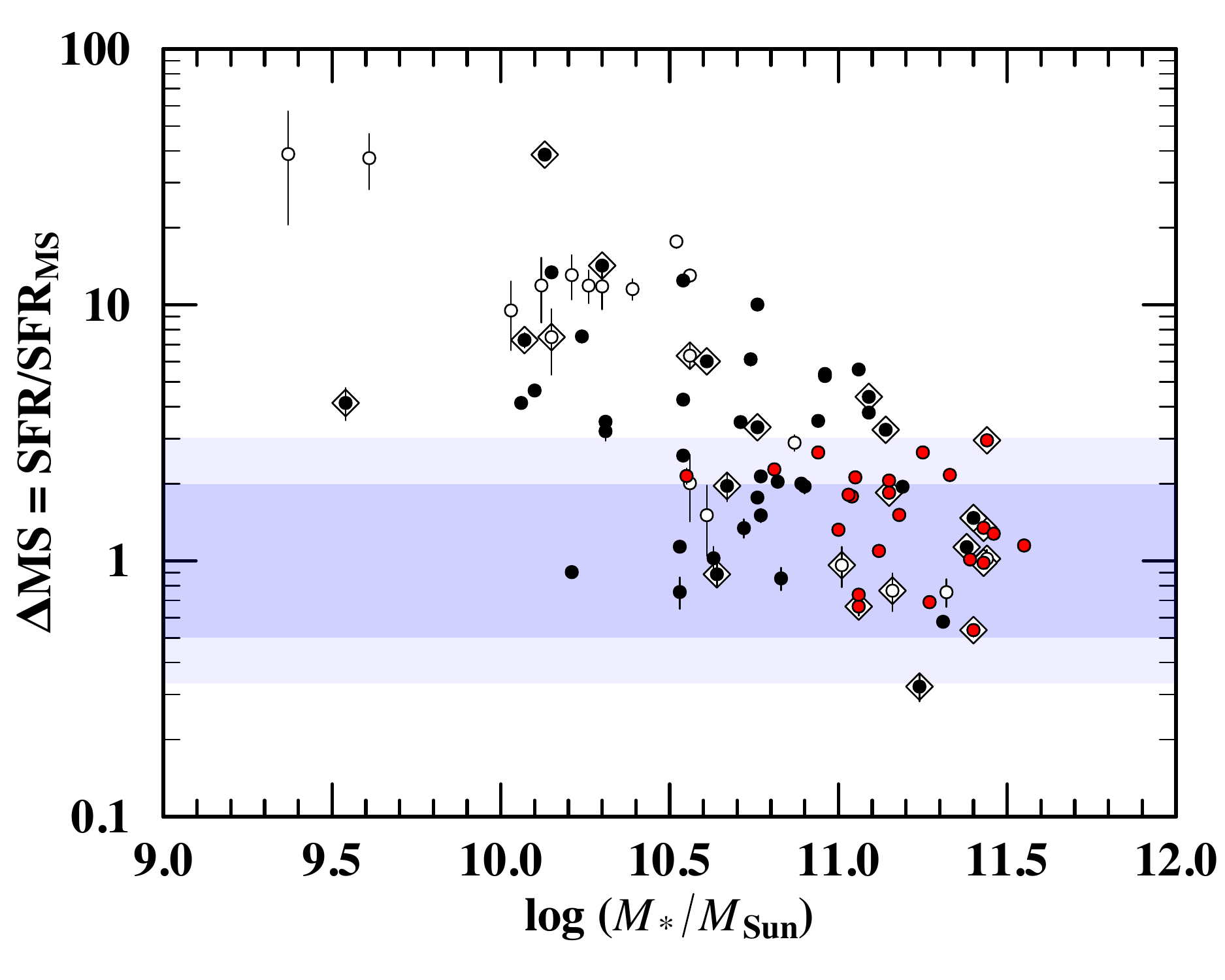}
\caption{Left panel: SFR-$M_{\rm{*}}$ plane with the MS from \citet{schreiber15} displayed as a solid blue line. Its 1$\sigma$ scatter associated with $0.5 < \Delta \rm{MS} < 2$ ($\sim 0.3$\,dex) is represented as a shaded blue area, with a more extended typical scatter of $0.33 < \Delta \rm{MS} < 3$ ($\sim 0.5$\,dex) in lighter blue. We note that the values are scaled to a common redshift ($z_{\rm{med}} = 2.46$) as explained in the main text. Right panel: distance to the MS defined as the ratio of the SFR to the SFR of the MS at a fixed stellar mass and redshift ($\Delta \rm{MS} = \rm{SFR/SFR_{MS}}$). Filled circles represent galaxies with a \textit{Herschel} counterpart and open circles galaxies without a \textit{Herschel} counterpart. SBs in the MS are highlighted in red: galaxies with $\tau_{\rm{dep}}$ below the scatter of the scaling relation and in the MS within a factor 3 (see Sect.~\ref{sec:scl_rel}). Galaxies identified as AGN are highlighted with diamonds.}
\label{fig:ms}
\end{center}
\end{figure*}

In Fig.~\ref{fig:ms} we place our sample in the MS context, showing the location of the galaxies in the SFR-$M_{\rm{*}}$ and $\Delta \rm{MS}$-$M_{\rm{*}}$ planes. The SFR-$M_{\rm{*}}$ plane is the most typical representation of the correlation between SFR and stellar mass of SFGs, so-called MS of SFGs \citep[e.g.,][]{brinchmann04,daddi07,elbaz07,noeske07,whitaker12}. However, given the rise of the MS normalization with increasing redshift \citep[e.g.,][]{noeske07,whitaker12,speagle14,schreiber15,leslie20}, this type of representation is disadvantageous for galaxy samples that span a wide redshift range like ours. In order to overcome this limitation, we scaled the SFR estimates to a common redshift, corresponding to the median value of the sample ($z_{\rm{med}} = 2.46$), using the MS from \citet{schreiber15}. While the scaled SFR values do not reflect the true SFR estimates of the individual galaxies, the overall distribution of the galaxy sample in the SFR-$M_{\rm{*}}$ plane maintains the $\Delta \rm{MS}$ that each galaxy would have if plotted against the MS associated with its redshift. An alternative representation is to simply plot the $\Delta \rm{MS}$-$M_{\rm{*}}$ plane, where the SFR of each galaxy is normalized by the SFR of the MS at its stellar mass and redshift. This type of representation offers a more direct view of the distribution of the galaxy sample in the MS framework without the limitation of the SFR-$M_{\rm{*}}$ plane for galaxy samples that cover a wide redshift range.

Our galaxy sample is located above and within the MS: 42\% above the MS ($\Delta \rm{MS} > 3$) and 58\% are in the MS within a factor 3 ($0.33 < \Delta \rm{MS} < 3$, 0.5\,dex).

In the following, we focused on the subset of galaxies with a \textit{Herschel} counterpart (69/88), so the properties presented in the different plots were obtained using the same methodology. This is particularly relevant in the case of $\tau_{\rm{dep}}$, for which a fixed MS or SB template from the dust SED libraries implies a constant $L_{\rm{IR}} / M_{\rm{dust}}$ ratio and, thus, $\tau_{\rm{dep}} = M_{\rm{gas}} / \rm{SFR} \propto M_{\rm{dust}}/L_{\rm{IR}}$ is also constant. It is very important in the case of $T_{\rm{dust}}$ too, for which there was not enough information to attempt a $T_{\rm{dust}}$ estimation in galaxies without a \textit{Herschel} counterpart. Galaxies with and without a \textit{Herschel} counterpart have on average similar $M_{\rm{gas}}$. Those without a \textit{Herschel} counterpart have $\times 1.5$ higher SFR and $\times 2.4$ lower stellar masses on average. Consequently, galaxies without a \textit{Herschel} counterpart have $\times 1.7$ lower $\tau_{\rm{dep}}$ and $\times 1.7$ higher $f_{\rm{gas}}$ on average compared to those with a \textit{Herschel} counterpart. Therefore, removing galaxies without a \textit{Herschel} counterpart partially removes the low stellar mass end of lower $\tau_{\rm{dep}}$ and higher $f_{\rm{gas}}$ galaxies.

\begin{figure*}
\begin{center}
\includegraphics[width=0.33\textwidth]{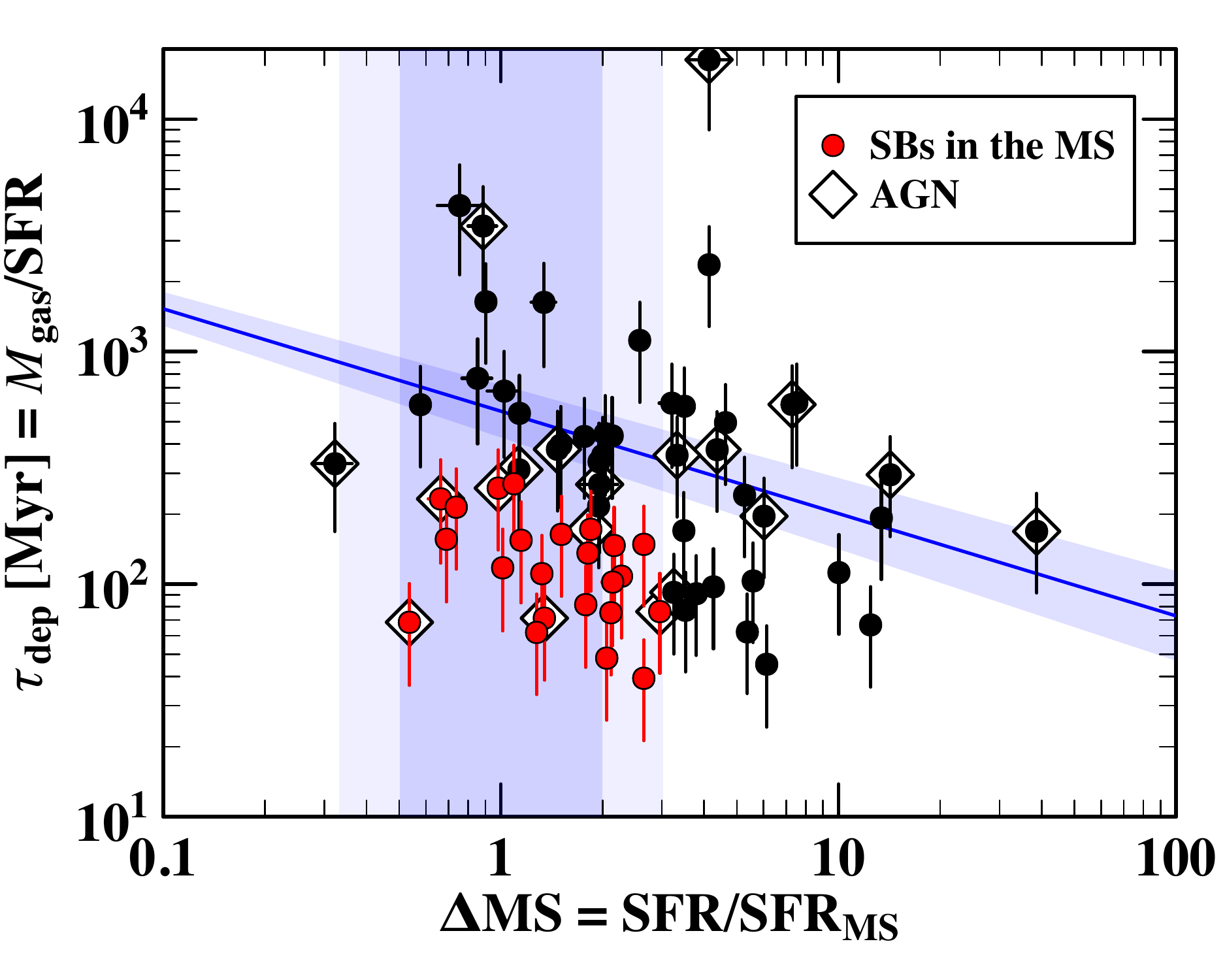}
\includegraphics[width=0.33\textwidth]{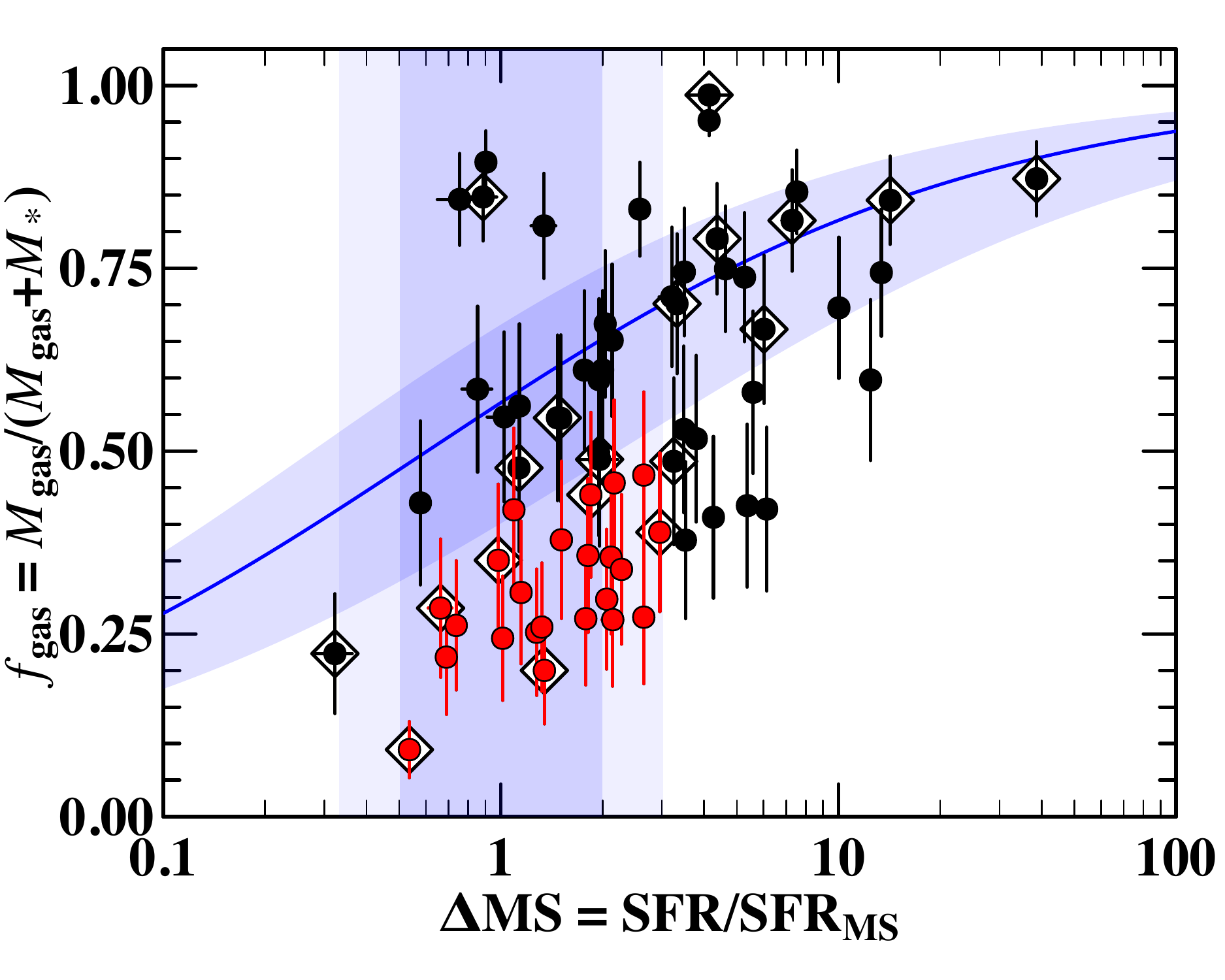}
\includegraphics[width=0.33\textwidth]{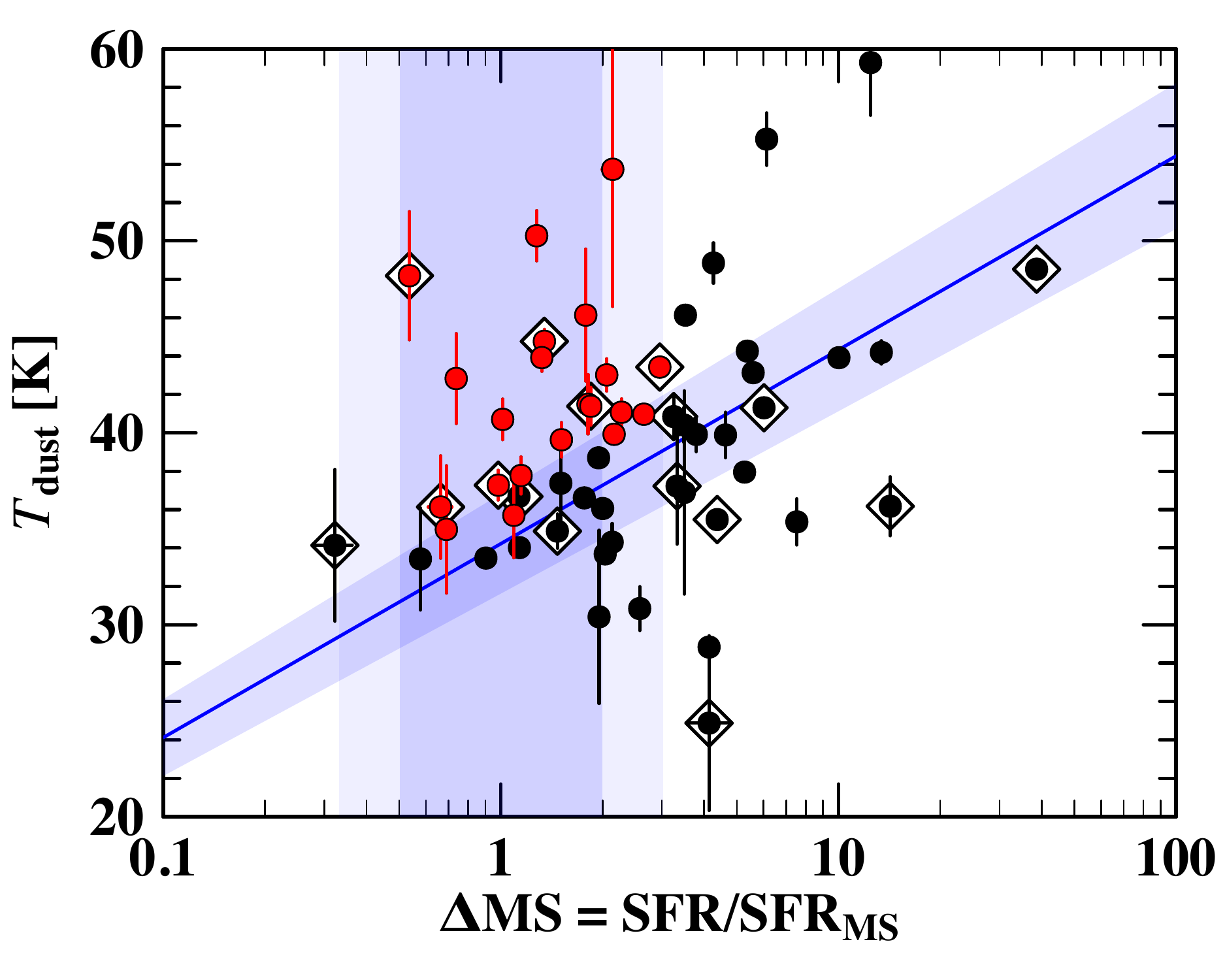}
\includegraphics[width=0.33\textwidth]{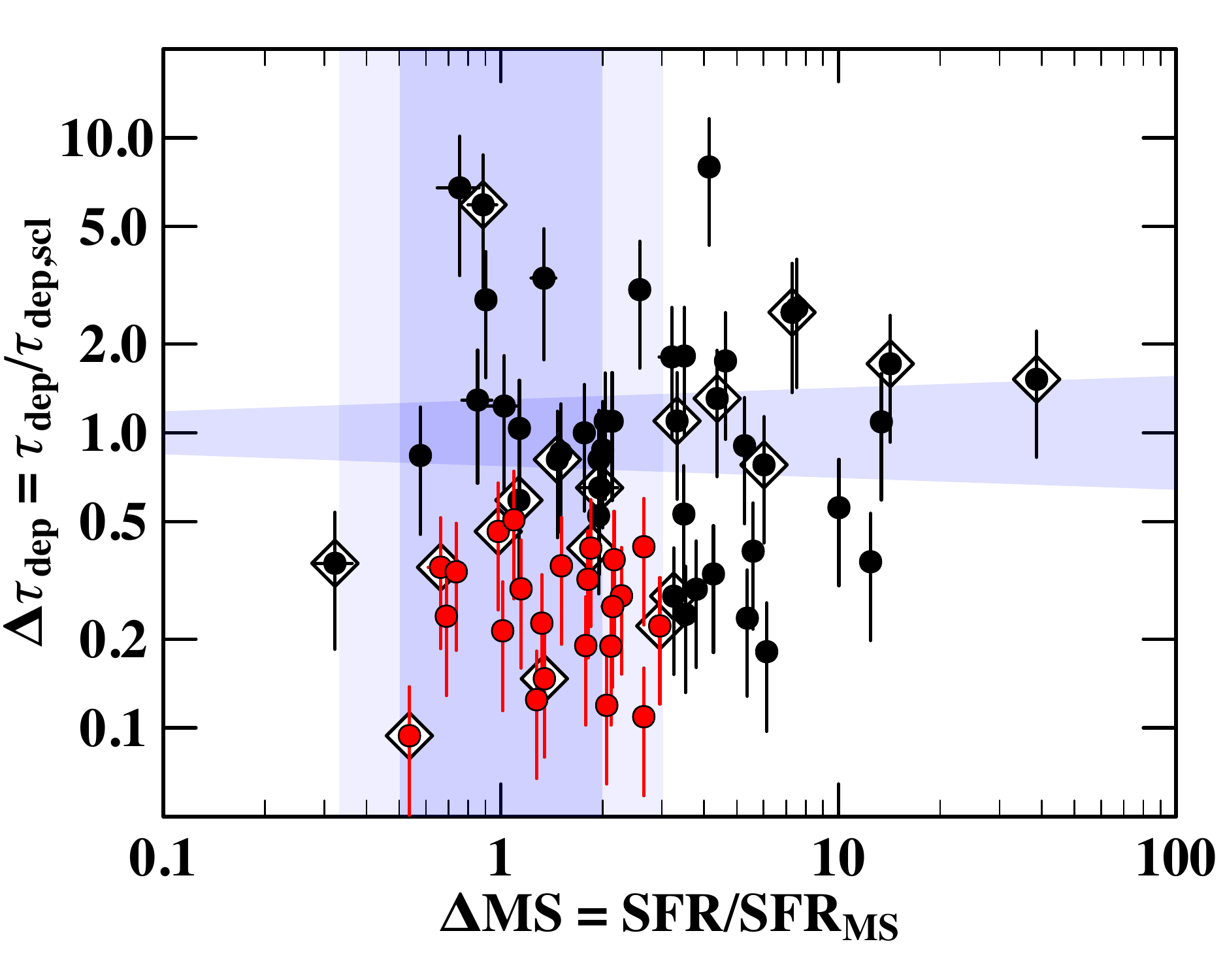}
\includegraphics[width=0.33\textwidth]{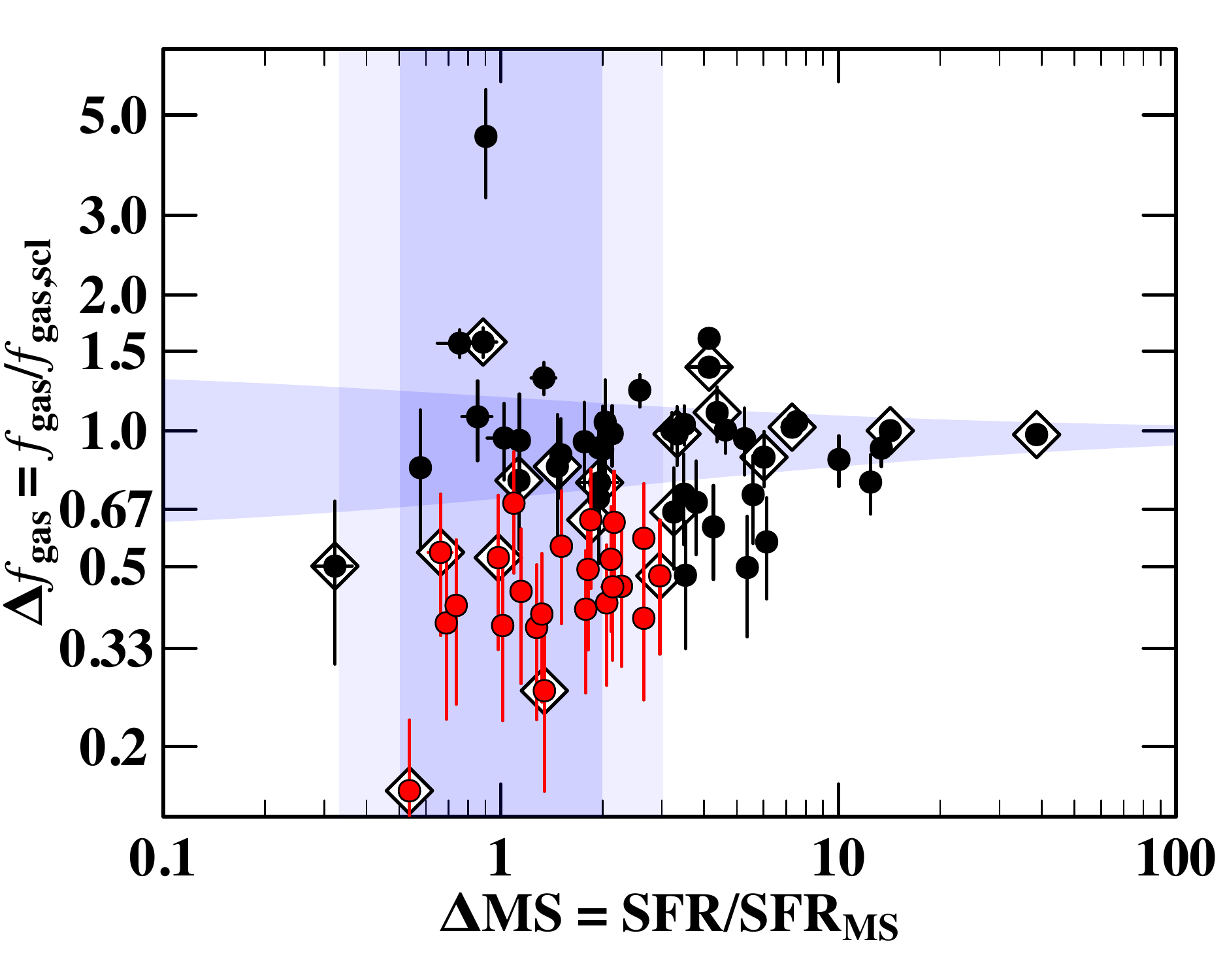}
\includegraphics[width=0.33\textwidth]{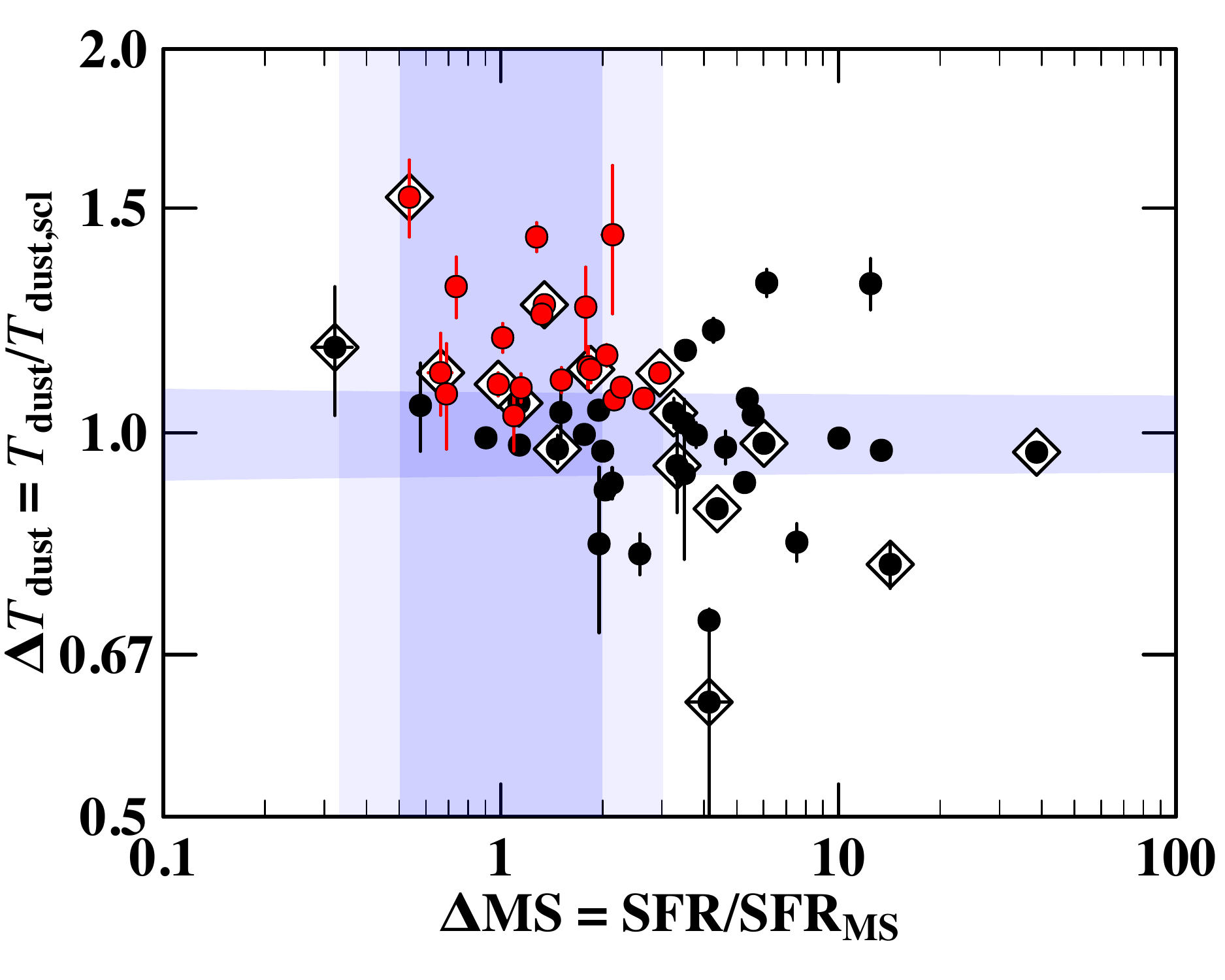}
\includegraphics[width=0.33\textwidth]{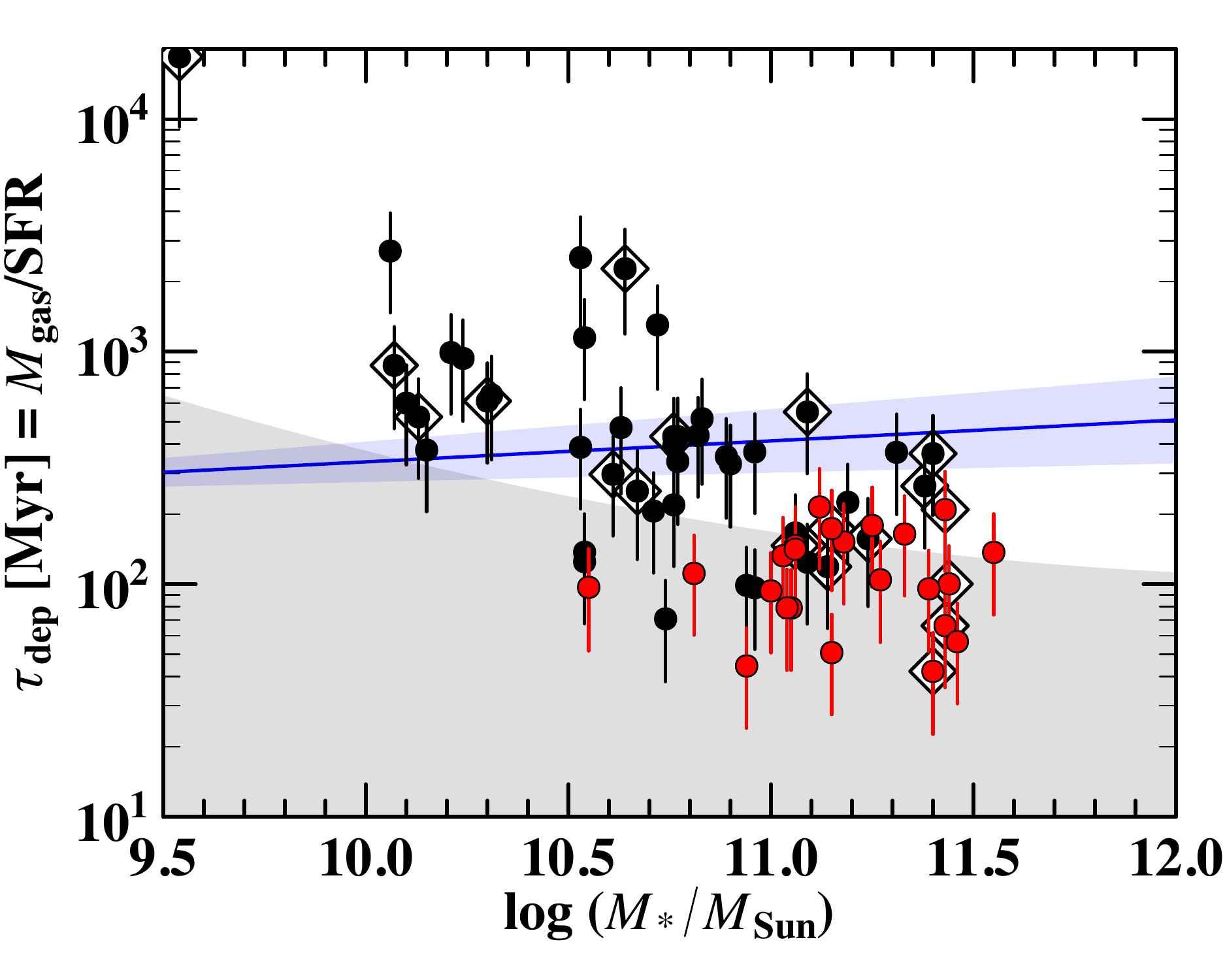}
\includegraphics[width=0.33\textwidth]{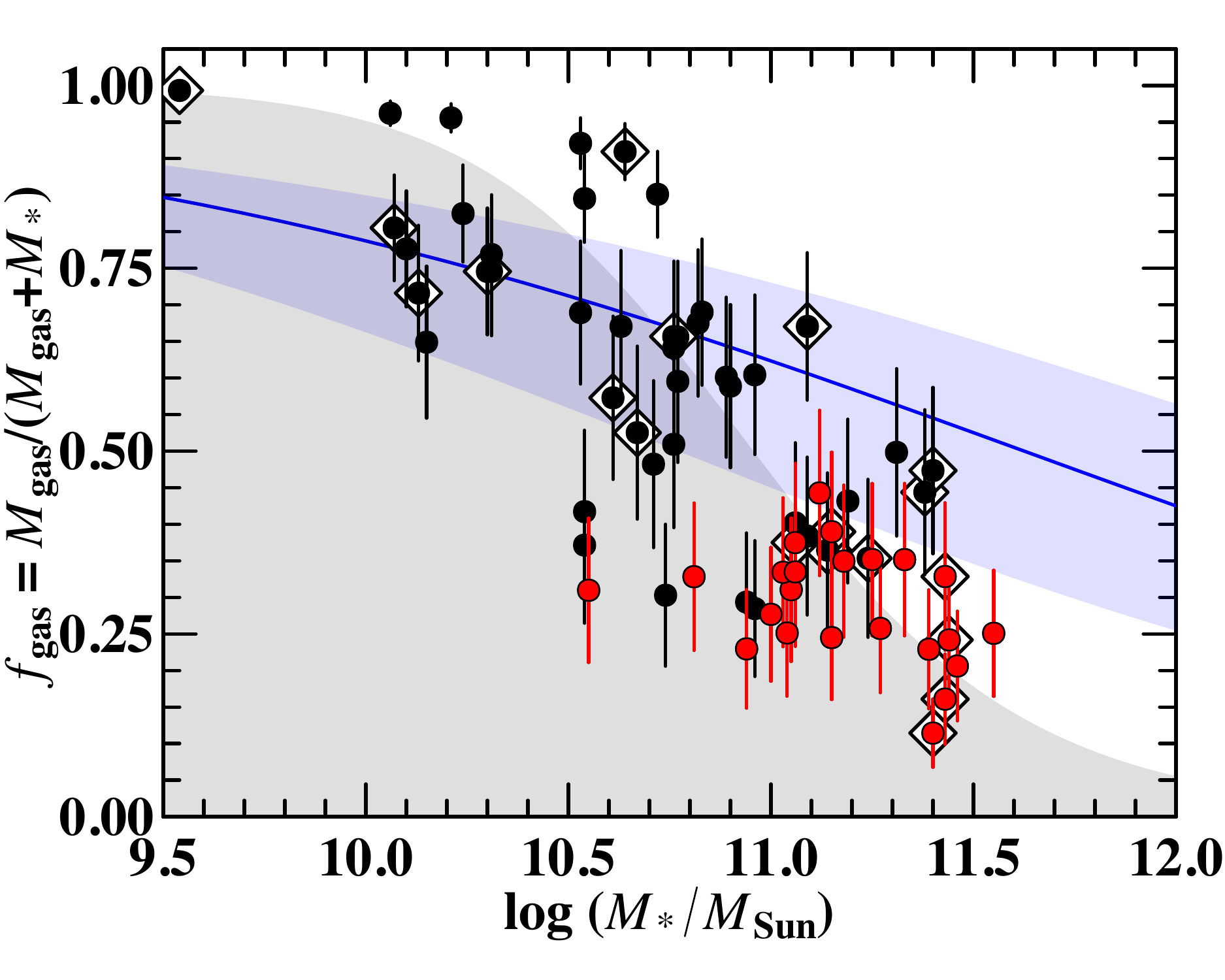}
\includegraphics[width=0.33\textwidth]{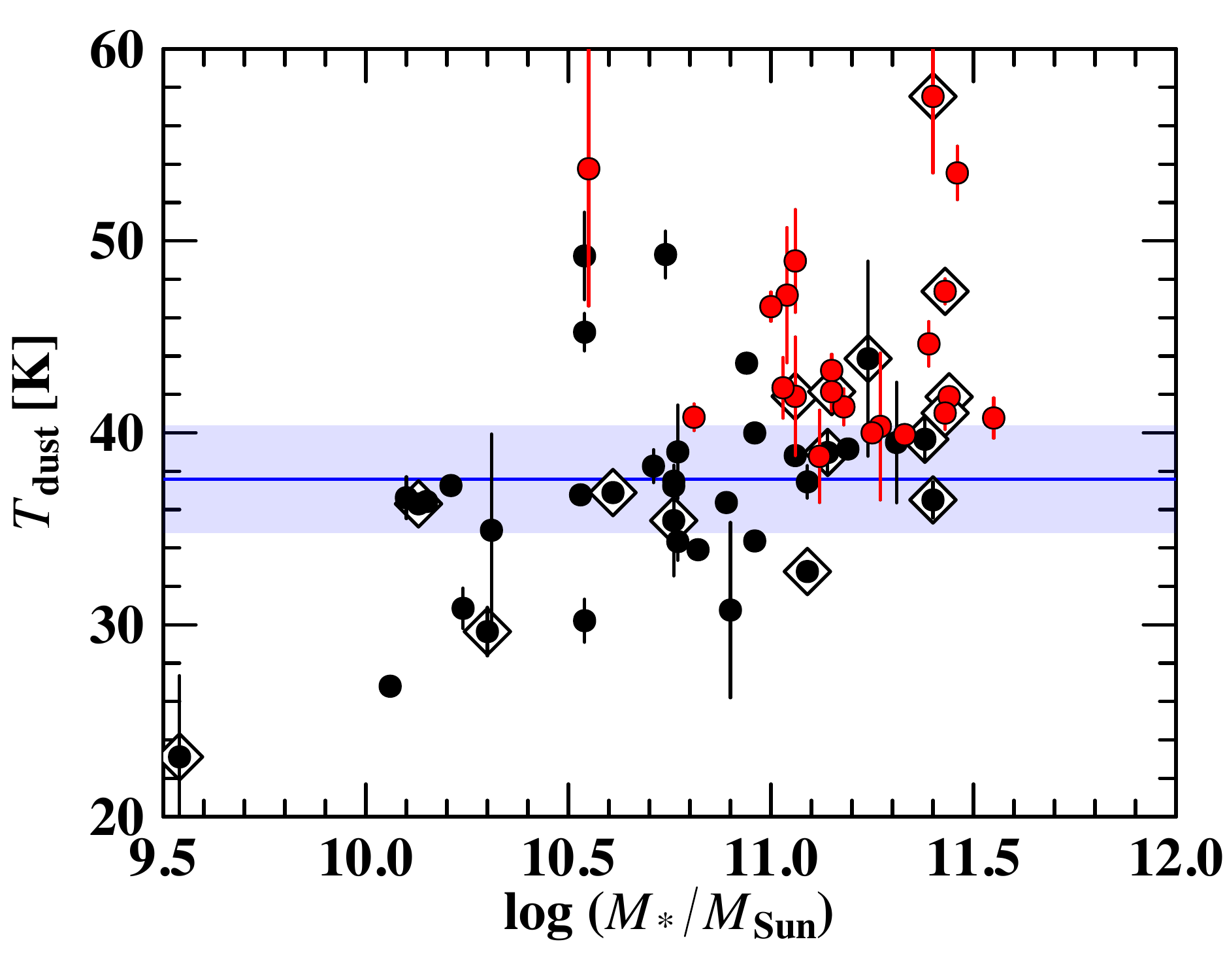}
\includegraphics[width=0.33\textwidth]{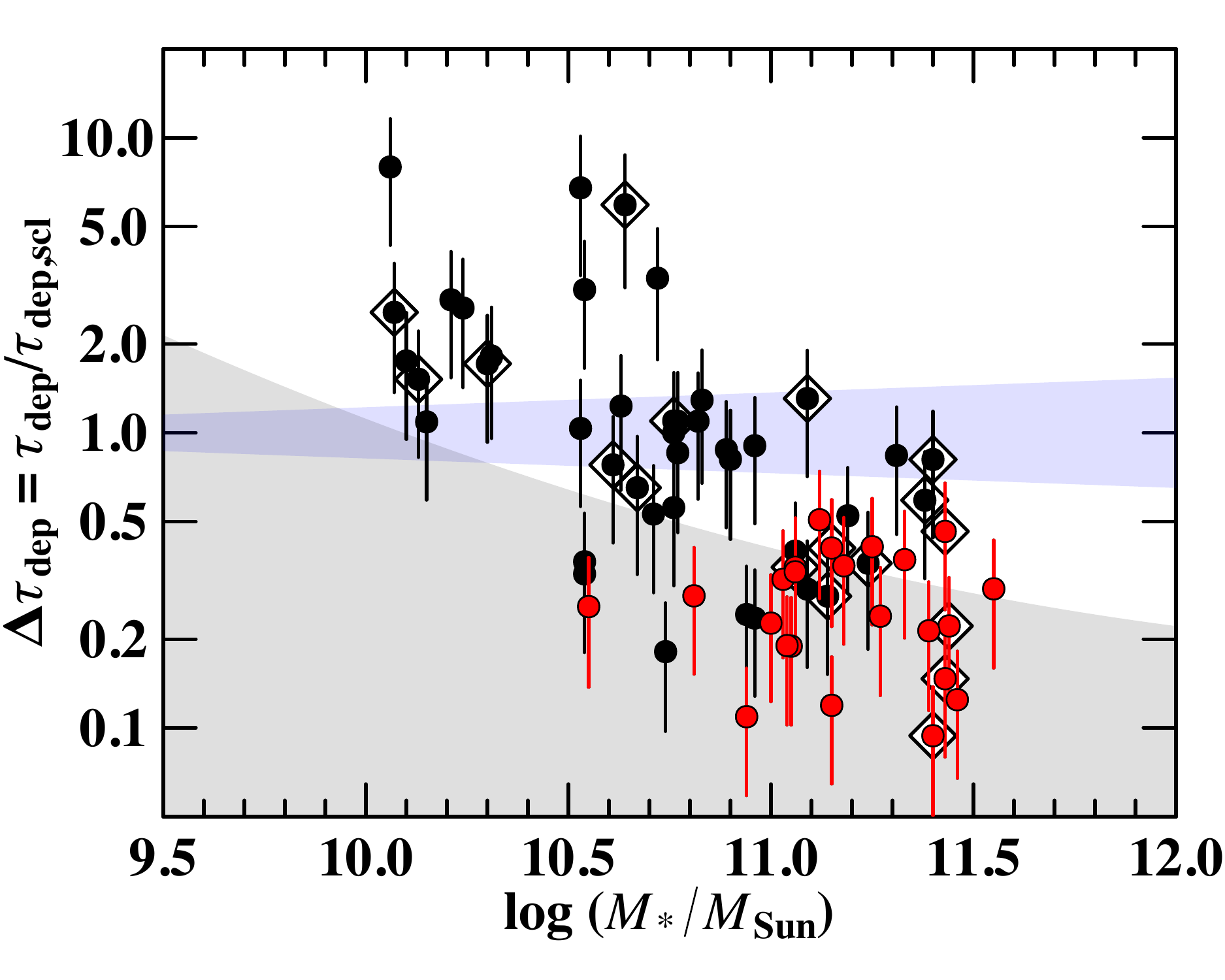}
\includegraphics[width=0.33\textwidth]{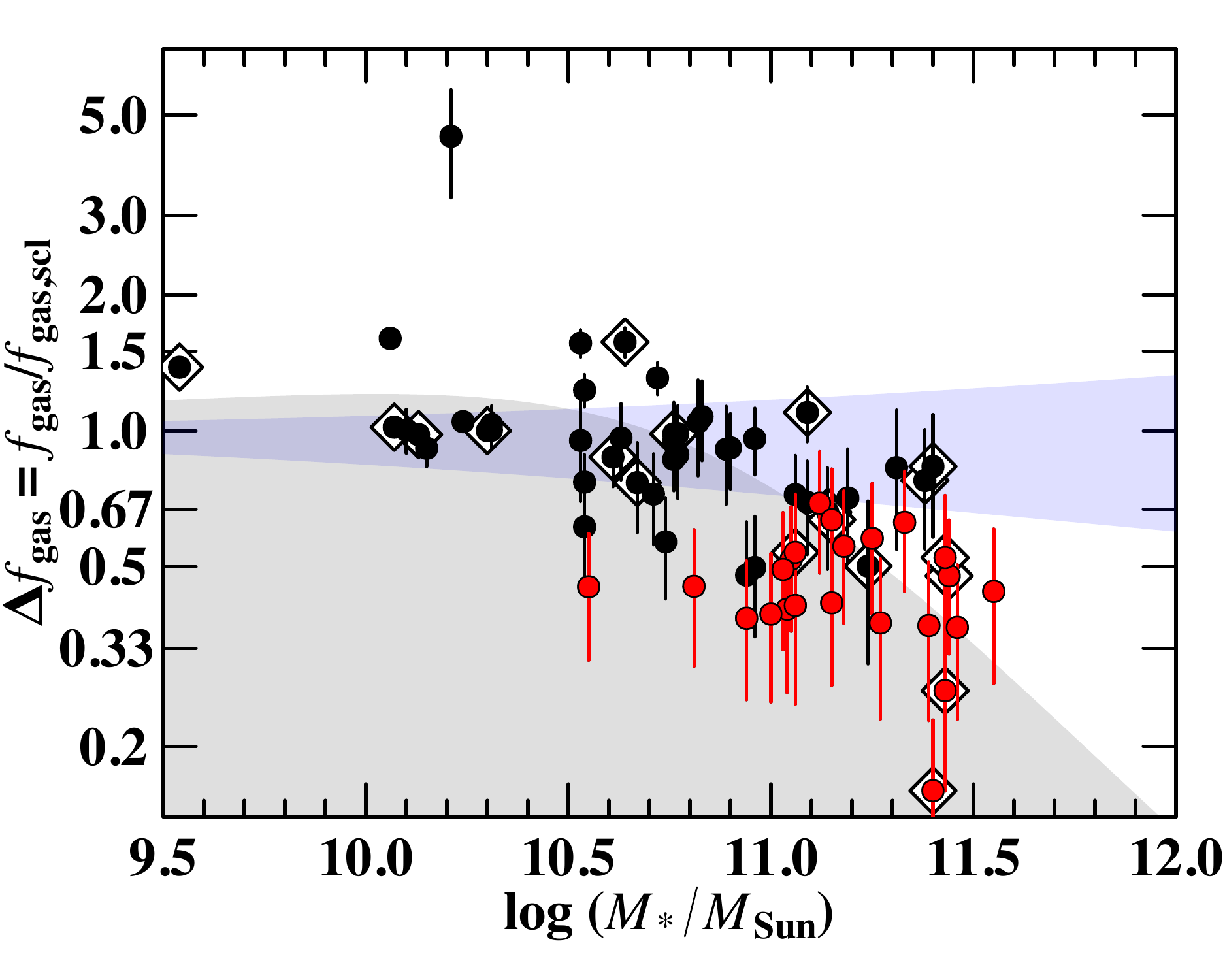}
\includegraphics[width=0.33\textwidth]{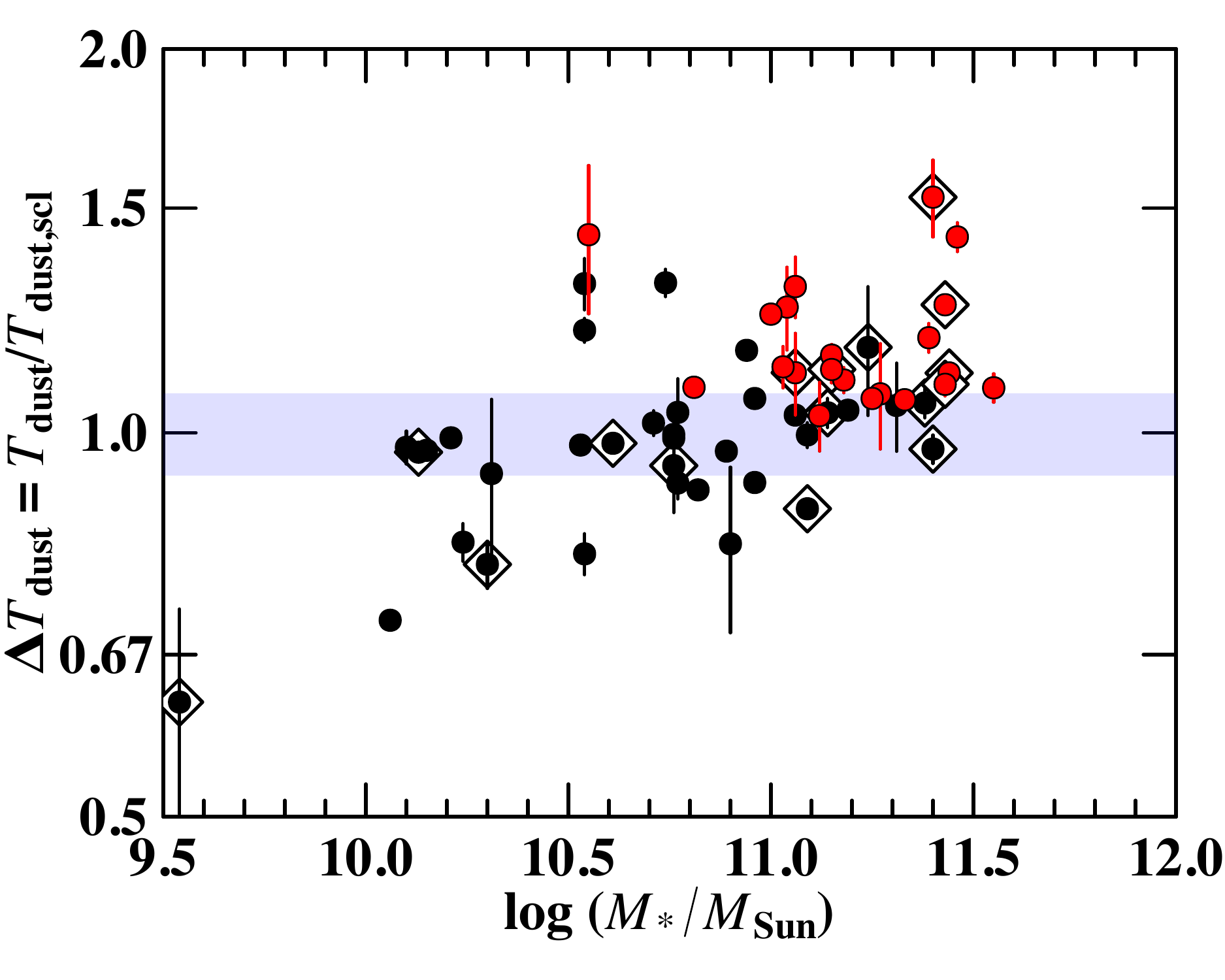}
\caption{$\tau_{\rm{dep}}$ (first column), $f_{\rm{gas}}$ (second column), and $T_{\rm{dust}}$ (third column) as a function of $\Delta \rm{MS}$ (first row) and stellar mass (third row). Scaling relations for $\tau_{\rm{dep}}(z,M_{\rm{*}},\Delta \rm{MS})$ and $f_{\rm{gas}}(z,M_{\rm{*}},\Delta \rm{MS})$ from \citet{tacconi18} and for $T_{\rm{dust}}(z,\Delta \rm{MS})$ from \citet{schreiber18} are shown as a solid blue line with their scatter as a shaded blue area. The 1$\sigma$ scatter of the MS ($0.5 < \Delta \rm{MS} < 2$, $\sim 0.3$\,dex) is also shown as a shaded blue area, with a more extended typical scatter of $0.33 < \Delta \rm{MS} < 3$ ($\sim 0.5$\,dex) in lighter blue (second row). We note that the values are scaled to a common redshift and stellar mass (first row) and to a common redshift and $\Delta \rm{MS}$ (third row) ($z_{\rm{med}} = 2.46$, $\log (M_{\rm{*med}}/M_{\odot}) = 10.79$, and $\Delta \rm{MS_{med}} = 2.15$) as explained in the main text. The second and fourth rows highlight the distance of a given property to its scaling relation, defined as the ratio of the property X to the property X in the scaling relation at a fixed redshift, stellar mass and $\Delta \rm{MS}$ ($\Delta \rm{X} = (X / X_{\rm{scl}})$, where $X = \tau_{\rm{dep}}$, $f_{\rm{gas}}$, and $T_{\rm{dust}}$). Only galaxies with a \textit{Herschel} counterpart are displayed. The shaded gray area shows the region where we are no longer $\sim 100\%$ complete (see main text). In all panels, SBs in the MS are in red: galaxies with $\tau_{\rm{dep}}$ below the scatter of the scaling relation and in the MS within a factor 3. Galaxies identified as AGN are highlighted with diamonds.}
\label{fig:scl_rel}
\end{center}
\end{figure*}

In Fig.~\ref{fig:scl_rel} our galaxy sample is placed in the context of the scaling relation for depletion timescales, gas fractions, and dust temperatures. We show $\tau_{\rm{dep}}$, $f_{\rm{gas}}$, and $T_{\rm{dust}}$ as a function of stellar mass and $\Delta \rm{MS}$ in comparison with the scaling relations established in the literature by \citet{tacconi18} for $\tau_{\rm{dep}}(z,M_{\rm{*}},\Delta \rm{MS})$ and $f_{\rm{gas}}(z,M_{\rm{*}},\Delta \rm{MS})$, and by \citet{schreiber18} for $T_{\rm{dust}}(z,\Delta \rm{MS})$. Similarly to Fig.~\ref{fig:ms}, we show two types of representation in Fig.~\ref{fig:scl_rel}. In the first and third rows, the values are scaled to a common redshift and stellar mass (first row) and to a common redshift and $\Delta \rm{MS}$ (third row) corresponding to the median values of the sample ($z_{\rm{med}} = 2.46$, $\log (M_{\rm{*med}}/M_{\odot}) = 10.79$, and $\Delta \rm{MS_{med}} = 2.15$), keeping constant the distance of the galaxies with respect to the scaling relations of interest. The second and fourth rows offer an alternative representation in terms of the distance of a given property to its scaling relation, defined as the ratio of the property X to the property X in the scaling relation at a fixed redshift, stellar mass and $\Delta \rm{MS}$ ($\Delta \rm{X} = (X / X_{\rm{scl}})$, where $X = \tau_{\rm{dep}}$, $f_{\rm{gas}}$, and $T_{\rm{dust}}$). While the first type of representation (first and third rows) is the most typical, it is again disadvantageous for galaxy samples that span a wide redshift range like ours and, thus, it requires the scaling so that the distance of a given property to the scaling relation is the same as if it was plotted against the scaling relation associated with its redshift, stellar mass, and $\Delta \rm{MS}$. The alternative representation (second and fourth rows) naturally solves this limitation by plotting directly the distance of a given property to the scaling relation at each galaxy redshift, stellar mass, and $\Delta \rm{MS}$. We note that in the alternative representation A2GS52 is not displayed in the panels involving $\Delta \tau_{\rm{dep}}$, as this galaxy is an outlier with respect to the scaling relation and modifying the $y$-axis range to include it would result in shrinking too much the rest of the plot to correctly visualize it.

From the $\tau_{\rm{dep}}$-$\Delta \rm{MS}$ panel in Fig.~\ref{fig:scl_rel}, there exists a subset of galaxies that exhibit low $\tau_{\rm{dep}}$, compared to the scaling relation of galaxies at the same redshift, stellar mass, and $\Delta \rm{MS}$, and located within the scatter of the MS. These characteristics were the ones outlined by \citet{elbaz18}, along with low $f_{\rm{gas}}$ and high SFR surface densities, for a population of SBs in the MS. We tagged these galaxies in red following the same definition as \citet{elbaz18}: SBs in the MS are galaxies with $\tau_{\rm{dep}}$ below the scatter of the scaling relation and in the MS within a factor 3 ($0.33 < \Delta \rm{MS} < 3$, 0.5\,dex). 23/69 galaxies follow this definition and were classified as SBs in the MS (where the total number of galaxies accounts only for galaxies with a \textit{Herschel} counterpart, as for the remaining galaxies without a \textit{Herschel} counterpart the assessment of whether they were or not SBs in the MS was not possible). We also see that these galaxies with low $\tau_{\rm{dep}}$ and within the scatter of the MS have naturally low $f_{\rm{gas}}$ compared to the scaling relation of galaxies at the same redshift, stellar mass, and $\Delta \rm{MS}$ (this is a direct consequence of having low $\tau_{\rm{dep}}$ while being within the scatter of the MS, as low $\tau_{\rm{dep}}$ at constant SFR means low $M_{\rm{gas}}$). In addition, they also exhibit anomalously high $T_{\rm{dust}}$ compared to the scaling relation of galaxies at the same redshift, stellar mass, and $\Delta \rm{MS}$. Another characteristic is their elevated stellar masses, as the subset of SBs in the MS is on average $\times 2.6$ more massive that the subset of remaining galaxies not labeled as SBs in the MS. Therefore, we conclude that the population of SBs in the MS is characterized by short depletion timescales, low gas fractions, and high dust temperatures in comparison with the scaling relations of galaxies at a fixed redshift, stellar mass, and $\Delta \rm{MS}$.

AGN as identified in Sect~\ref{subsec:agn} are highlighted with black diamonds in Fig.~\ref{fig:scl_rel}. There is no evidence for correlations between $\tau_{\rm{dep}}$, $f_{\rm{gas}}$, or $T_{\rm{dust}}$ and AGN activity. Although, a number of galaxies are classified as AGN and it could be a sign of coevolution between AGN and star formation activity, their location in the panels do not follow any particular trend. This is in line with studies indicating the absence of suppression or enhancement of the SFR, dust, or gas masses in AGN hosts at a fixed stellar mass, except for a handful extreme cases \citep[e.g.,][]{valentino21}.

The distribution of our galaxy sample in Fig.~\ref{fig:scl_rel} is subject to selection effects. In Sect.~\ref{subsec:compl} we introduced the aspects of completeness and selection limits in the GOODS-ALMA 2.0 survey that we now project into Fig.~\ref{fig:scl_rel}. In order to do so, we used the flux density completeness threshold of $S_{1.1\rm{mm}} > 1$\,mJy for sources with dust continuum sizes up to 1\arcsec~FWHM that defines the GOODS-ALMA 2.0 survey $\sim 100\%$ completeness \citep[see][]{gomezguijarro21}. We converted this flux density completeness threshold into a $M_{\rm{dust}}$ limit by using the dust SED template corresponding to the median redshift and $\Delta \rm{MS}$ of our galaxy sample ($z_{\rm{med}} = 2.46$ and $\Delta \rm{MS_{med}} = 2.15$), as the values in the panels of the first row were scaled to these medians. Then, the $M_{\rm{dust}}$ limit was converted into a $M_{\rm{gas}}$ limit through the $\delta_{\rm{GDR}}$--$Z$ technique where, in this case, the range of stellar masses represented in the $x$-axis was used to calculate the associated metallicity through MZR and, thus, the associated $\delta_{\rm{GDR}}$. Therefore, the border of the shaded gray area in the $f_{\rm{gas}}$--$M_{\rm{*}}$ panel represents the $\sim 100\%$ completeness threshold and our sample is located alongside it. We see that in the low stellar mass end ($\log (M_{\rm{*}}/M_{\odot}) < 10.5$), the selection effects play against finding galaxies below the scaling relation. However, in the high stellar mass end ($\log (M_{\rm{*}}/M_{\odot}) > 10.5$) galaxies above and below the scaling relation are within the reach of the survey. At the most massive end ($\log (M_{\rm{*}}/M_{\odot}) > 11.0$), there is a significant excess of galaxies below the scaling relation, which constitute the subset of SBs in the MS, with respect to galaxies above the scaling relation. This is not due to selection effects as galaxies above the scaling relation are favored than galaxies below it in terms of selection effects. Similarly, we project the selection effects into the $\Delta f_{\rm{gas}}$--$M_{\rm{*}}$ panel, offering an alternative view.

In the case of $\tau_{\rm{dep}}$, selection effects are more elusive to be described accurately, as the 1.1\,mm flux density completeness threshold better correlates with $M_{\rm{gas}}$ than with SFR ($\tau_{\rm{dep}} = M_{\rm{gas}} / \rm{SFR}$) for the redshift range of our galaxy sample. $M_{\rm{gas}}$ is very sensitive to the data points in the RJ side of the IR SED, while SFR is more sensitive to variations in the peak of the dust SED and, thus, $T_{\rm{dust}}$. At 1.1\,mm, for the redshift range spanned in our galaxy sample, data points are located in the RJ side of the IR SED. Therefore, the limits should be treated as an indication of the regions of the panels subject to incompleteness, but not as an accurate selection function. We converted the flux density completeness threshold into $M_{\rm{dust}}$ and $L_{\rm{IR}}$ limits by using the dust SED template corresponding to the median redshift and $\Delta \rm{MS}$ of our galaxy sample. Then, the $M_{\rm{dust}}$ limit was converted into a $M_{\rm{gas}}$ limit through the $\delta_{\rm{GDR}}$--$Z$ technique and $L_{\rm{IR}}$ into SFR to finally get the $\tau_{\rm{dep}}$ limit as a function of stellar mass. We see that in the low stellar mass end ($\log (M_{\rm{*}}/M_{\odot}) < 10.5$), the selection effects play against finding galaxies below the scaling relation. As in the case of $f_{\rm{gas}}$, at the most massive end ($\log (M_{\rm{*}}/M_{\odot}) > 11.0$) there exist a significant excess of galaxies below the scaling relation, not due to selection effects, that constitute the subset of SBs in the MS.

Last, for $T_{\rm{dust}}$ selection effects are more complicated to be described as the 1.1\,mm flux density does not correlate with $T_{\rm{dust}}$. Nevertheless, selection is more favorable toward colder dust SEDs. A warmer dust SED would require either higher $M_{\rm{dust}}$ (higher $M_{\rm{gas}}$, for a fixed $\delta_{\rm{GDR}}$) or higher $L_{\rm{IR}}$ to be detected for a given flux density threshold in the RJ side of the IR SED. We see that in the low stellar mass end ($\log (M_{\rm{*}}/M_{\odot}) < 10.5$), the selection effects play against finding warmer galaxies above the scaling relation. In the most massive end ($\log (M_{\rm{*}}/M_{\odot}) > 11.0$), there exist a significant excess of warmer galaxies above the scaling relation, that constitute the subset of SBs in the MS. This is again not due to selection effects, as colder galaxies below the scaling relation are more favored than warmer galaxies above it.

Appendix~\ref{sec:appendix_c} tests the impact of the metallicity assumptions for $M_{\rm{gas}}$ estimates in the distribution of our galaxy sample in Fig.~\ref{fig:scl_rel}. Figs.~\ref{fig:scl_rel_fmr} and \ref{fig:scl_rel_solar} reproduce Fig.~\ref{fig:scl_rel} for $M_{\rm{gas}}$ estimates obtained by using the $\delta_{\rm{GDR}}$--$Z$ with FMR and fixed solar metallicity, respectively. For FMR (solar) the whole distribution shifts to higher (lower) $M_{\rm{gas}}$ regimes with respect to the gas scaling relations, as expected given the increase (decrease) of $\delta_{\rm{GDR}}$ for metallicities on average lower (higher) through FMR (solar) compare to MZR. We note that both the gas scaling relations and the assessment of selection effects were calculated based on MZR. Nonetheless, the conclusion of the existence of the subset of SBs in the MS remains unchanged. These galaxies are characterized by short depletion timescales, low gas fractions, and high dust temperatures in comparison with the scaling relations of galaxies at a fixed redshift, stellar mass, and $\Delta \rm{MS}$, regardless of the metallicity assumption for $M_{\rm{gas}}$ estimates. We note that depending on the metallicity assumption, the subset of SBs in the MS would have slightly varied, as some galaxies enter or exit the selection criteria, that is galaxies with $\tau_{\rm{dep}}$ below the scatter of the scaling relation and in the MS within a factor 3.

\section{Compact star formation as a physical driver of depletion timescales, gas fractions, and dust temperatures} \label{sec:compact}

We investigated the role of the spatial extent of the ongoing star formation, traced by the dust continuum emission at 1.1\,mm linked to the star formation episode and gas reservoir, as a physical driver of the galaxy sample behavior in relation with the scaling relations shown in the previous section. In particular, we focused on the driver of the anomalous characteristics of the SBs in the MS compared to the scaling relations.

Dust continuum sizes at 1.1\,mm for the galaxy sample were presented in \citet{gomezguijarro21}. Briefly, sizes were measured in the $uv$ plane of the GOODS-ALMA 2.0 combined dataset employing the CASA task \texttt{UVMODELFIT} to fit single component models to single sources. We fit a Gaussian model with fixed circular axis ratio, since the scope of the work was to get global size measurements. They were obtained for the galaxies in the 100\% pure main catalog (corresponding to ID from A2GS1 to A2GS44), which have a detection $\rm{S/N_{peak}} \geq 5$. Measurements below this S/N with a $\rm{S/N_{peak}} \leq 5$ were unreliable (corresponding to ID from A2GS45 to A2GS88). In \citet{gomezguijarro21} we concluded that the galaxy sample dust continuum sizes are generally compact, with a median effective (half-light) radius of $R_{\rm{e}} = 0\farcs10 \pm 0\farcs05$ and a median physical size of $R_{\rm{e}} = 0.73 \pm 0.29$\,kpc, calculated at the redshift of each galaxy (where the uncertainties are given by the median absolute deviation). In addition, We concluded that for sources with flux densities $S_{\rm{1.1mm}} > 1$\,mJy compact dust continuum emission at 1.1\,mm prevails, and sizes as extended as typical star-forming stellar disks are rare. For sources with $S_{\rm{1.1mm}} < 1$\,mJy compact dust continuum emission appears slightly more extended at 1.1\,mm, although still generally compact below the sizes of typical star-forming stellar disks.

Among the galaxies for which size measurements were reported in \citet{gomezguijarro21}, a special distinction was made to galaxies with a detection $\rm{S/N_{peak}} \geq 6.5$, for which the size measurements were the most reliable (corresponding to ID from A2GS1 to A2GS26). In the following we restricted the analysis to the latter subset. As the size measurements are more reliable with increasing S/N, this choice is a good compromise between using the best size measurements and a sizable sample that includes the sources with flux densities consistent with $S_{\rm{1.1mm}} > 1$\,mJy, the flux density threshold for which the GOODS-ALMA 2.0 survey reaches a $\sim 100\%$ completeness for sources with dust continuum sizes up to 1\arcsec~FWHM. Therefore, removing galaxies with a detection peak $\rm{S/N_{peak}} \leq 6.5$ restricts the analysis to the bright end of the 1.1\,mm flux densities where the survey is complete, typically associated with higher $M_{\rm{gas}}$ galaxies.

\begin{figure*}
\begin{center}
\includegraphics[width=0.33\textwidth]{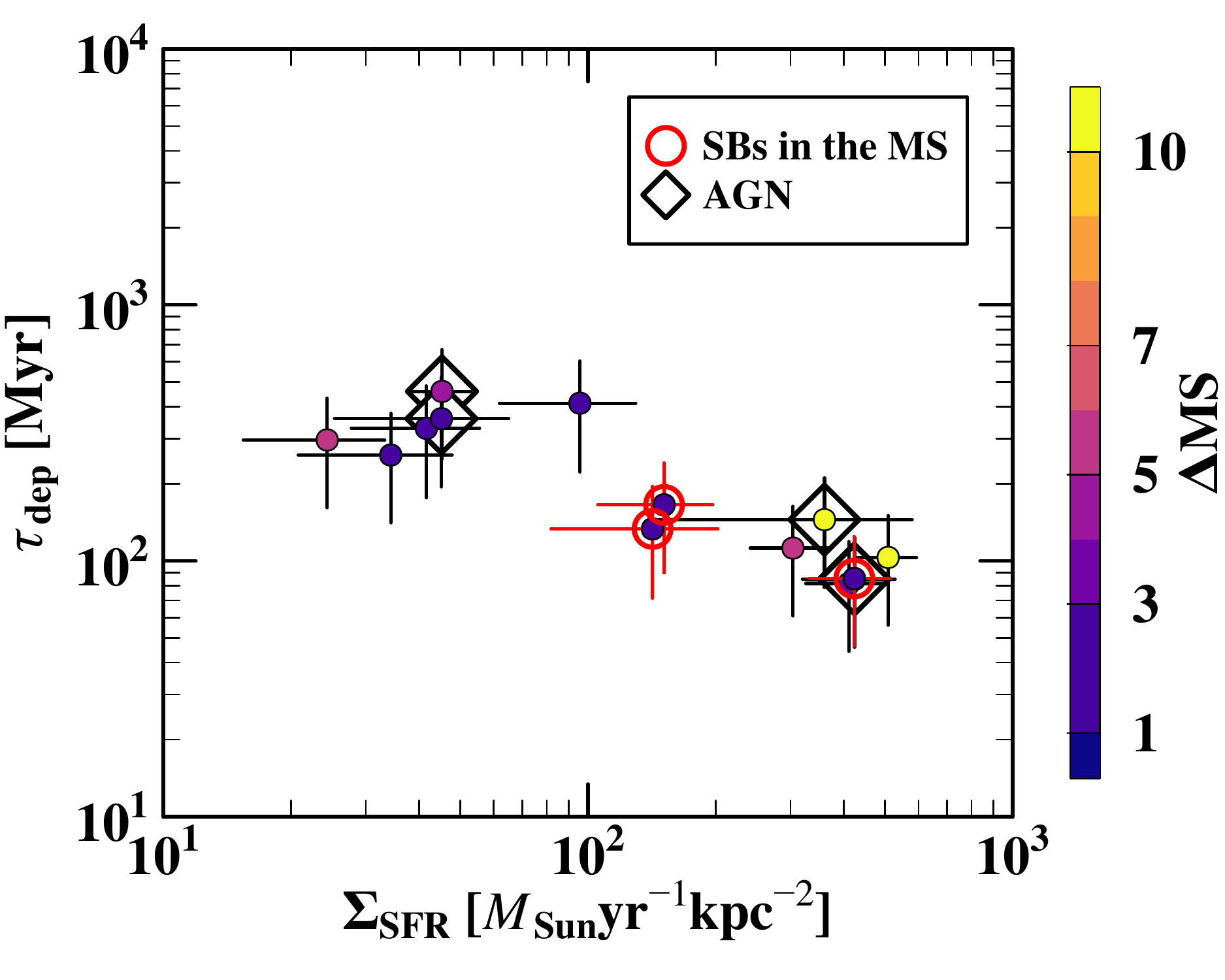}
\includegraphics[width=0.33\textwidth]{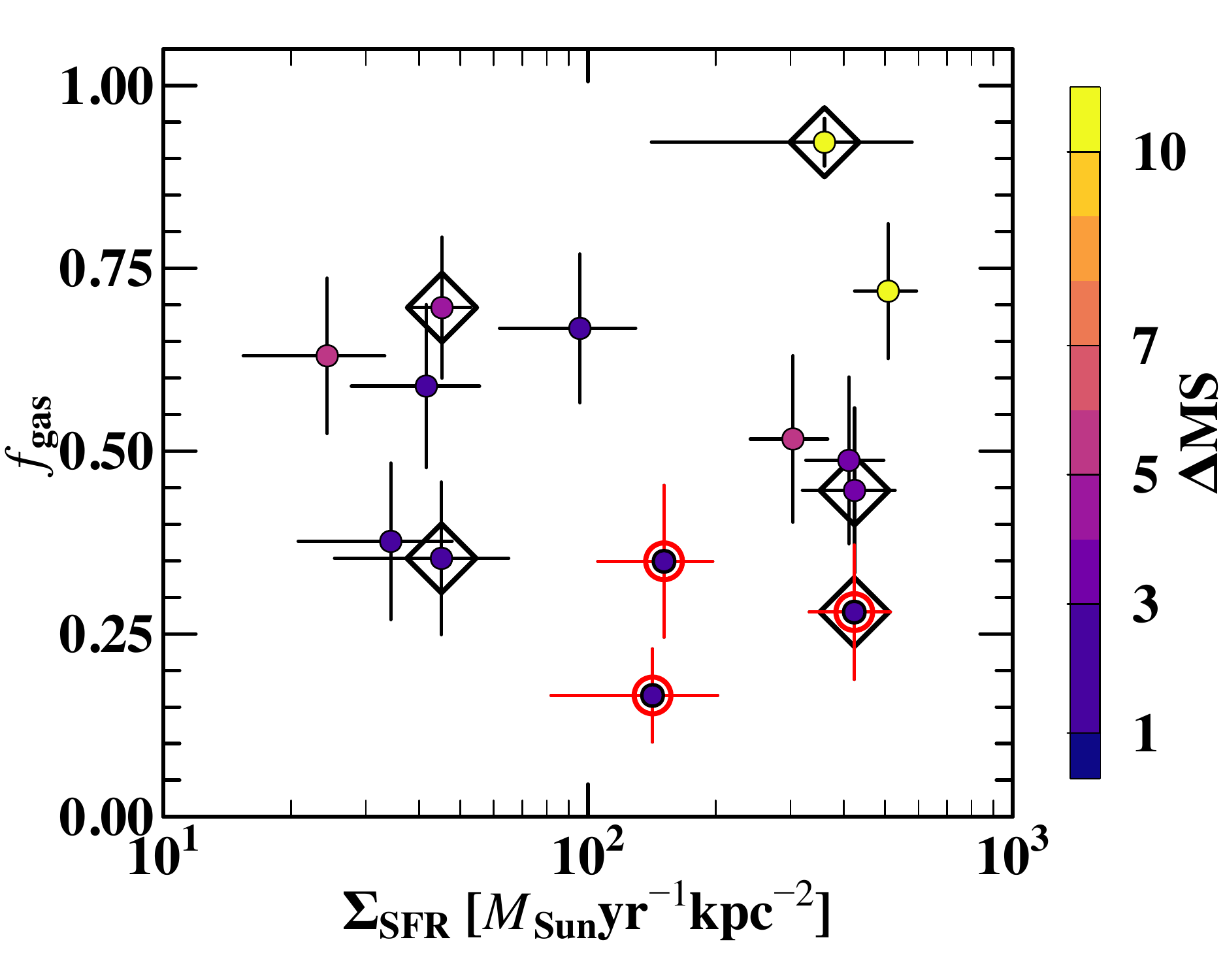}
\includegraphics[width=0.33\textwidth]{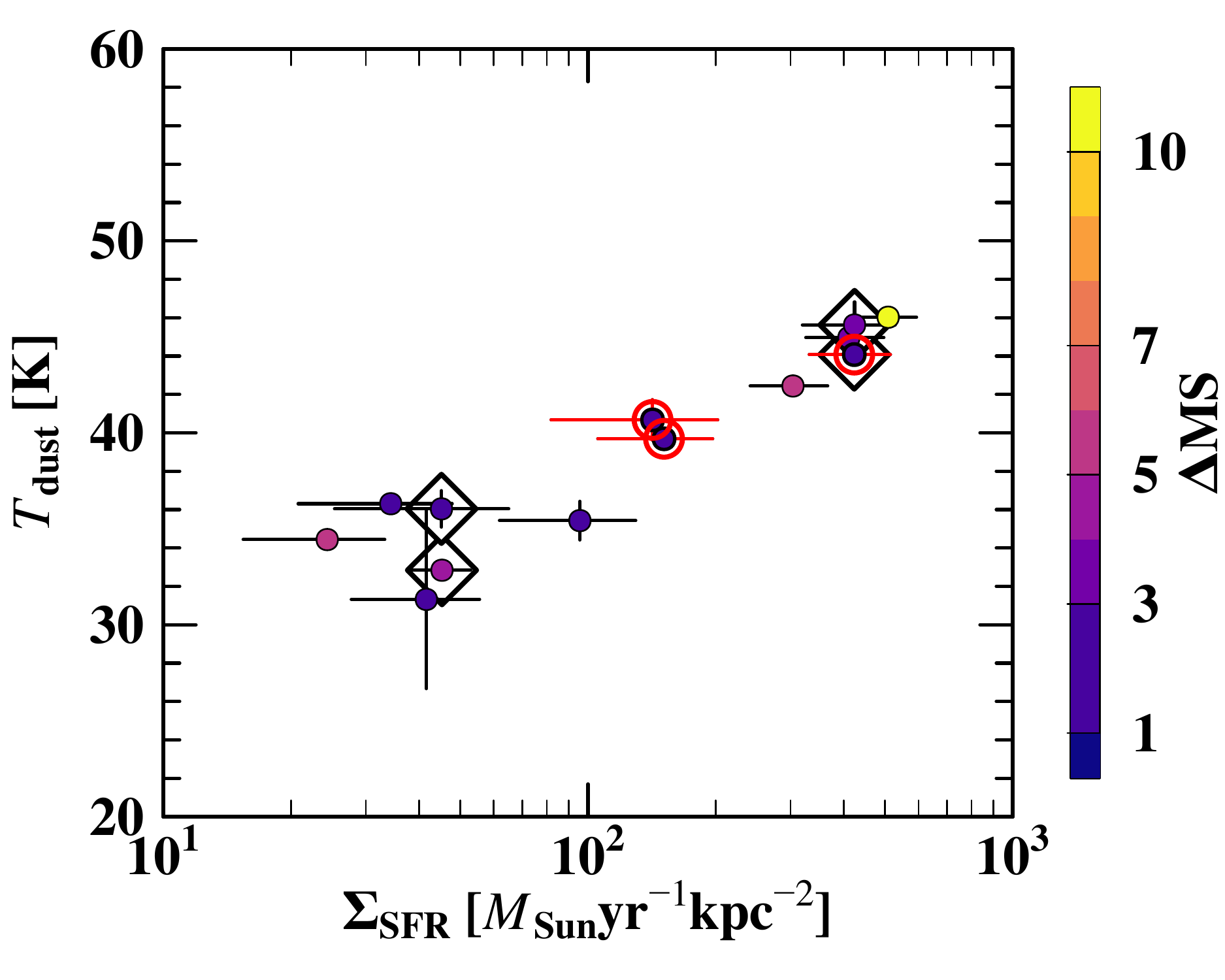}
\includegraphics[width=0.33\textwidth]{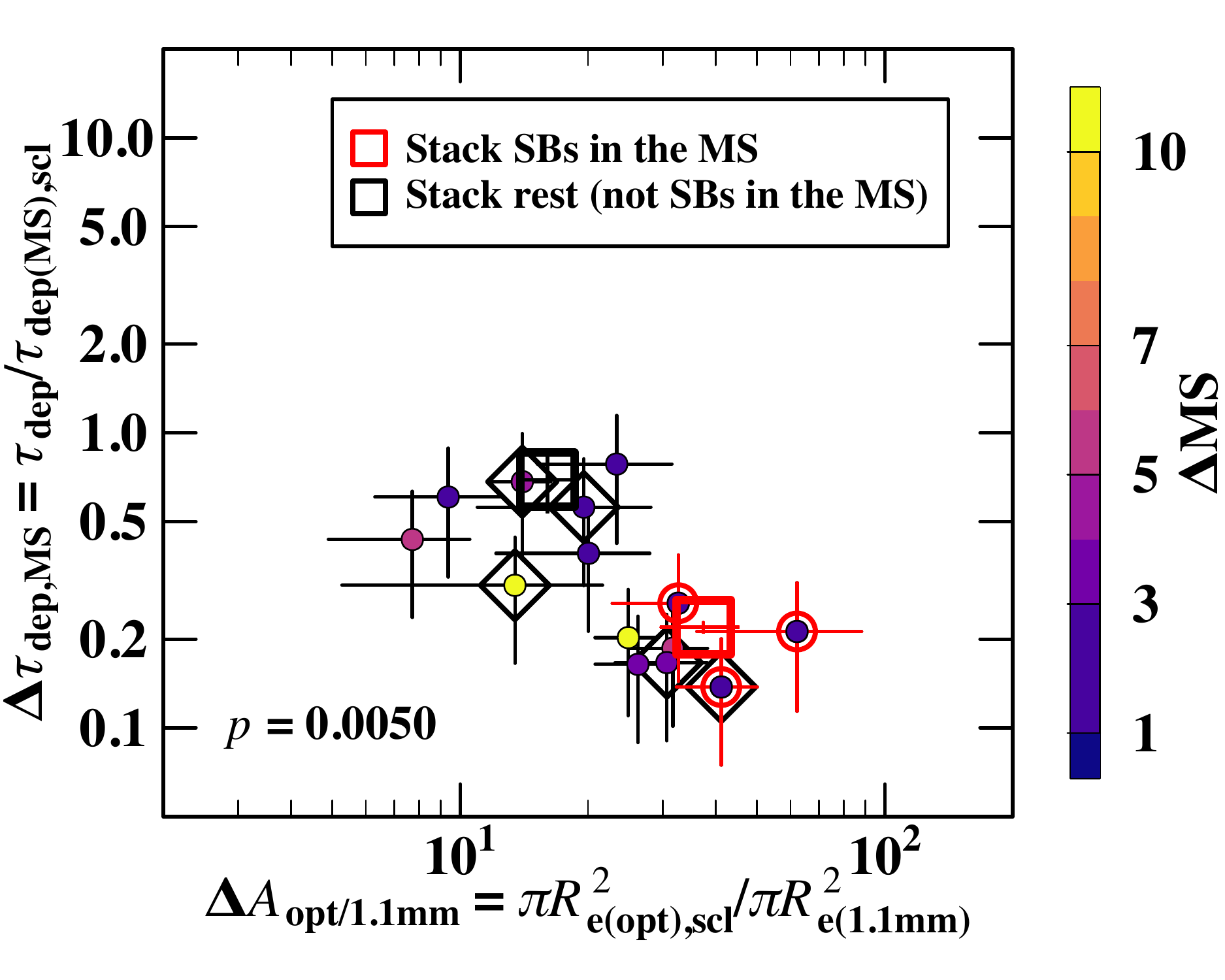}
\includegraphics[width=0.33\textwidth]{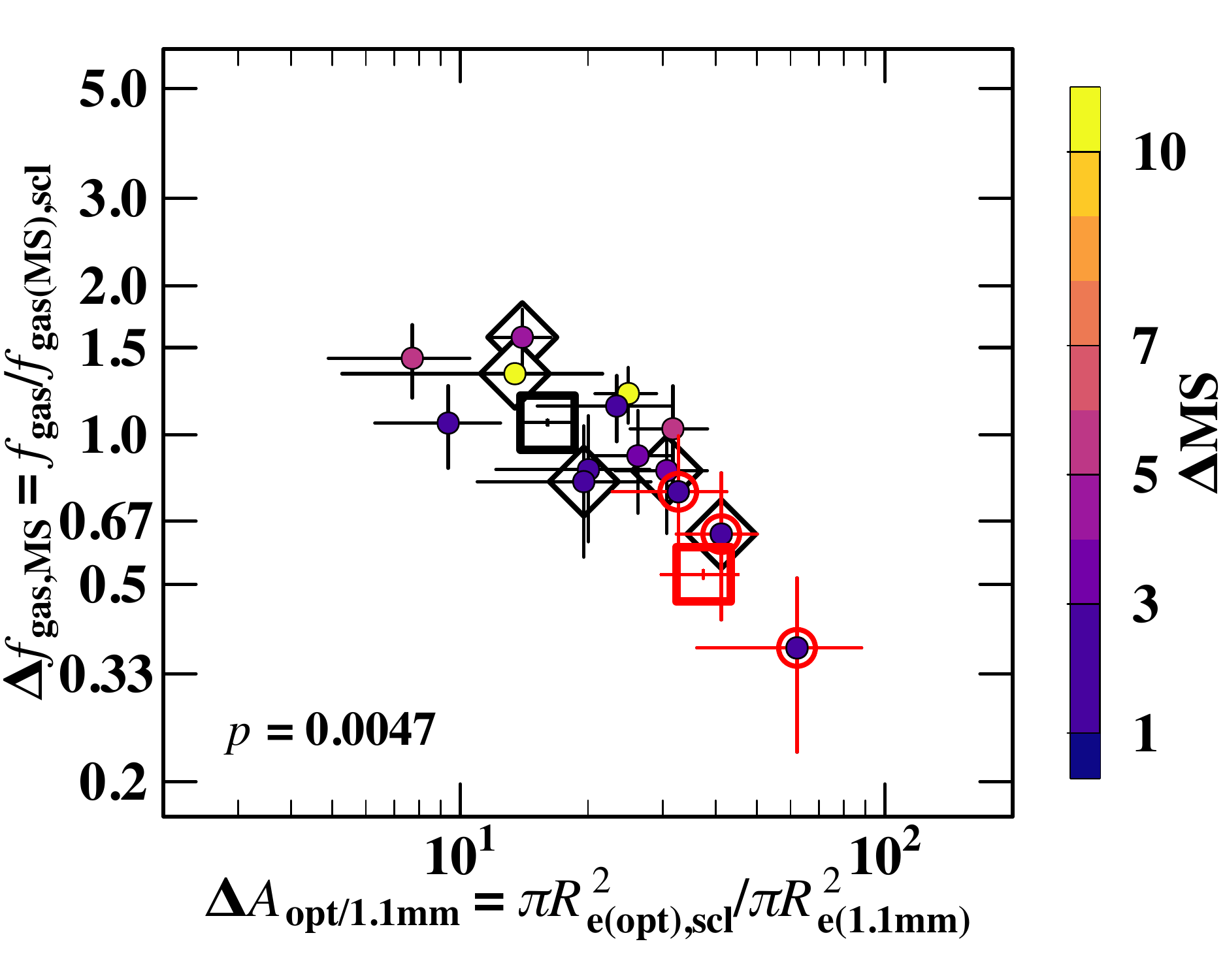}
\includegraphics[width=0.33\textwidth]{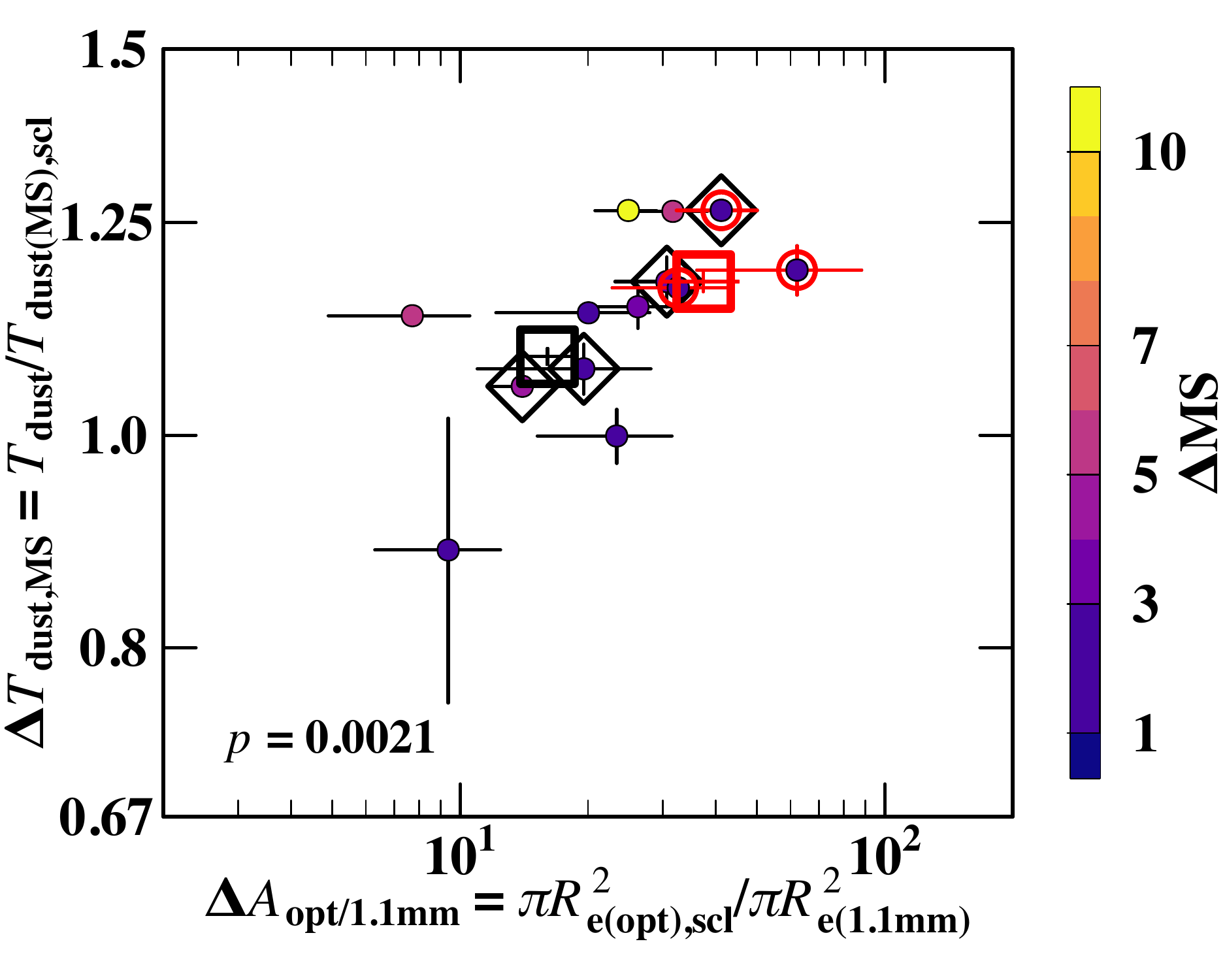}
\caption{Top row: $\tau_{\rm{dep}}$ (first column), $f_{\rm{gas}}$ (second column), and $T_{\rm{dust}}$ (thrid column) as a function of $\Sigma_{\rm{SFR}}$. Only galaxies with a \textit{Herschel} counterpart with the most reliable size measurements in \citet{gomezguijarro21} are shown (corresponding to a detection $\rm{S/N_{peak}} \geq 6.5$). SBs in the MS are in red. Galaxies identified as AGN are highlighted with diamonds. All galaxies are color coded according to their $\Delta \rm{MS}$. Bottom row: deviations of $\tau_{\rm{dep}}$ (first column), $f_{\rm{gas}}$ (second column), and $T_{\rm{dust}}$ (third column) compared to the gas scaling relations of normal MS SFGs ($\Delta \rm{MS} = 1$) in relation with the deviations of the dust continuum areas compared to the structural scaling relation of normal SFGs at a fixed stellar mass and redshift. The gas scaling relations for $\tau_{\rm{dep}}(z,M_{\rm{*}},\Delta \rm{MS})$ and $f_{\rm{gas}}(z,M_{\rm{*}},\Delta \rm{MS})$ are from \citet{tacconi18}, and for $T_{\rm{dust}}(z,\Delta \rm{MS})$ from \citet{schreiber18}, where $\Delta \rm{MS} = 1$. In the case of the structural scaling relation, we compared the dust continuum areas with the typical areas of star-forming stellar disks measured at optical wavelengths for late-type galaxies from \citet{vanderwel14}, at a fixed stellar mass and redshift. The stack of the 23/69 galaxies that were classified as SBs in the MS originally in Fig.~\ref{fig:scl_rel} is shown as an empty red square (where the total number of galaxies accounts only for galaxies with a \textit{Herschel} counterpart). Similarly, the stack of the remaining galaxies that were not labeled as SBs in the MS in Fig.~\ref{fig:scl_rel} are shown as an empty black square.}
\label{fig:trends}
\end{center}
\end{figure*}

In the top row of Fig.~\ref{fig:trends} we show how $\tau_{\rm{dep}}$, $f_{\rm{gas}}$, and $T_{\rm{dust}}$, as derived in Sect.~\ref{sec:properties}, behave as a function of the SFR surface density ($\Sigma_{\rm{SFR}} = \rm{SFR} / (2 \pi R_{\rm{e}}^2)$). It is clear that $\tau_{\rm{dep}}$ and $T_{\rm{dust}}$ correlate with $\Sigma_{\rm{SFR}}$, with higher $\Sigma_{\rm{SFR}}$ related to lower $\tau_{\rm{dep}}$ (negative correlation) and higher $T_{\rm{dust}}$ (positive correlation). Conversely, $f_{\rm{gas}}$ does not seem to correlate with $\Sigma_{\rm{SFR}}$. These type of relations have been reported by several studies in the literature, both in the case of $\tau_{\rm{dep}}$ \citep[e.g.,][]{elbaz18,franco20b} and $T_{\rm{dust}}$ \citep{burnham21}, establishing a physical connection between the increase in $\Sigma_{\rm{SFR}}$ and the origin of such relations. However, in order to better establish such physical connections, it is necessary to account for the stellar mass and redshift evolution of the studied properties. In addition, the correlations are at least partially driven by the fact that the studied properties and $\Sigma_{\rm{SFR}}$ are related to each other through $L_{\rm{IR}}$. Therefore, the drivers of $\tau_{\rm{dep}}$, $f_{\rm{gas}}$, and $T_{\rm{dust}}$ remain to be understood at a fixed stellar mass and redshift and without invoking related physical quantities.

In the bottom row of Fig.~\ref{fig:trends} we show the deviations of $\tau_{\rm{dep}}$, $f_{\rm{gas}}$, and $T_{\rm{dust}}$ compared to the gas scaling relations of normal MS SFGs ($\Delta \rm{MS} = 1$) in relation to the deviations of the dust continuum sizes compared to the structural scaling relation of normal SFGs at the same stellar mass and redshift. In the case of the gas scaling relations, we compared them with those established in the literature for $\tau_{\rm{dep}}(z,M_{\rm{*}},\Delta \rm{MS})$ and $f_{\rm{gas}}(z,M_{\rm{*}},\Delta \rm{MS})$ from \citet{tacconi18}, and for $T_{\rm{dust}}(z,\Delta \rm{MS})$ from \citet{schreiber18}, as in Sect.~\ref{sec:scl_rel}, where $\Delta \rm{MS} = 1$ in this case. Regarding the structural scaling relation, we compared the dust continuum sizes with the typical size of star-forming stellar disks measured at optical wavelengths for late-type galaxies from \citet{vanderwel14} at the same stellar mass and redshift. In \citet{gomezguijarro21} we found that dust continuum sizes evolve with redshift and stellar mass resembling the trends of the stellar sizes measured at optical wavelengths, albeit a lower normalization compared to those of late-type galaxies. The outlined approach permits to study the relation of a geometrical property like the sizes of the ongoing star formation regions with the physical properties of the galaxies in terms of their $\tau_{\rm{dep}}$, $f_{\rm{gas}}$, and $T_{\rm{dust}}$, at a fixed stellar mass and redshift. We note that in the case of the sizes we used areas instead of one-dimensional sizes. Also we note that the definition of the deviations of the dust continuum areas compared to the structural scaling relation of normal SFGs at the fixed stellar mass and redshift is given by $\Delta A = (\pi R_{\rm{e(opt),scl}}^2 / \pi R_{\rm{e(1.1mm)}}^2)$, so that increasing compact star formation with respect to typical sizes of star-forming stellar disks is associated with increasing values in the $x$-axis.

Dust continuum areas at 1.1\,mm (dust compactness) with respect to the typical size of star-forming stellar disks (measured at optical wavelengths) appear to be correlated with $\tau_{\rm{dep}}$, $f_{\rm{gas}}$, and $T_{\rm{dust}}$. As dust continuum becomes more compact with respect to typical sizes of star-forming stellar disks, $\tau_{\rm{dep}}$ and $f_{\rm{gas}}$ become lower and $T_{\rm{dust}}$ higher compared to the gas scaling relations of normal MS SFGs, at a fixed stellar mass and redshift.

In order to study whether there exist formal correlations, we performed statistical simulations of the nonparametric Spearman rank correlation test. We followed a bootstrapping approach generating 100\,000 simulations of the galaxy sample. In other words, for each of the statistical simulations we generated a sample of galaxies allowing replacements of the same number of galaxies as the original sample and calculated the Spearman's rank correlation coefficient ($r_s$). This approach ensures that single points are not the sole drivers of a potential correlation, as points are included and removed randomly in the different realizations studying the existence of a correlation each time. The probability of no correlation ($p$-value) is given by the number of simulations for which $r_s < 0$, for a positive correlation, or $r_s > 0$, for a negative correlation (multiplied by two since it is a bilateral test). The probabilities of no correlation are 0.50\%, 0.47\%, and 0.21\% ($p$-value) for $\tau_{\rm{dep}}$, $f_{\rm{gas}}$, and $T_{\rm{dust}}$, respectively. This confirms correlations for $\tau_{\rm{dep}}$, $f_{\rm{gas}}$, and $T_{\rm{dust}}$ (considering rejections at 5\%).

\begin{figure*}
\begin{center}
\includegraphics[width=\textwidth]{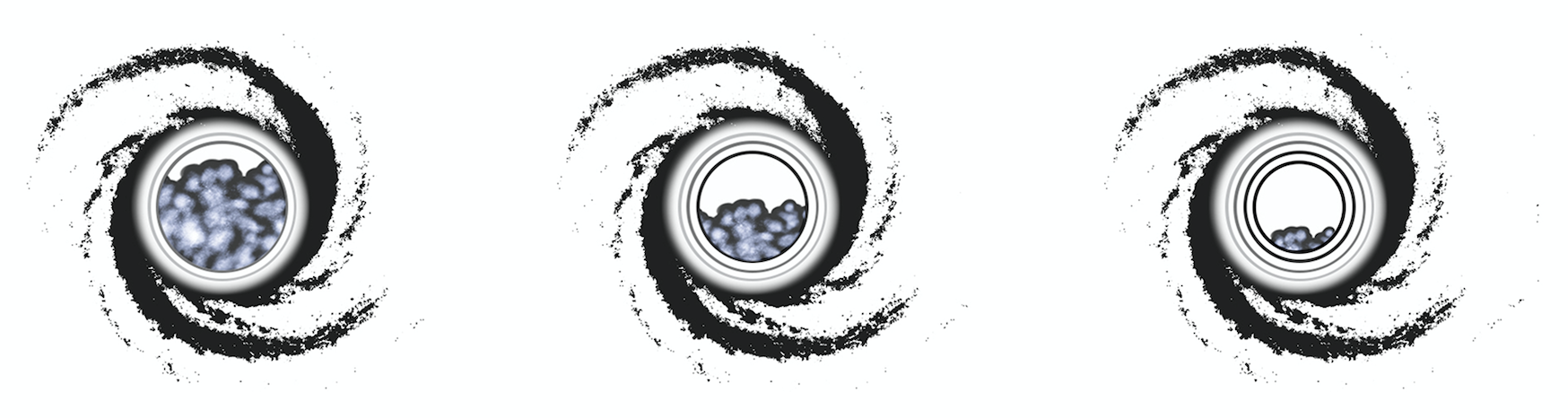}
\caption{Drawing summarizing the results from our observations in terms of the gas fraction. From left to right, the less amount of gas in the galaxy (represented as a shrinking blue region), the more compact the star-forming region becomes (represented as a shrinking inner black circle).}
\label{fig:sum_gas}
\end{center}
\end{figure*}

Fig.~\ref{fig:sum_gas} summarizes in a simple drawing the results from our observations in terms of the gas fraction. The less amount of gas in the galaxy (represented as the shrinking blue region), the more compact the star-forming region becomes (represented as the shrinking inner black circle). We note that this drawing is aimed to be a simple representation of the trend involving $f_{\rm{gas}}$ in Fig.~\ref{fig:sum_gas} in the $\Delta f_{\rm{gas,MS}}$--$\Delta A_{\rm{opt/1.1mm}}$ plane, but it is not aimed to be a representation of what could be the physical processes necessary for this compression to happen or about the externally and/or internally-driven morphological transformations involved in the process.

Another interesting aspect in Fig.~\ref{fig:trends} is that, when comparing the top and bottom rows, SBs in the MS move to the extremes of the trends. SBs in the MS appear to be the extreme cases where the dust continuum areas are the most compact ones compared to the structural scaling relation for typical star-forming stellar disks. They are associated with the lowest $\tau_{\rm{dep}}$ and $f_{\rm{gas}}$, and the highest $T_{\rm{dust}}$ compared to the gas scaling relations for normal MS SFGs at a fixed stellar mass and redshift.

However, our galaxy sample of SBs in the MS in Fig.~\ref{fig:trends} is scarce. Given that SBs in the MS are characterized by depleted $f_{\rm{gas}}$, it could be that they had a low surface brightness extension of the dust continuum emission not properly captured in the size measurements for individual galaxies. In order to verify the latter, we stacked the 1.1\,mm data for all the 23/69 galaxies that were classified as SBs in the MS originally in Fig.~\ref{fig:scl_rel} (where the total number of galaxies accounts only for galaxies with a \textit{Herschel} counterpart). In addition, we stacked the 1.1\,mm data for the remaining galaxies that were not labeled as SBs in the MS in Fig.~\ref{fig:scl_rel}. The stacks were performed in the $uv$ plane following the methodology described in \citet{gomezguijarro21} for $uv$ plane stacking. We measured the sizes of the stacks employing the Common Astronomy Software Applications \citep[CASA;][]{mcmullin07} task \texttt{UVMODELFIT}, fitting a Gaussian model with fixed circular axis ratio as in \citet{gomezguijarro21}. In the bottom row of Fig.~\ref{fig:trends} we show in the $x$-axis dust continuum areas resulting from these stacks and in the $y$-axis median values for $\tau_{\rm{dep}}$, $f_{\rm{gas}}$, and $T_{\rm{dust}}$ for both SBs in the MS (red) and the remaining galaxies (black). We note that the deviations compared to the scaling relations were evaluated at the median redshift and stellar mass of each subset. These values are in agreement with the correlations drawn from the individual galaxies as explained above.

In \citet{gomezguijarro21} we found that dust continuum sizes evolve with redshift and stellar mass resembling the trends of the stellar sizes measured at optical wavelengths. As mentioned in Sect.~\ref{sec:scl_rel}, the subset of SBs in the MS is on average $\times 2.6$ more massive than the subset of remaining galaxies not labeled as SBs in the MS. The median redshifts and stellar masses of the SBs in the MS are $z = 2.36 \pm 0.05$ and $\log (M_{\rm{*}}/M_{\odot}) = 11.15 \pm 0.01$, while for the remaining galaxies are $z = 2.17 \pm 0.02$ and $\log (M_{\rm{*}}/M_{\odot}) = 10.73 \pm 0.01$. The representation of the bottom row of Fig.~\ref{fig:trends} naturally accounts for these differences by plotting directly the deviation of a given property to the scaling relation at a fixed redshift and stellar mass. But if sizes as given by the effective radius were to be compared, they need to be expressed in the same terms of redshift and stellar mass. We corrected the dust continuum sizes of the stacks to a common redshift and stellar mass, as given by the median values of our galaxy sample ($z_{\rm{med}} = 2.46$ and $\log (M_{\rm{*med}}/M_{\odot}) = 10.79$), by using the $R_{\rm{e}}$($z,M_{\rm{*}}$) dependency of late-type galaxies of \citet{vanderwel14}. The stack of SBs in the MS with a dust continuum size of $R_{\rm{e}}^{\rm{cor}} = 0.52 \pm 0.05$\,kpc is indeed more compact than the stack of the remaining galaxies that were not labeled as SBs in the MS, with a dust continuum size of $R_{\rm{e}}^{\rm{cor}} = 0.99 \pm 0.06$\,kpc.

Selection effects are at play in Fig.~\ref{fig:trends} and could potentially affect the correlations found. In general, as GOODS-ALMA 2.0 is a flux limited survey, for a fixed flux density threshold detections are biased against more extended sources. Therefore, the selection effects in Fig.~\ref{fig:trends} would miss more extended and gas poor galaxies that could potentially populate the bottom-left area in the $\tau_{\rm{dep}}$ and $f_{\rm{gas}}$ panels. Similarly, as cold galaxies are also favored, selection effects would miss more extended and warm galaxies that could populate the top-left area in the $T_{\rm{dust}}$ panel. Nonetheless, as the opposite corners of the three panels are not subject to these selection effects, the most compact galaxies would still have lower $\tau_{\rm{dep}}$ and $f_{\rm{gas}}$, and higher $T_{\rm{dust}}$ compared to the gas scaling relations of normal MS SFGs, at a fixed stellar mass and redshift. In addition, it is important to note that the galaxies represented in Fig.~\ref{fig:trends} have flux densities $S_{1.1\rm{mm}} > 1$\,mJy, the flux density threshold for which the GOODS-ALMA 2.0 survey reaches a $\sim 100\%$ completeness for sources with dust continuum sizes up to 1\arcsec~FWHM. In \citet{gomezguijarro21} we concluded that for sources with flux densities $S_{\rm{1.1mm}} > 1$\,mJy compact dust continuum emission at 1.1\,mm prevails, and sizes as extended as typical star-forming stellar disks are rare. Therefore, it is not expected that in the $S_{\rm{1.1mm}} > 1$\,mJy flux density regime included in Fig.~\ref{fig:trends} the selection function depends on the size. In \citet{gomezguijarro21}, for sources with $S_{\rm{1.1mm}} < 1$\,mJy, compact dust continuum emission appeared slightly more extended at 1.1\,mm, although still generally compact below the sizes of typical star-forming stellar disks. In this flux density regime, it would be plausible to find more extended galaxies populating the areas toward lower values in the $x$-axis in all the panels of Fig.~\ref{fig:trends}. It is also plausible to image some cases of outliers, while not violating the causes outlined by the correlations. For example, two merging galaxies that have not yet reached coalescence would appear more extended that what would correspond to the correlations, but not necessarily reflect variations in $\tau_{\rm{dep}}$, $f_{\rm{gas}}$, or $T_{\rm{dust}}$. Another case also related would be an already ignited starburst episode, reflecting shorter $\tau_{\rm{dep}}$ and higher $T_{\rm{dust}}$ than the scaling relations, which could also appear more extended and more gas rich than what would correspond to the correlations if it was receiving an extra gas supply from another merging galaxy that has not yet reached coalescence.

Recent studies by \citet{jin19} and \citet{cortzen20} have shown evidence for deceptively cold $T_{\rm{dust}}$ in massive starbursts at $z > 3$ derived under the common assumption of an optically thin far-IR dust emission. Conversely, when a general opacity solution was applied, the authors found warmer temperatures fully consistent with the excitation temperatures from far-IR lines. The authors concluded that the presence of optically thick dust up to longer wavelengths than the classical thin assumptions lead to, not only a systematic underestimation of $T_{\rm{dust}}$, but also an overestimation of $M_{\rm{dust}}$ and, thus, $M_{\rm{gas}}$ for a fixed $\delta_{\rm{GDR}}$. While the more compact the dust emitting region the more likely is to be optically thick up to longer wavelengths, the high $T_{\rm{dust}}$ estimates for our subset of SBs in the MS point to a correct opacity assumption. Even if the latter was not the case, a general opacity solution would lead to lower $M_{\rm{gas}}$ estimates, strengthening our conclusions.

Appendix~\ref{sec:appendix_c} tests the impact of the metallicity assumptions for $M_{\rm{gas}}$ estimates in the distribution of our galaxy sample in Fig.~\ref{fig:scl_rel}. Figs.~\ref{fig:scl_rel_fmr} and \ref{fig:scl_rel_solar} reproduce Fig.~\ref{fig:scl_rel} for $M_{\rm{gas}}$ estimates obtained by using the $\delta_{\rm{GDR}}$--$Z$ with FMR and fixed solar metallicity, respectively. For FMR (solar) the whole distribution shifts to higher (lower) $M_{\rm{gas}}$ regimes with respect to the gas scaling relations, as expected given the increase (decrease) of $\delta_{\rm{GDR}}$ for metallicities on average lower (higher) through FMR (solar) compare to MZR. The correlations still hold regardless of the metallicity assumption for $M_{\rm{gas}}$ estimates (except in the case of $\tau_{\rm{dep}}$ for fixed solar metallicity, although it is only due to one outlier). Besides, in all cases SBs in the MS move to the extremes of the trends.

Last, in the case of AGN there is no evidence for a correlation between the properties displayed in Fig.~\ref{fig:trends} and AGN activity, as their location in the panels is not particularly linked to a specific region of the plot or to the SBs in the MS preferentially respect to the remaining galaxies that were not labeled as SBs in the MS.

\section{Discussion} \label{sec:discussion}

In this section we interpret the results presented in the previous sections in the global picture of galaxy evolution. The scope of this discussion is to compare the expectations of common scenarios for galaxy evolution with our observations to verify up to what degree they are able to explain our measurements and whether updated scenarios need to be introduced to explain all the results presented in previous sections.

As outlined in Sect.~\ref{sec:intro}, the MS of SFGs and, in particular, its small scatter suggests that secular evolution is the dominant mode of stellar assembly. In this picture, gas inflows, outflows, and consumption in star formation are in a steady equilibrium regulating galaxy evolution \citep[e.g.,][]{daddi10,genzel10,tacconi10,dekel13,feldmann15}. As a result, SFGs would spend most of their time evolving as extended star-forming disks. Conversely, QGs are located below the MS and are typically more compact than SFGs at a fixed stellar mass and redshift \citep[e.g.,][]{shen03,vanderwel14,barro17,suess19,mowla19}. Therefore, the quenching of star formation appears to involve the build-up of a central stellar core \citep[e.g.,][]{kauffmann03,whitaker17,barro17,gomezguijarro19,suess21}. Optically compact MS SFGs have been proposed as the missing link between the SFGs and the more compact QGs, as they exhibit star formation consistent with the trends of more extended MS SFGs, but with a stellar structure comparable to QGs \citep[e.g.,][]{barro13,nelson14,williams14,vandokkum15,gomezguijarro19}. Other recent works have also indicated the existence of galaxies that exhibit compact emission in several far-IR tracers, such as dust continuum at submm/mm and radio wavelengths or CO lines, while located within the scatter of the MS \citep[e.g.,][]{elbaz18,jimenezandrade19,puglisi19,puglisi21,franco20b}. However, the origin, nature, and role in galaxy evolution of these different types of compact MS SFGs galaxies remains to be understood.

\begin{figure*}
\begin{center}
\includegraphics[width=0.33\textwidth]{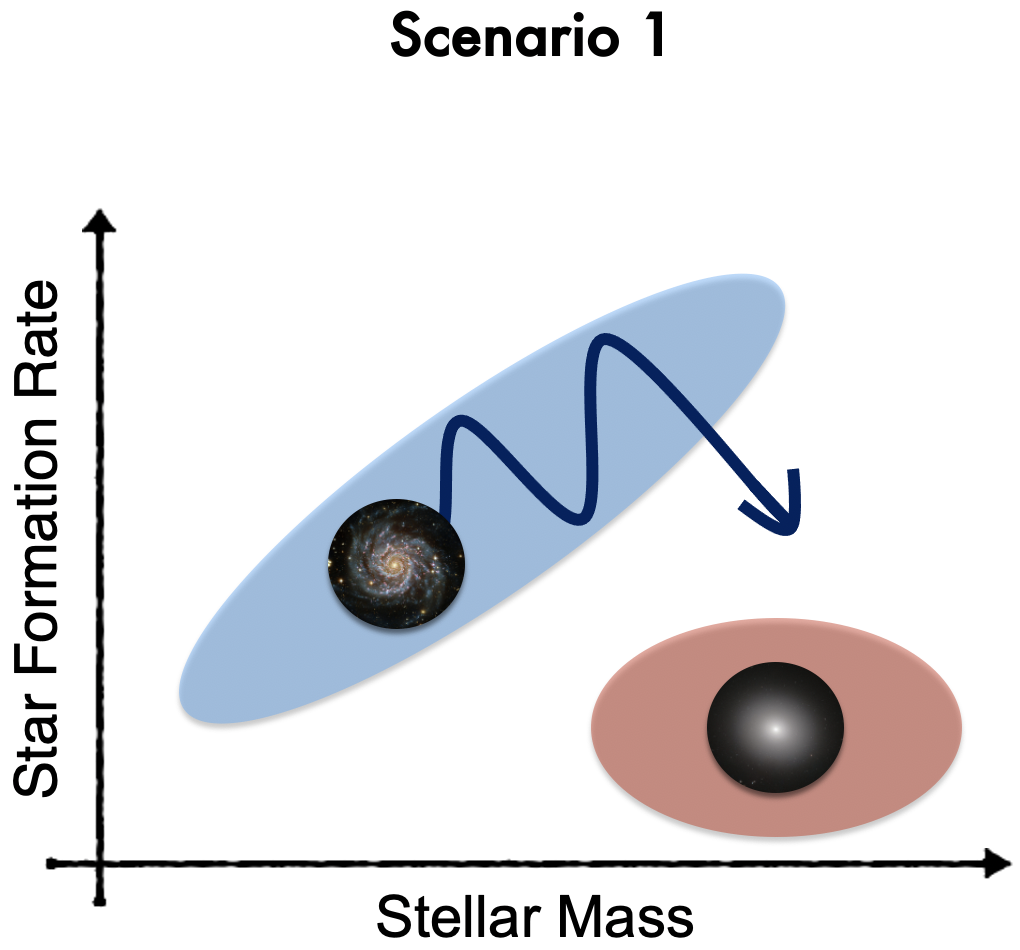}
\includegraphics[width=0.33\textwidth]{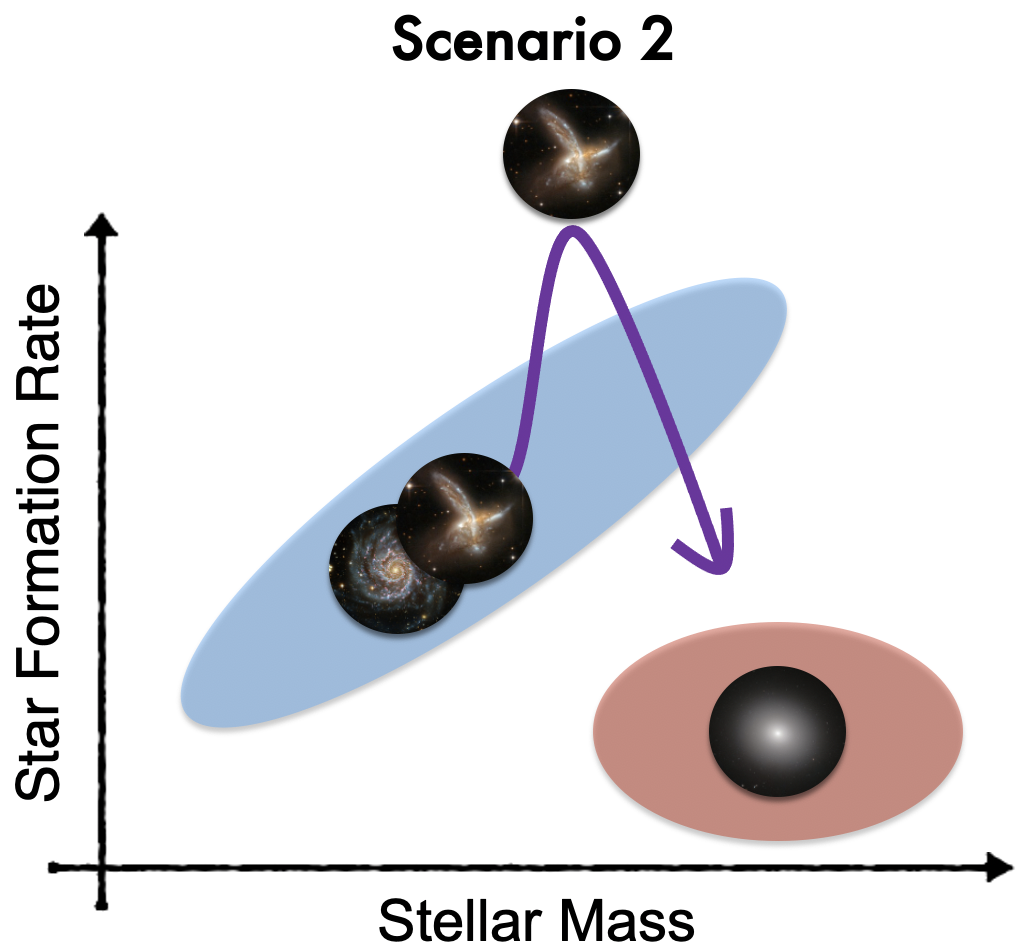}
\includegraphics[width=0.33\textwidth]{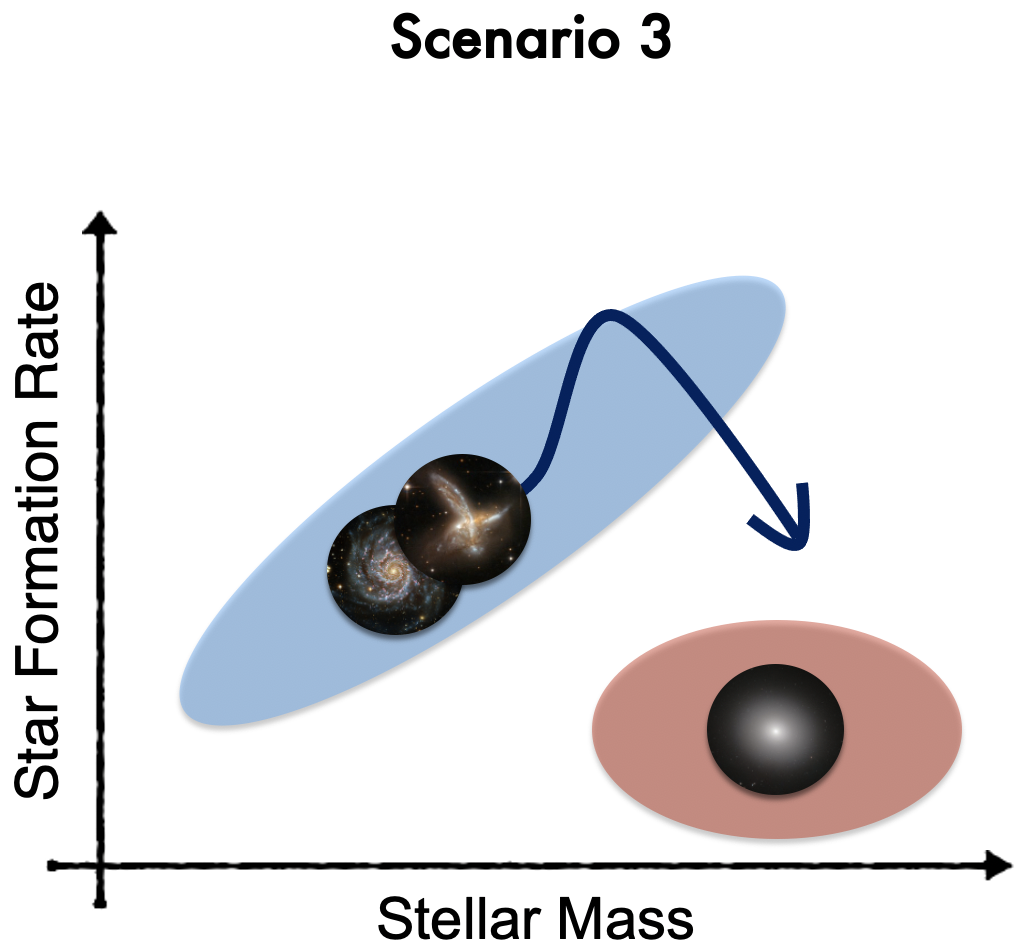}
\caption{Drawing depicting possible scenarios of galaxy evolution in the main sequence framework. Scenario 1 (left panel): external gas inflow triggers a compaction event. The gas is funneled to the center of the galaxy and it moves to the upper bound of the MS. This phase is characterized by a compact core with a high $f_{\rm{gas}}$, high SFR, and low $\tau_{\rm{dep}}$, compared to normal MS SFGs at a fixed stellar mass and redshift. Central gas depletion leads to inside-out quenching and the galaxy moves to the lower bound of the MS. This implies that when MS SFGs have low $f_{\rm{gas}}$, they exhibit a more extended star-forming region in comparison to the previous phase. Scenario 2 (middle panel): external gas supply from gas-rich mergers. The gas is funneled to the center of collision, enhances SFR, and the galaxy moves well above the MS. As the gas reservoir is consumed, the SFR declines and the galaxy crosses, incidentally, the MS. Scenario 3 (right panel): angular momentum loss is driven externally (accretion through mergers or counter-rotating streams) or internally (clump migration). The gas is funneled to the center of the galaxy in either case (represented by a merger in the panel), enhances SFR (although not necessarily directly or well above the MS in all cases). As the gas reservoir is consumed the star-forming region becomes more compact sustaining the SFR (outside-in gas consumption). In all three scenarios, the assumption is that at high masses the hot dark matter halo prevents further gas inflow and the galaxy eventually quenches.}
\label{fig:sum_evol}
\end{center}
\end{figure*}

On one hand, several studies have advocated for compaction events predicted by galaxy formation models in which extended MS SFGs can secularly evolve into compact MS SFGs by funneling gas to their central regions yielding the build up of their stellar cores \citep[e.g.,][]{dekel13,zolotov15,tacchella16}. In particular \citet{tacchella16} describes this scenario as the following (Scenario 1, Fig.~\ref{fig:sum_evol} left panel): an intense gas inflow event triggers the so-called compaction event, involving counter-rotating streams or minor mergers, and is commonly associated with violent disk instabilities. The inflow rate is more efficient than the SFR. The gas is funneled to the center of the galaxy and a compact massive core of gas and star formation rate develops. This phase is characterized by a short depletion time and a high gas fraction. Subsequently, the galaxy moves to the upper bound of the MS scatter and the downturn is associated with the peak of SFR and outflow and the suppression of the inflow. Consequently, central gas depletion leads to inside-out quenching and the galaxy moves to the lower bound of the MS scatter. At low masses, the dark matter halo is still cold and gas inflow is resumed and subsequent compaction events follow. At high masses, the hot dark matter halo prevents further gas inflow, leading to gas depletion and full quenching.

On the other hand, other studies have advocated that SBs dominated by a violent episode of star formation typical of gas-rich mergers are also capable of funneling gas to the center of collision and quickly build up compact stellar cores \citep[e.g.,][]{mihos96,hopkins06,toft14,gomezguijarro18,puglisi21}. In this case (Scenario 2, Fig.~\ref{fig:sum_evol} middle panel): an extra supply of gas comes from gas-rich galaxy mergers that efficiently dissipate angular momentum. The gas is funneled to the center of collision, becomes denser, and star formation is enhanced. As a result, the galaxy moves well above the MS. Subsequently, it falls toward the MS when the SFR declines as the gas reservoir is consumed. Finally, the galaxy crosses the MS toward quiescence. SBs in the MS would be galaxies on their way to quiescence, incidentally in the MS, but without being part of it \citep[e.g.,][]{elbaz18,gomezguijarro19,puglisi21}.

Our findings are more in line with Scenario 2. In particular, the characteristics of the SBs in the MS, exhibiting the most compact star formation, shortest depletion times, lowest gas fractions (and highest dust temperatures) of our galaxy sample. These characteristics point toward SBs in the MS as a stage prior to passivization that could potentially evolve into optically compact MS SFGs once the central stellar core is built up and eventually quench forming QGs. Conversely, in Scenario 1, gas-depleted galaxies in the lower bound of the MS should be more extended than those in the upper bound as the gas is centrally depleted. Besides, we do not find galaxies characterized simultaneously by ongoing compact star formation, short depletion times, high gas fractions, and located within the scatter of the MS, as predicted in Scenario 1 in the upper bound of the MS.

Selection effects described in previous sections are at play. In the low stellar mass end ($\log (M_{\rm{*}}/M_{\odot}) < 10.5$), the selection effects act in favor of finding galaxies above the scaling relation in the $f_{\rm{gas}}$--$M_{\rm{*}}$ plane, but against finding galaxies below the scaling relation in the $\tau_{\rm{dep}}$--$M_{\rm{*}}$ plane. In the high stellar mass end ($\log (M_{\rm{*}}/M_{\odot}) > 10.5$) galaxies above and below the scaling relations are within the reach of the survey. Therefore, galaxies characterized simultaneously by ongoing compact star formation, short depletion times, high gas fractions, and located within the scatter of the MS (as predicted in Scenario 1 in the upper bound of the MS), although not present in the high mass end, they could still exist in the low stellar mas end but remain out of reach to our survey. We note that the interpretation outlined in this discussion is, therefore, to be suitable in the high stellar mass end. In this massive regime, we assumed that quenching proceeds providing that gas replenishment is halted or strongly suppressed, following studies that indicated that the hot dark matter halo at high masses prevents further gas inflow and/or gas cooling leading to gas depletion and full quenching \citep[e.g.,][]{rees77,birnboim03,feldmann15}.

An aspect to consider is that in both Scenario 1 and Scenario 2 the loss of angular momentum to compact the galaxy comes from an externally-driven process, invoking counter rotating streams or mergers. Some studies have indicated that internally-driven processes can also compact the galaxy, such as violent instabilities in gas-rich systems at high redshift that create giant clumps that migrate to the center of the galaxy \citep[e.g.,][]{elmegreen08}. In addition, feedback-induced turbulence in a SB is a proposed mechanism to self-regulate star formation in these systems \citep[e.g.,][]{ostriker11,lehnert13,brucy20}, so the extra supply of gas from gas-rich galaxy mergers could not always result in enhanced star formation or the enhancement could be delayed. In Fig.~\ref{fig:scl_rel} we see that there are some galaxies above the scaling relation in the $f_{\rm{gas}}$--$M_{\rm{*}}$ within the scatter of the MS and, simultaneously, above the scaling relation in the $\tau_{\rm{dep}}$--$M_{\rm{*}}$ plane. Galaxies of this kind were found by \citet{kokorev21}, so-called gas giants, and their nature is still to be confirmed. They could be candidates to galaxies in which SFR is not enhanced, either caught in an early phase before doing so or self-regulated by their own feedback-driven turbulence.

Another important aspect to be considered are our findings in Sect~\ref{sec:compact} depicting correlations between the physical extent of the dust continuum areas and the deviations from the gas scaling relations. In Fig~\ref{fig:trends} we showed that as dust continuum areas become more compact compared to the structural scaling relation of typical star-forming stellar disks, $\tau_{\rm{dep}}$ and $f_{\rm{gas}}$ become lower and $T_{\rm{dust}}$ higher compared to the gas scaling relations of normal MS SFGs, at a fixed stellar mass and redshift. In addition, SBs in the MS appear to be the extreme cases where the dust continuum areas are the most compact ones, associated with the lowest $\tau_{\rm{dep}}$ and $f_{\rm{gas}}$, and the highest $T_{\rm{dust}}$. In particular, it is important to note the result in terms of the gas fraction depicted in Fig~\ref{fig:sum_gas}, reflecting that the less amount of gas in the galaxy, the more compact the star-forming region becomes. Moreover, SBs in the MS form a high fraction of the subset of the most massive galaxies in the sample ($\log (M_{\rm{*}}/M_{\odot}) > 11.0$), accounting for 59\% (19/32, where the total number of galaxies accounts only for galaxies with a \textit{Herschel} counterpart in that mass regime).

These aspects appear to imply that the SFR is somehow sustained, keeping galaxies within the MS even when the $f_{\rm{gas}}$ is low in very massive systems, presumably on their way to quiescence. It seems that gas and star formation compression allows to hold the SFR. While the specific physical mechanism at play is beyond the scope of this paper, it could be plausible to think that as the gas is funneled to the central regions and stars are formed at the core, the potential well gets steeper and, provided angular momentum is lost, gas could be funneled progressively more and more to the core.

In the latter case, there is one point to be considered for galaxies with enhanced star formation that move well above the MS and, subsequently, fall toward the MS when the SFR is declined as their gas reservoir is consumed. They would slow down in their descent as the gas and star formation are compressed and the SFR held, in comparison to the case in which the gas is consumed but without compression. In Fig.~\ref{fig:trends} we see that this could be the case judging from the relative locations of SBs well above the MS ($\Delta \rm{MS} > 10$, yellow points) compared to SBs in the MS in the $\Delta f_{\rm{gas,MS}}$--$\Delta A_{\rm{opt/1.1mm}}$ plane. SBs well above the MS appear more extended compared to SBs in the MS. However, at the moment the sample size for both SBs in the MS and SBs above the MS in this diagram is scarce to draw strong conclusions.

All together, an alternative scenario could be at play (Scenario 3, Fig.~\ref{fig:sum_evol} right panel): angular momentum loss is driven either externally (accretion through mergers or counter-rotating streams) or internally (clump migration). The gas is funneled to the center in either case (represented by a merger in Fig.~\ref{fig:sum_evol} right panel) and star formation enhanced, although not necessarily directly or well above the MS in all cases. In any case, as the gas reservoir is consumed, the star-forming region becomes more compact sustaining the SFR (outside-in gas consumption). The galaxy would spend more time on average within the MS compared to that of Scenario 2. SBs in the MS would be galaxies on their way to quiescence, but still experiencing the last phases of regulation implied by the MS.

Scenario 3 is consistent with a slow downfall \citep[e.g.,][]{schreiber16,gomezguijarro19,franco20b}, in which galaxies seem to progressively reduce their $f_{\rm{gas}}$ and SFR as they assemble their stellar masses and build-up their compact stellar cores. Last, far-IR compact MS SFGs exhibit properties that resemble those of SBs, such us higher gas densities and more excited spectral line energy distributions compared to normal MS SFGs \citep{puglisi21}. Optically compact SFGs exhibit far-IR and radio properties consistent with being old SBs winding down compared to normal MS SFGs \citep{gomezguijarro19}. Both Scenario 2 and 3 could explain these characteristics. In particular, in the case of Scenario 3 it would mean that as star formation becomes more compact it acts effectively as a central starburst.

\begin{table}
\scriptsize
\caption{Summary of expectations and observations for the different scenarios of galaxy evolution in the main sequence framework}
\label{tab:sum_evol}
\centering
\begin{tabular}{p{1.5cm}p{3cm}p{3cm}}
\hline\hline
 & Expected & Observed \\
\hline
Scenario 1 & In the MS higher gas content related to more compact star-forming region than lower gas content & In the MS the lowest gas content related to the most compact star-forming region \\
Scenario 2 & Compact star-forming region well above the MS related to higher gas content and in the MS related to lower gas content & In the MS the lowest gas content related to more compact star-forming region than well above the MS with higher gas content \\
Scenario 3 & \multicolumn{2}{p{6cm}}{In the MS higher gas content related to more extended star-forming region than lower gas content. The less amount of gas in the galaxy, the more compact the star-forming region becomes} \\
\hline
\end{tabular}
\tablefoot{This table summarizes the main expectations and observational aspects with respect to the gas content and the sizes of the star-forming regions. We note that for Scenario 3 expectations and observations are the same, as this scenario is motivated to match the observations this work.}
\end{table}

A summary of each scenario is provided in Fig.~\ref{fig:sum_evol} and its caption. Table~\ref{tab:sum_evol} provides a summary of what the expectations are of the different scenarios compared to the observations in this work. Note also that not all aspects of the scenarios are mutually exclusive. In terms of the low gas fraction phase, Scenario 1 would lead to a more extended star-forming region configuration excluded in Scenario 2 and 3. In the case of Scenarios 2 and 3, both lead to a more compact star-forming region associated with the low gas fraction phase, but Scenario 3 includes the fact that it would be the most compact, even more compact than SBs above the MS to reproduce the correlation found. In terms of the mechanism triggering compact star formation, both Scenario 1 and 2 invoke external processes. In the case of Scenario 1, these are predominately accretion-driven counter rotating streams or minor mergers. Scenario 2 instead alludes to the importance of gas-rich mergers. Scenario 3 alludes to the possibility of both external and internal processes, as the observables studied in this work do not preferentially point to either of them, and both seem plausible according to literature studies.

\section{Summary and conclusions} \label{sec:summary}

In this work we follow up on the galaxy sample presented in \citet{gomezguijarro21} from the GOODS-ALMA 2.0 survey, an ALMA blind survey at 1.1\,mm covering a contiguous area of 72.42\,arcmin$^2$ using two array configurations at a similar and homogeneous depth over the whole field. The combined mosaic with both configurations reaches an average sensitivity of 68.4$\,\mu$Jy beam$^{-1}$ at an average angular resolution of 0\farcs447\,$\times$\,0\farcs418. We present the dust and stellar-based properties of the galaxy sample. In particular, we study the physical properties of the galaxies in the framework of the main sequence and the scaling relations for depletion timescales, gas fractions, and dust temperatures. We focus on the role of compact star formation traced by the dust continuum emission as a physical driver of the galaxies properties, and on the origin, nature, and role of starburst galaxies located within the main sequence of SFGs. In summary we find:

\begin{itemize}

\item There exist a subset of galaxies that exhibit short depletion timescales compared to typical SFGs at the same redshift, stellar mass, and distance to the main sequence. At the same time, these galaxies are located within the scatter of the main sequence. Therefore, we describe these galaxies as starbursts in the main sequence.

\item Starbursts in the main sequence are characterized by short depletion timescales, low gas fractions, and high dust temperatures in comparison with typical main sequence galaxies at the same stellar mass and redshift. They also exhibit high stellar masses, accounting for 59\% of the most massive galaxies in the sample ($\log (M_{\rm{*}}/M_{\odot}) > 11.0$).

\item Dust continuum areas at 1.1\,mm appear to correlate with depletion timescales, gas fractions, and dust temperatures. As galaxies become more compact compared to typical SFGs stellar disks, depletion timescales and gas fractions become lower, and dust temperatures higher, compared to typical main sequence SFGs, at the same stellar mass and redshift. These trends manifest a direct link between a geometrical property like the sizes of the ongoing star formation regions and other physical properties of the galaxies.

\item Starbursts in the main sequence appear to be located at the extremes of the trends, reflecting that they are extreme cases where the dust continuum areas are the most compact ones, associated with the shortest depletion timescales, lowest gas fractions, and the highest dust temperatures, compared to typical main sequence SFGs at the same stellar mass and redshift.

Our findings suggest that the star formation rate is somehow sustained in very massive SFGs, keeping them within the main sequence even when their gas fractions are low and they are presumably on their way to quiescence. It seems that gas and star formation compression allows to hold their star formation rate.

\end{itemize}

\begin{acknowledgements}

V.I.K. and G.E.M. acknowledge the Villum Fonden research grant 13160 “Gas to stars, stars to dust: tracing star formation across cosmic time” and the Cosmic Dawn Center of Excellence funded by the Danish National Research Foundation under the grant No. 140. M.T.S. acknowledges support from a Scientific Exchanges visitor fellowship (IZSEZO\_202357) from the Swiss National Science Foundation. M.F. acknowledges the support from STFC (grant number ST/R000905/1). H.I. acknowledges support from JSPS KAKENHI Grant Number JP19K23462 and JP21H01129. This paper makes use of the following ALMA data: ADS/JAO.ALMA\#2015.1.00543.S and ADS/JAO.ALMA\#2017.1.00755.S. ALMA is a partnership of ESO (representing its member states), NSF (USA) and NINS (Japan), together with NRC (Canada), MOST and ASIAA (Taiwan), and KASI (Republic of Korea), in cooperation with the Republic of Chile. The Joint ALMA Observatory is operated by ESO, AUI/NRAO and NAOJ.

\end{acknowledgements}

\bibliographystyle{aa}
\bibliography{A2GS_pp.bib}

\begin{appendix}

\onecolumn

\section{Infrared SED fits} \label{sec:appendix_a}

\begin{figure*}[h!]
\begin{center}
\includegraphics[width=\textwidth]{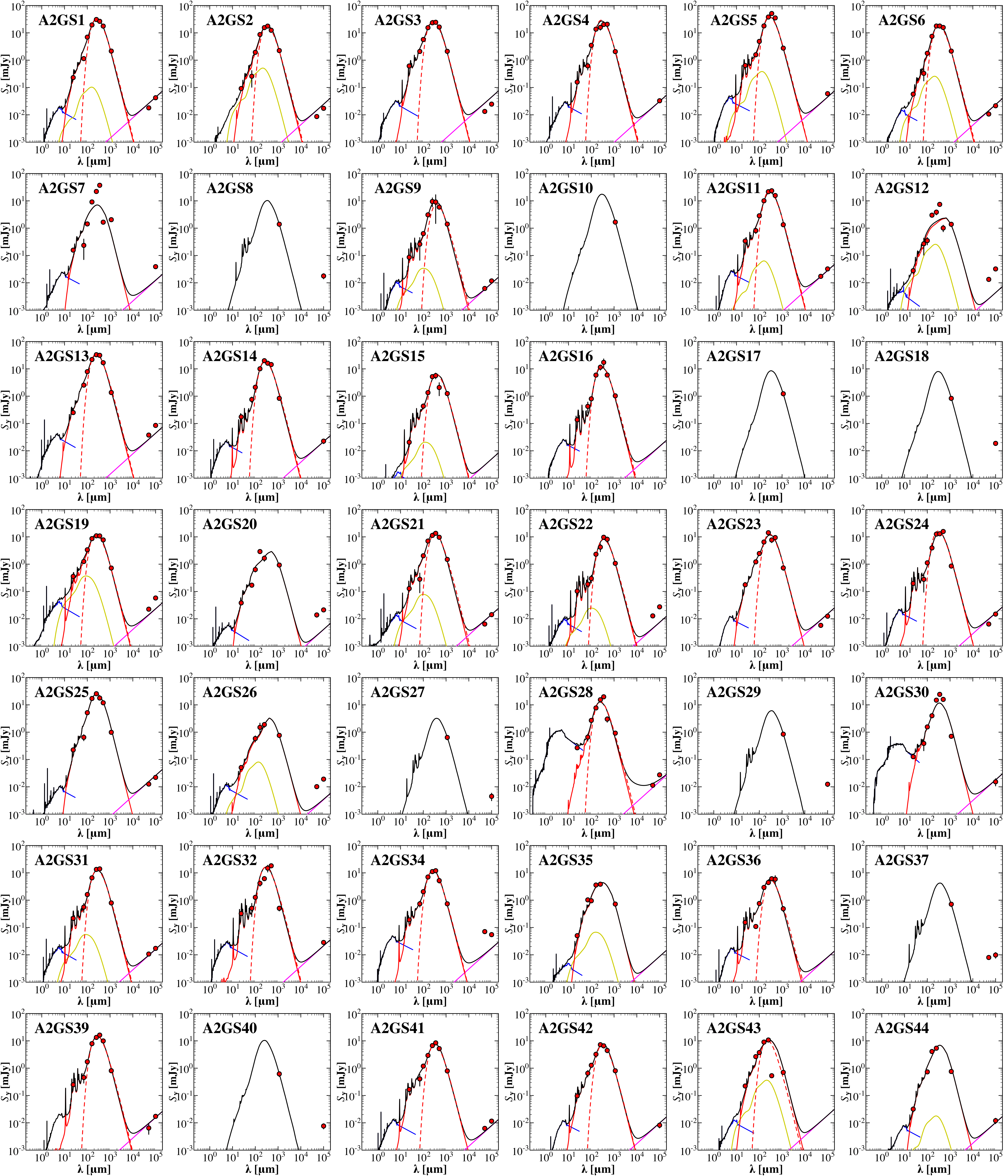}
\caption{Mid-IR to mm SEDs for the galaxy sample. A2GS1 to A2GS44 correspond to the 100\% pure source catalog in \citet{gomezguijarro21}. \textit{Herschel}-detected galaxies were fit with the code \texttt{Stardust} accounting for the stellar emission (blue), AGN contribution (yellow), dust emission from star formation (red), and radio emission (magenta). MBB fit is shown in dashed red. The galaxies without \textit{Herschel} counterpart were fit following an iterative approach with the dust SED libraries by \citet{schreiber18} (see Sect.~\ref{subsec:fir_sed}).}
\label{fig:fir_seds_1}
\end{center}
\end{figure*}

\begin{figure*}
\begin{center}
\includegraphics[width=\textwidth]{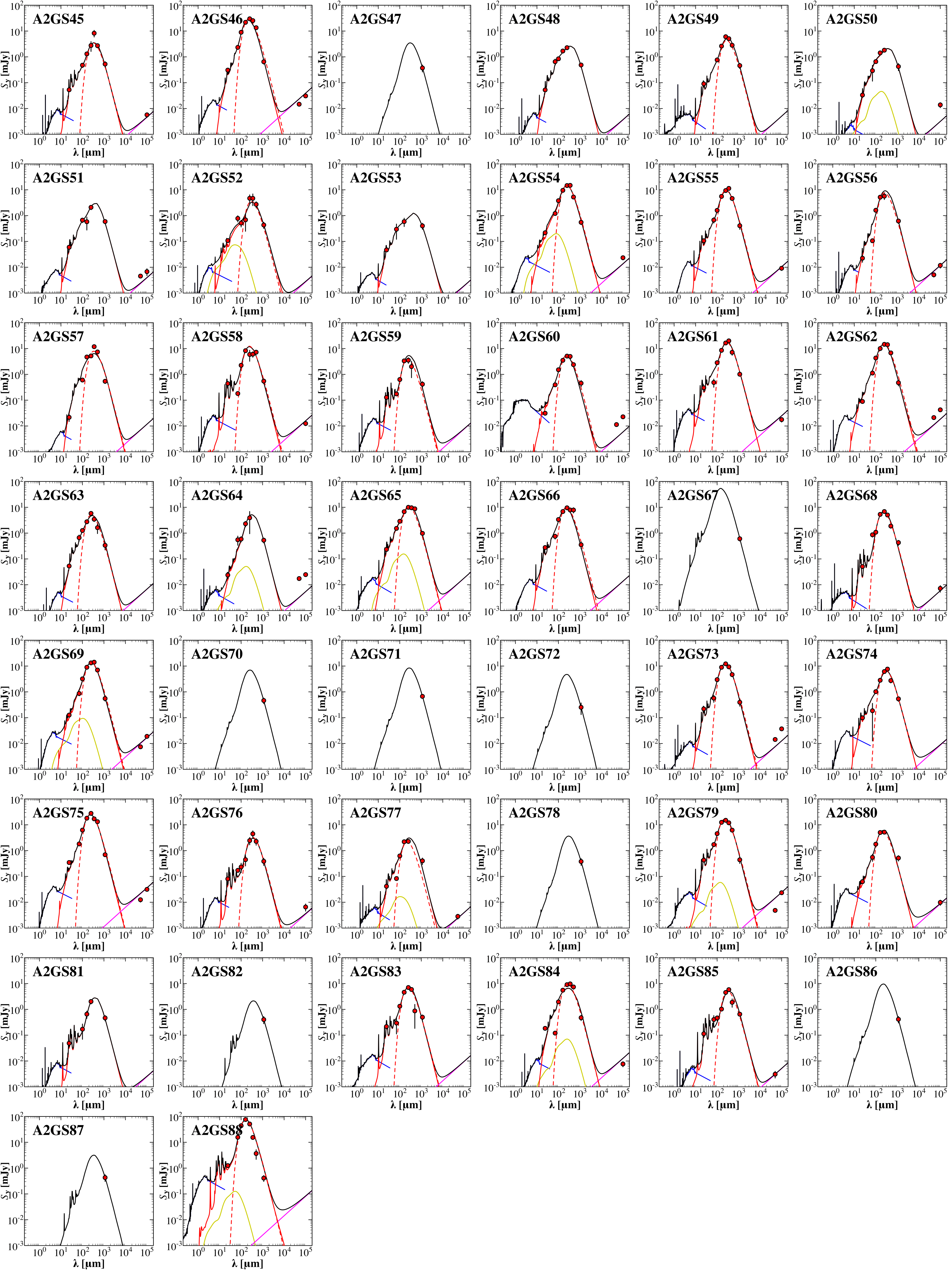}
\caption{Mid-IR to mm SEDs for the galaxy sample as in Fig.~\ref{fig:fir_seds_1}. A2GS45 to A2GS88 correspond to the prior-based source catalog in \citet{gomezguijarro21}.}
\label{fig:fir_seds_2}
\end{center}
\end{figure*}

\FloatBarrier

\section{Infrared flux predictions} \label{sec:appendix_b}

\begin{figure}[h!]
\begin{center}
\includegraphics[width=0.5\textwidth]{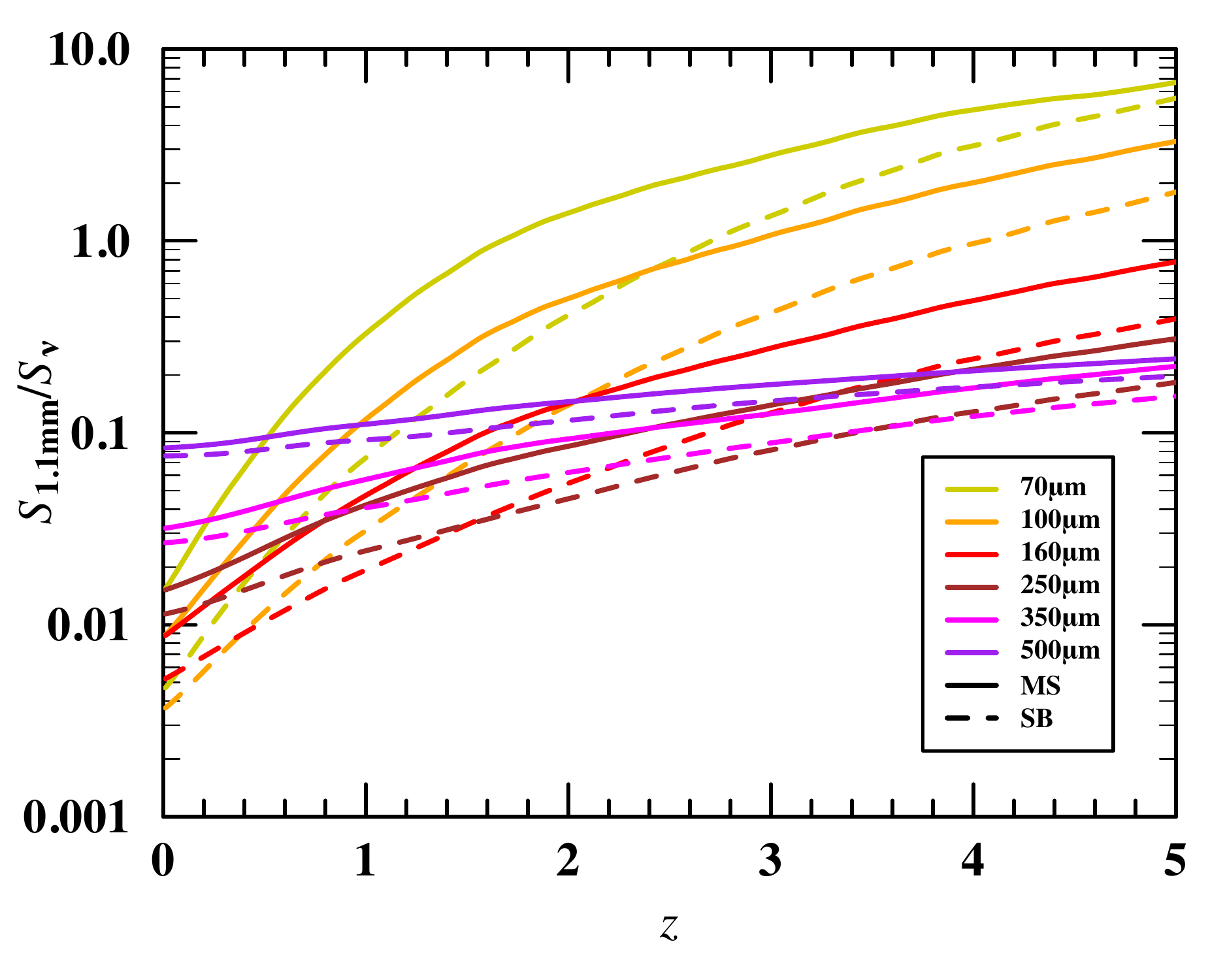}
\caption{Predictions of flux ratios between ALMA 1.1\,mm and \textit{Herschel} PACS (70, 100, 160\,$\mu$m) and SPIRE (250, 350, 500\,$\mu$m) bands as a function of redshift for both MS ($\Delta \rm{MS} = 1$) and SB ($\Delta \rm{MS} = 5$) templates. See Table~\ref{tab:fluxratio_pred} for these flux ratio predictions tabulated.}
\label{fig:fluxratio_pred}
\end{center}
\end{figure}

\begin{table*}
\scriptsize
\caption{Flux ratios between ALMA 1.1\,mm and \textit{Herschel} bands for a MS (SB) template as a function of redshift.}
\label{tab:fluxratio_pred}
\centering
\begin{tabular}{lcccccc}
\hline\hline
$z$ & $S_{\rm{1.1mm}}/S_{\rm{70\mu m}}$ & $S_{\rm{1.1mm}}/S_{\rm{100\mu m}}$ & $S_{\rm{1.1mm}}/S_{\rm{160\mu m}}$ & $S_{\rm{1.1mm}}/S_{\rm{250\mu m}}$ & $S_{\rm{1.1mm}}/S_{\rm{350\mu m}}$ & $S_{\rm{1.1mm}}/S_{\rm{500\mu m}}$ \\  &  &  &  &  &  &  \\
\hline
0.1 & 0.0214 (0.0065) & 0.0112 (0.0046) & 0.0103 (0.0060) & 0.0164 (0.0121) & 0.0329 (0.0274) & 0.0844 (0.0763) \\
0.2 & 0.0322 (0.0089) & 0.0154 (0.0058) & 0.0125 (0.0069) & 0.0181 (0.0129) & 0.0347 (0.0282) & 0.0862 (0.0769) \\
0.3 & 0.0472 (0.0119) & 0.0208 (0.0072) & 0.0150 (0.0078) & 0.0202 (0.0138) & 0.0367 (0.0292) & 0.0884 (0.0779) \\
0.4 & 0.0676 (0.0169) & 0.0282 (0.0094) & 0.0182 (0.0092) & 0.0227 (0.0152) & 0.0393 (0.0308) & 0.0915 (0.0800) \\
0.5 & 0.0877 (0.0222) & 0.0354 (0.0117) & 0.0210 (0.0104) & 0.0249 (0.0166) & 0.0415 (0.0323) & 0.0942 (0.0821) \\
0.6 & 0.1207 (0.0286) & 0.0467 (0.0143) & 0.0251 (0.0118) & 0.0280 (0.0179) & 0.0444 (0.0338) & 0.0978 (0.0842) \\
0.7 & 0.1622 (0.0364) & 0.0610 (0.0174) & 0.0301 (0.0134) & 0.0315 (0.0195) & 0.0478 (0.0355) & 0.1018 (0.0865) \\
0.8 & 0.2125 (0.0481) & 0.0781 (0.0217) & 0.0356 (0.0153) & 0.0352 (0.0212) & 0.0512 (0.0375) & 0.1055 (0.0888) \\
0.9 & 0.2707 (0.0626) & 0.0977 (0.0268) & 0.0415 (0.0175) & 0.0388 (0.0230) & 0.0544 (0.0393) & 0.1084 (0.0906) \\
1.0 & 0.3408 (0.0755) & 0.1215 (0.0314) & 0.0482 (0.0195) & 0.0425 (0.0245) & 0.0576 (0.0409) & 0.1112 (0.0921) \\
1.1 & 0.3943 (0.0898) & 0.1402 (0.0363) & 0.0536 (0.0214) & 0.0456 (0.0258) & 0.0603 (0.0423) & 0.1135 (0.0933) \\
1.2 & 0.4795 (0.1125) & 0.1692 (0.0439) & 0.0613 (0.0242) & 0.0494 (0.0276) & 0.0637 (0.0443) & 0.1164 (0.0953) \\
1.3 & 0.5809 (0.1312) & 0.2046 (0.0506) & 0.0704 (0.0268) & 0.0538 (0.0294) & 0.0675 (0.0463) & 0.1200 (0.0974) \\
1.4 & 0.6952 (0.1524) & 0.2461 (0.0576) & 0.0813 (0.0292) & 0.0590 (0.0308) & 0.0721 (0.0480) & 0.1244 (0.0993) \\
1.5 & 0.7745 (0.1864) & 0.2746 (0.0689) & 0.0885 (0.0330) & 0.0622 (0.0332) & 0.0750 (0.0505) & 0.1274 (0.1024) \\
1.6 & 0.9146 (0.2257) & 0.3257 (0.0819) & 0.1009 (0.0374) & 0.0676 (0.0358) & 0.0796 (0.0533) & 0.1319 (0.1056) \\
1.7 & 1.0020 (0.2560) & 0.3574 (0.0919) & 0.1091 (0.0407) & 0.0711 (0.0377) & 0.0827 (0.0553) & 0.1350 (0.1080) \\
1.8 & 1.1623 (0.3050) & 0.4149 (0.1071) & 0.1223 (0.0452) & 0.0766 (0.0402) & 0.0869 (0.0577) & 0.1391 (0.1109) \\
1.9 & 1.2595 (0.3612) & 0.4504 (0.1246) & 0.1309 (0.0503) & 0.0802 (0.0429) & 0.0896 (0.0601) & 0.1419 (0.1137) \\
2.0 & 1.3622 (0.4014) & 0.4869 (0.1371) & 0.1395 (0.0540) & 0.0839 (0.0451) & 0.0921 (0.0619) & 0.1445 (0.1159) \\
2.1 & 1.5330 (0.4675) & 0.5522 (0.1576) & 0.1544 (0.0598) & 0.0900 (0.0482) & 0.0962 (0.0645) & 0.1485 (0.1190) \\
2.2 & 1.6287 (0.5422) & 0.5885 (0.1804) & 0.1636 (0.0660) & 0.0942 (0.0516) & 0.0989 (0.0672) & 0.1517 (0.1223) \\
2.3 & 1.7244 (0.6267) & 0.6255 (0.2059) & 0.1722 (0.0727) & 0.0979 (0.0551) & 0.1011 (0.0698) & 0.1543 (0.1254) \\
2.4 & 1.9218 (0.6836) & 0.7047 (0.2231) & 0.1901 (0.0775) & 0.1051 (0.0577) & 0.1054 (0.0717) & 0.1586 (0.1279) \\
2.5 & 2.0105 (0.7789) & 0.7410 (0.2521) & 0.2000 (0.0852) & 0.1097 (0.0617) & 0.1083 (0.0746) & 0.1618 (0.1314) \\
2.6 & 2.2131 (0.8860) & 0.8266 (0.2843) & 0.2197 (0.0933) & 0.1175 (0.0656) & 0.1130 (0.0775) & 0.1664 (0.1347) \\
2.7 & 2.3127 (1.0014) & 0.8693 (0.3185) & 0.2294 (0.1018) & 0.1212 (0.0698) & 0.1149 (0.0804) & 0.1685 (0.1380) \\
2.8 & 2.3925 (1.1285) & 0.9038 (0.3574) & 0.2397 (0.1112) & 0.1259 (0.0741) & 0.1179 (0.0834) & 0.1715 (0.1412) \\
2.9 & 2.6165 (1.2073) & 0.9971 (0.3803) & 0.2596 (0.1169) & 0.1332 (0.0768) & 0.1222 (0.0853) & 0.1754 (0.1432) \\
3.0 & 2.8484 (1.3432) & 1.0964 (0.4210) & 0.2820 (0.1265) & 0.1415 (0.0812) & 0.1270 (0.0885) & 0.1795 (0.1463) \\
3.1 & 2.9566 (1.5005) & 1.1395 (0.4683) & 0.2916 (0.1373) & 0.1452 (0.0859) & 0.1293 (0.0918) & 0.1812 (0.1494) \\
3.2 & 3.1958 (1.6649) & 1.2412 (0.5174) & 0.3147 (0.1480) & 0.1538 (0.0904) & 0.1345 (0.0950) & 0.1853 (0.1522) \\
3.3 & 3.2979 (1.8409) & 1.2844 (0.5708) & 0.3236 (0.1596) & 0.1570 (0.0953) & 0.1366 (0.0984) & 0.1867 (0.1550) \\
3.4 & 3.5444 (2.0242) & 1.3964 (0.6260) & 0.3482 (0.1716) & 0.1655 (0.1003) & 0.1417 (0.1019) & 0.1902 (0.1578) \\
3.5 & 3.8083 (2.1329) & 1.5140 (0.6580) & 0.3727 (0.1785) & 0.1740 (0.1033) & 0.1469 (0.1042) & 0.1939 (0.1596) \\
3.6 & 3.8906 (2.3364) & 1.5527 (0.7207) & 0.3840 (0.1919) & 0.1788 (0.1089) & 0.1502 (0.1082) & 0.1962 (0.1628) \\
3.7 & 4.1455 (2.5523) & 1.6728 (0.7873) & 0.4108 (0.2057) & 0.1881 (0.1144) & 0.1558 (0.1121) & 0.2001 (0.1658) \\
3.8 & 4.4072 (2.7798) & 1.8005 (0.8593) & 0.4391 (0.2203) & 0.1974 (0.1200) & 0.1612 (0.1160) & 0.2037 (0.1688) \\
3.9 & 4.6629 (3.0191) & 1.9278 (0.9356) & 0.4684 (0.2352) & 0.2073 (0.1255) & 0.1670 (0.1195) & 0.2075 (0.1713) \\
4.0 & 4.7213 (3.1304) & 1.9610 (0.9686) & 0.4775 (0.2425) & 0.2111 (0.1287) & 0.1697 (0.1220) & 0.2092 (0.1732) \\
4.1 & 4.9384 (3.3752) & 2.0904 (1.0503) & 0.5069 (0.2592) & 0.2206 (0.1351) & 0.1751 (0.1262) & 0.2126 (0.1763) \\
4.2 & 5.1402 (3.4927) & 2.2222 (1.0856) & 0.5365 (0.2650) & 0.2304 (0.1372) & 0.1807 (0.1276) & 0.2163 (0.1772) \\
4.3 & 5.3284 (3.7507) & 2.3636 (1.1726) & 0.5696 (0.2813) & 0.2411 (0.1431) & 0.1864 (0.1313) & 0.2200 (0.1799) \\
4.4 & 5.5201 (4.0051) & 2.5058 (1.2588) & 0.6019 (0.2972) & 0.2515 (0.1488) & 0.1919 (0.1348) & 0.2234 (0.1824) \\
4.5 & 5.7285 (4.2821) & 2.6485 (1.3540) & 0.6354 (0.3156) & 0.2629 (0.1557) & 0.1982 (0.1391) & 0.2276 (0.1857) \\
4.6 & 5.7424 (4.4140) & 2.6869 (1.3974) & 0.6430 (0.3234) & 0.2660 (0.1589) & 0.2002 (0.1412) & 0.2290 (0.1874) \\
4.7 & 5.9467 (4.7096) & 2.8400 (1.4997) & 0.6766 (0.3413) & 0.2766 (0.1653) & 0.2057 (0.1451) & 0.2326 (0.1903) \\
4.8 & 6.1782 (4.9964) & 2.9968 (1.6030) & 0.7115 (0.3601) & 0.2878 (0.1719) & 0.2113 (0.1489) & 0.2361 (0.1931) \\
4.9 & 6.4334 (5.3042) & 3.1509 (1.7130) & 0.7450 (0.3788) & 0.2987 (0.1783) & 0.2168 (0.1525) & 0.2396 (0.1956) \\
5.0 & 6.7310 (5.4396) & 3.3228 (1.7577) & 0.7834 (0.3858) & 0.3110 (0.1811) & 0.2230 (0.1542) & 0.2436 (0.1970) \\
\hline
\end{tabular}
\end{table*}

\FloatBarrier

\clearpage

\section{Impact of metallicity assumptions} \label{sec:appendix_c}

\begin{figure*}[h!]
\begin{center}
\includegraphics[width=0.33\textwidth]{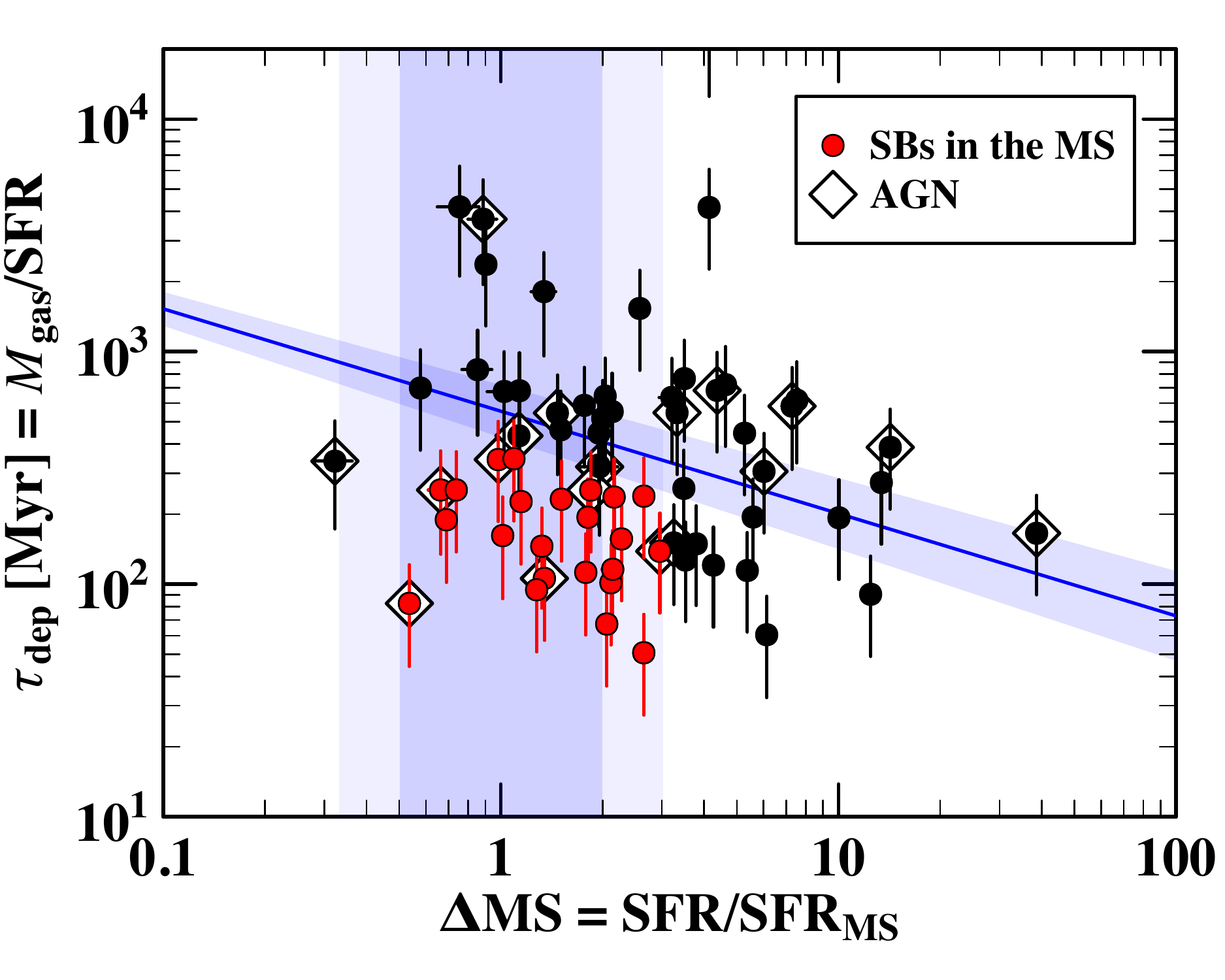}
\includegraphics[width=0.33\textwidth]{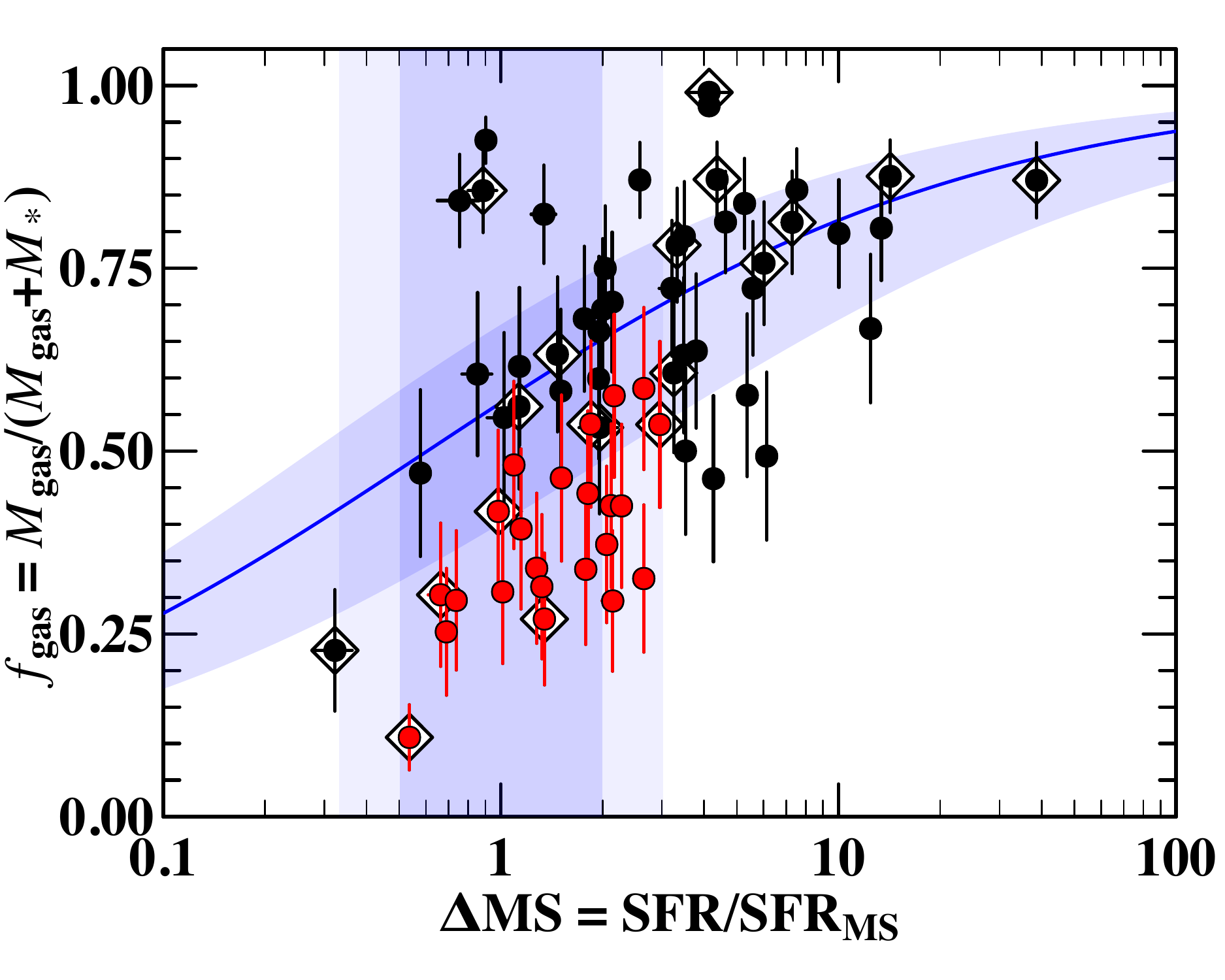}
\includegraphics[width=0.33\textwidth]{tdust_dms.pdf}
\includegraphics[width=0.33\textwidth]{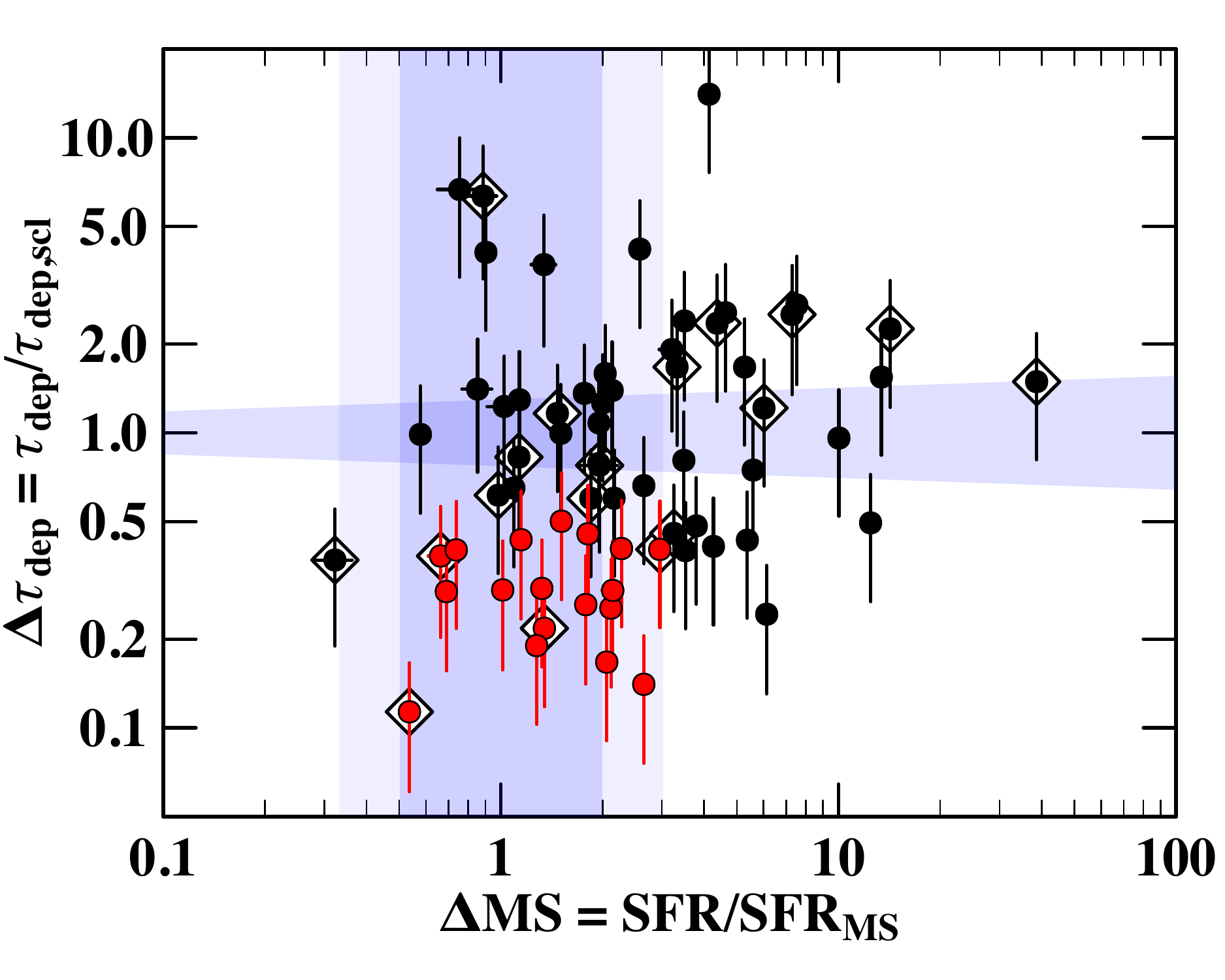}
\includegraphics[width=0.33\textwidth]{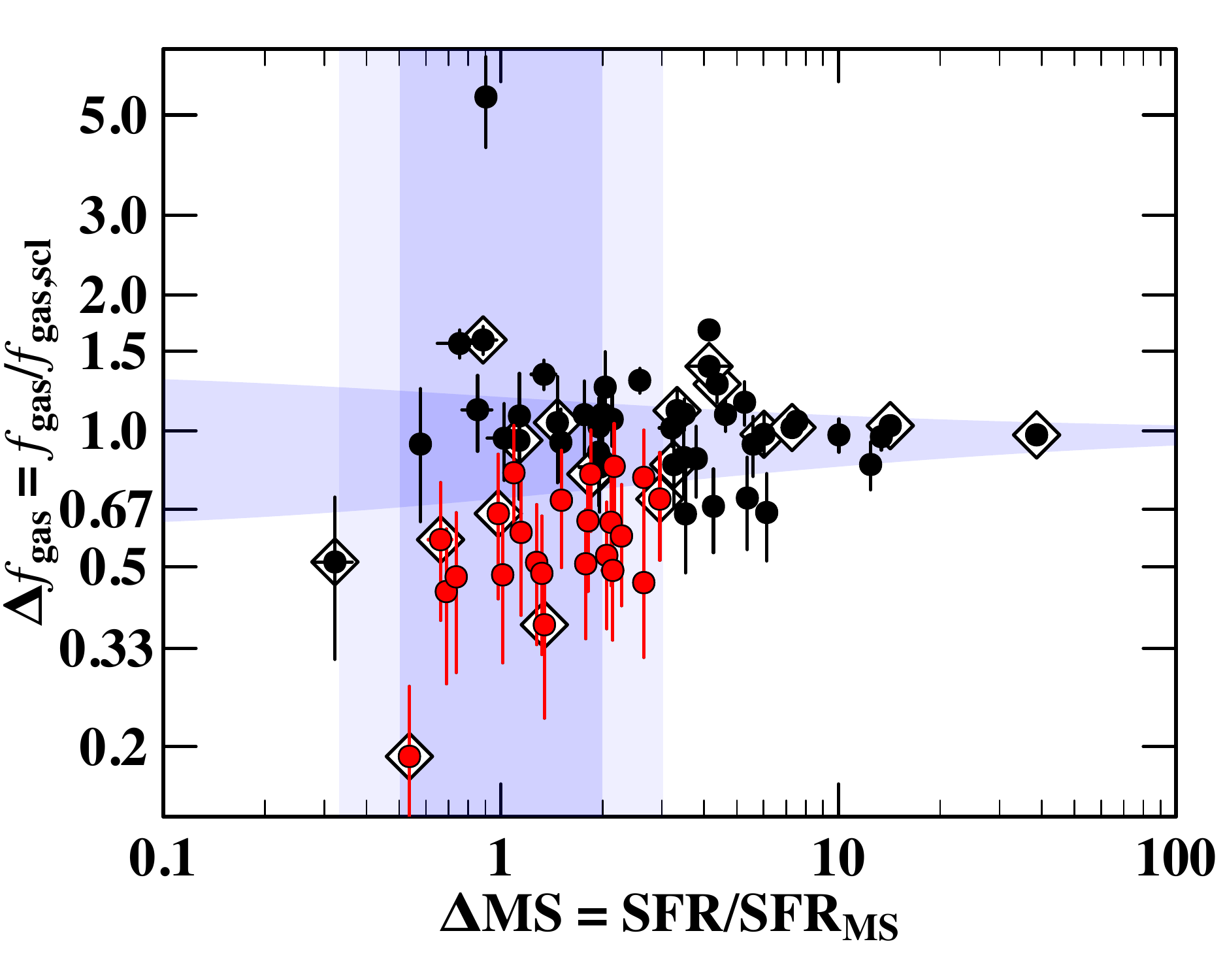}
\includegraphics[width=0.33\textwidth]{dtdust_dms.pdf}
\includegraphics[width=0.33\textwidth]{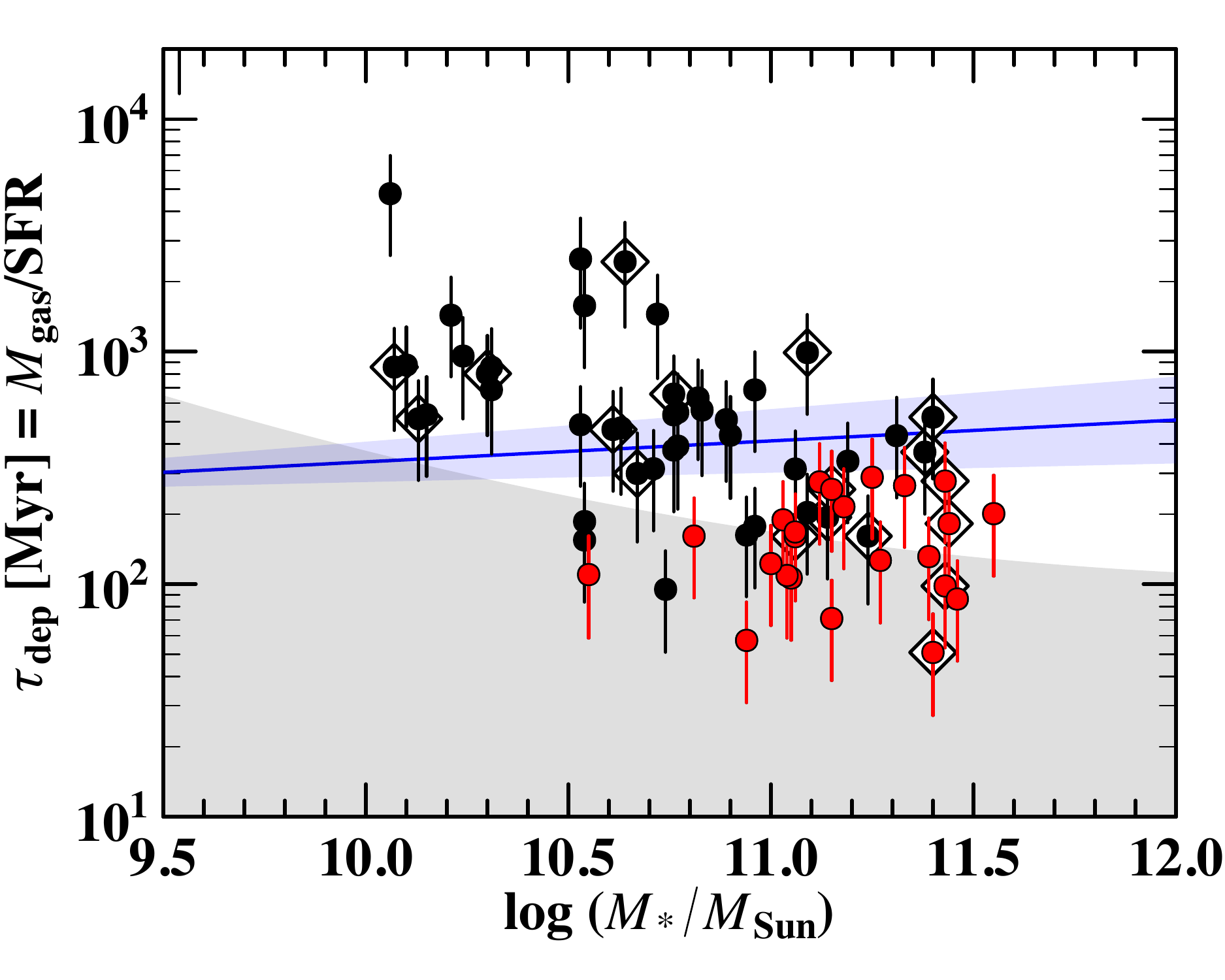}
\includegraphics[width=0.33\textwidth]{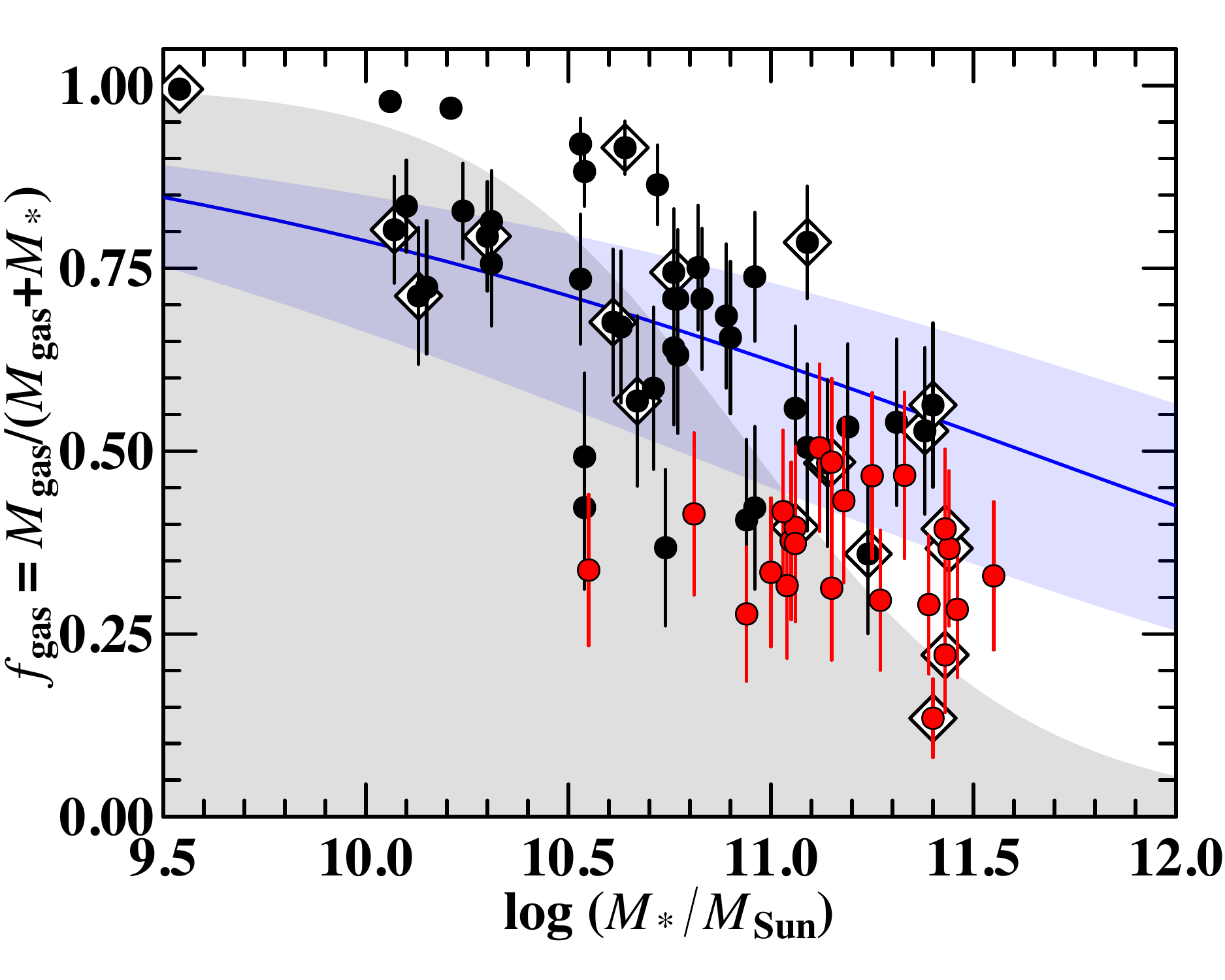}
\includegraphics[width=0.33\textwidth]{tdust_mstar.pdf}
\includegraphics[width=0.33\textwidth]{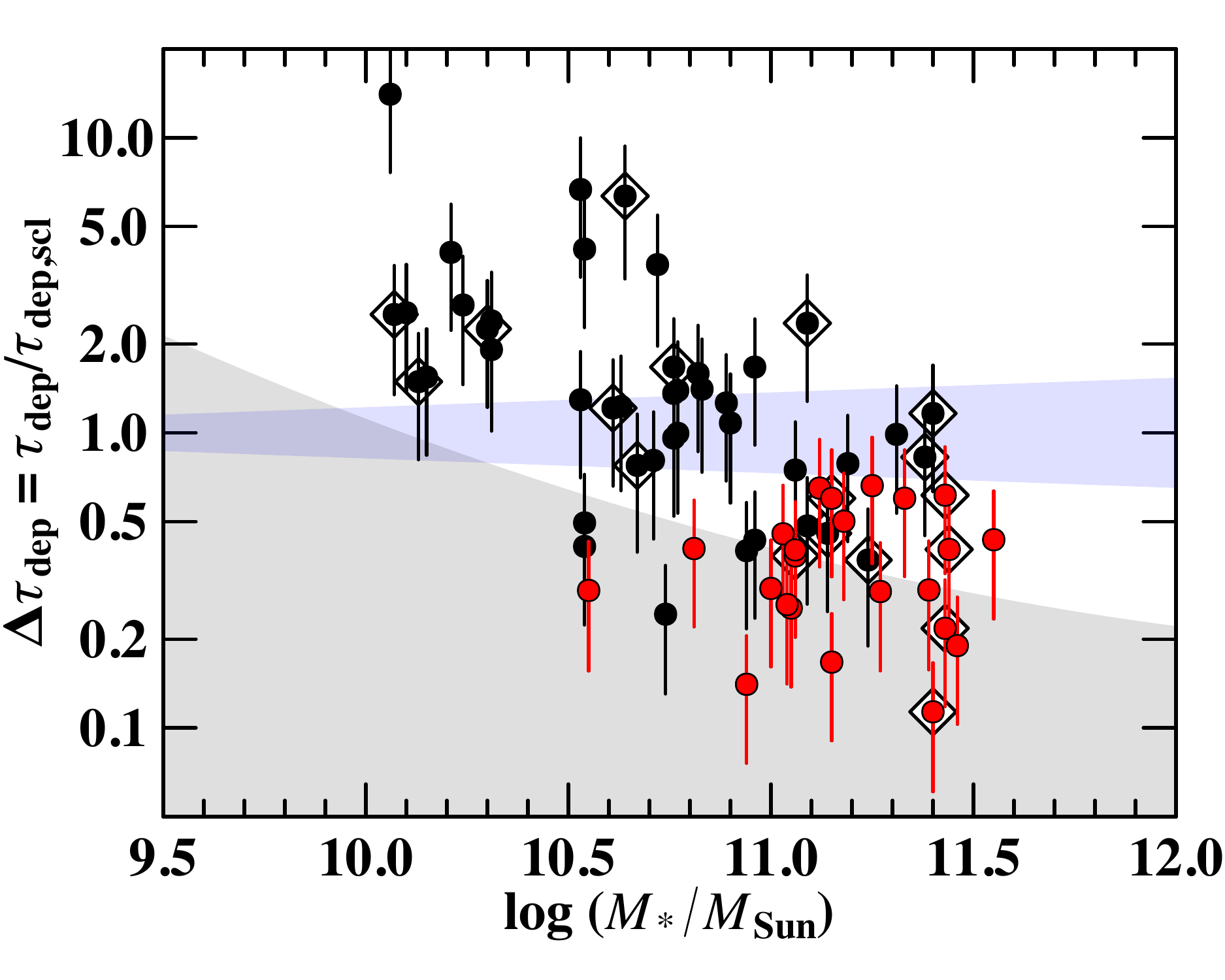}
\includegraphics[width=0.33\textwidth]{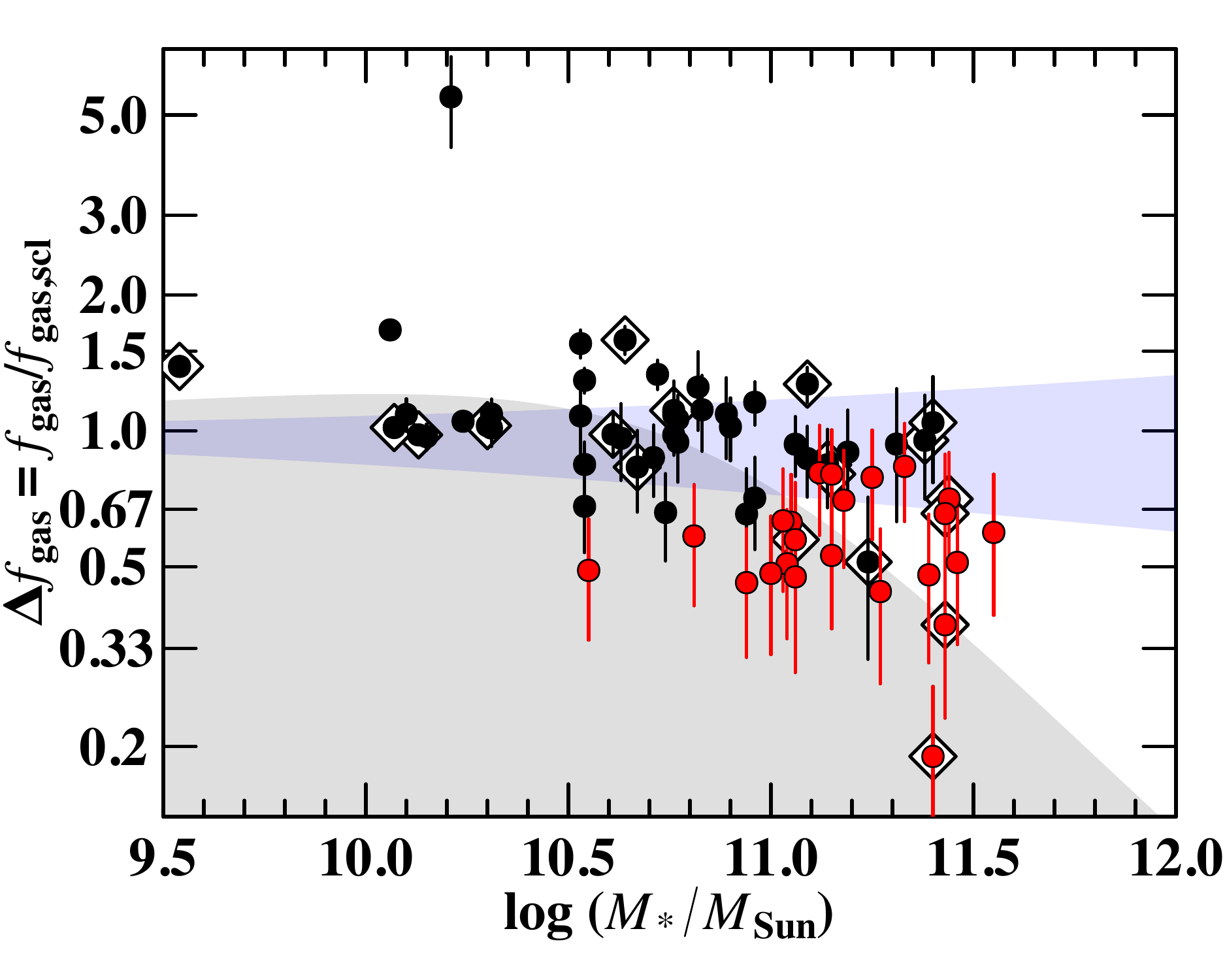}
\includegraphics[width=0.33\textwidth]{dtdust_mstar.pdf}
\caption{Same as Fig~\ref{fig:scl_rel}, but where $M_{\rm{gas}}$ estimates were obtained by using the $\delta_{\rm{GDR}}$--$Z$ with FMR instead of MZR. We note that as $T_{\rm{dust}}$ estimates are independent of the metallicity assumption, the third column remains unchanged. The subset of SBs in the MS highlighted in red is the same as in Fig~\ref{fig:scl_rel} selected from MZR.}
\label{fig:scl_rel_fmr}
\end{center}
\end{figure*}

\begin{figure*}
\begin{center}
\includegraphics[width=0.33\textwidth]{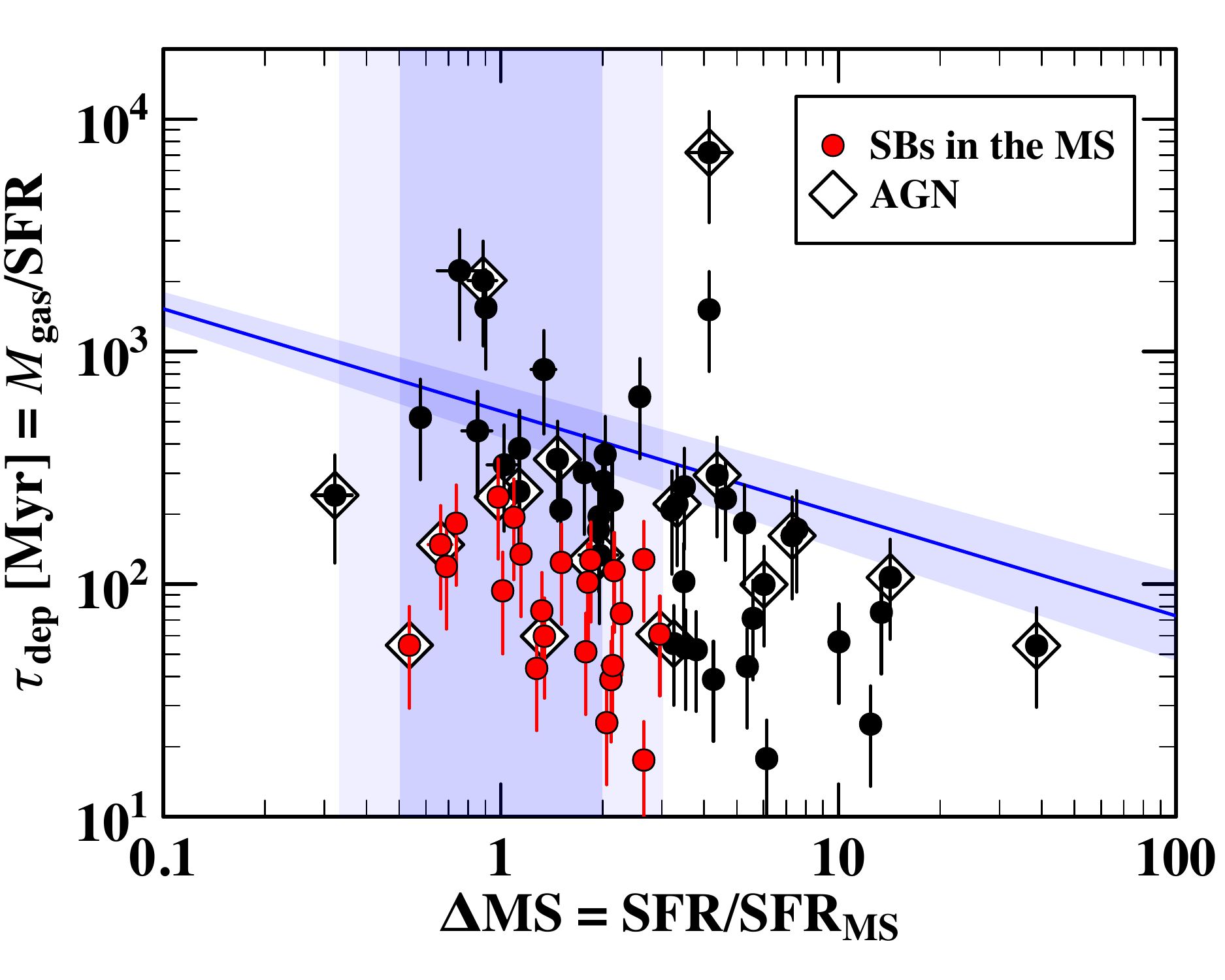}
\includegraphics[width=0.33\textwidth]{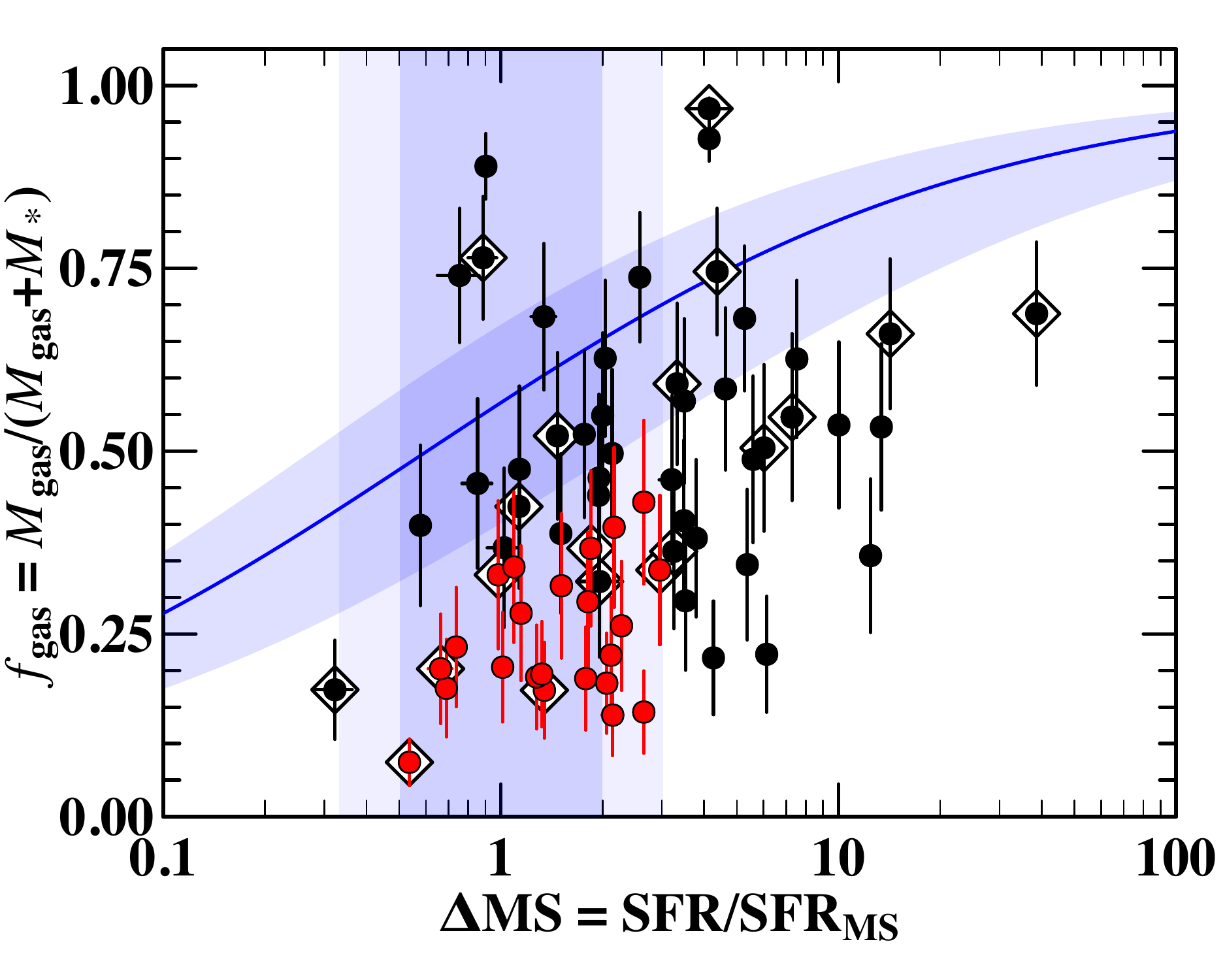}
\includegraphics[width=0.33\textwidth]{tdust_dms.pdf}
\includegraphics[width=0.33\textwidth]{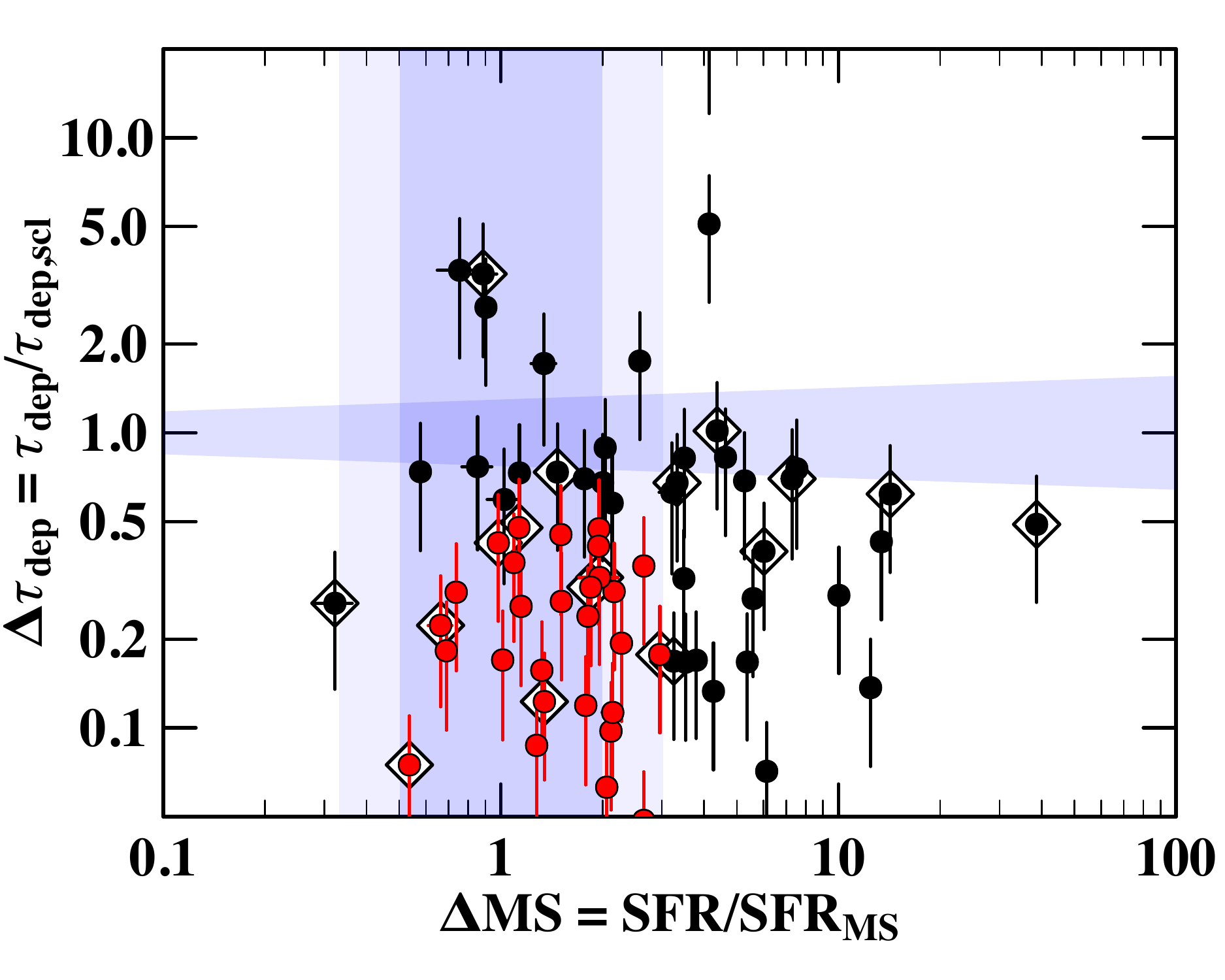}
\includegraphics[width=0.33\textwidth]{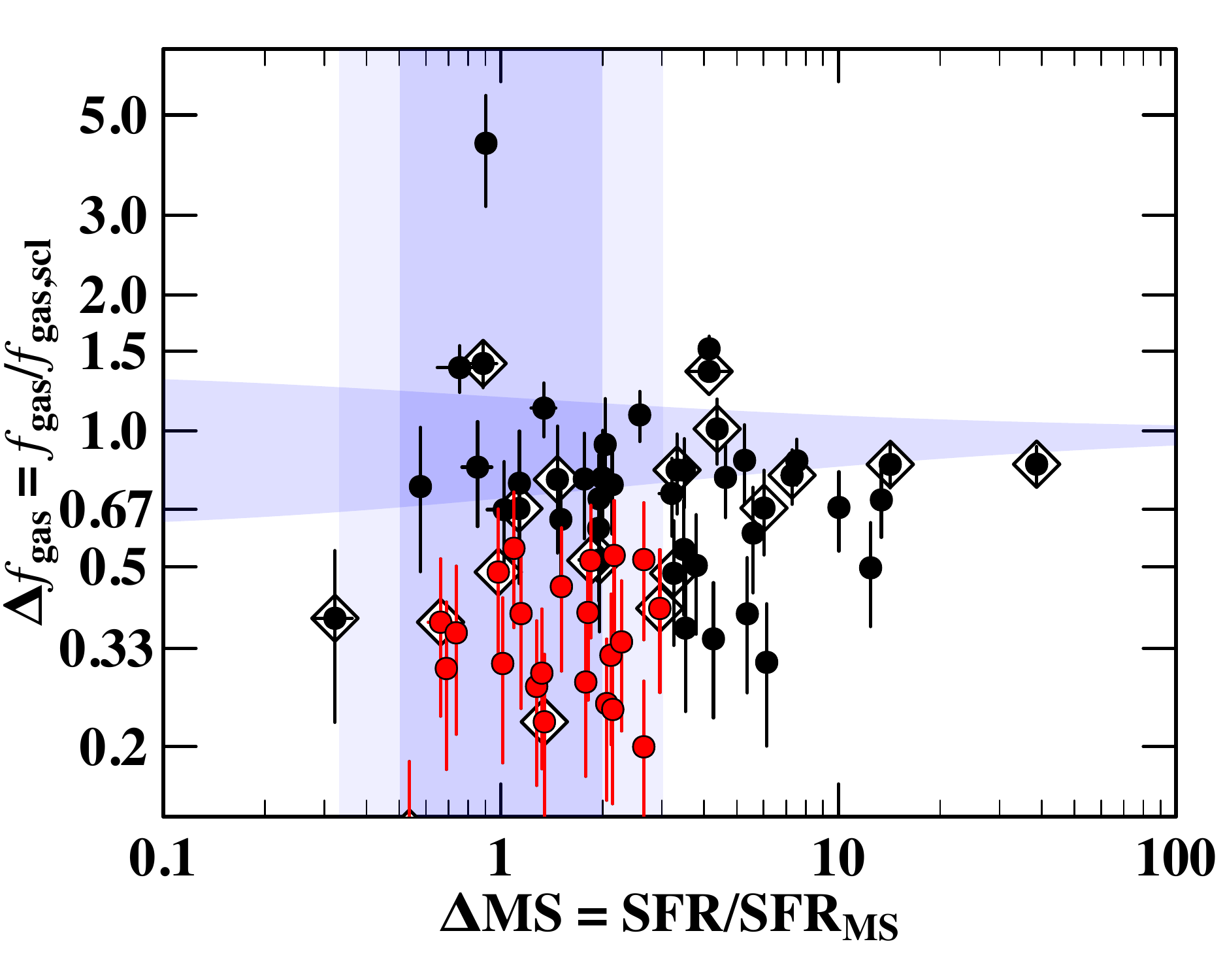}
\includegraphics[width=0.33\textwidth]{dtdust_dms.pdf}
\includegraphics[width=0.33\textwidth]{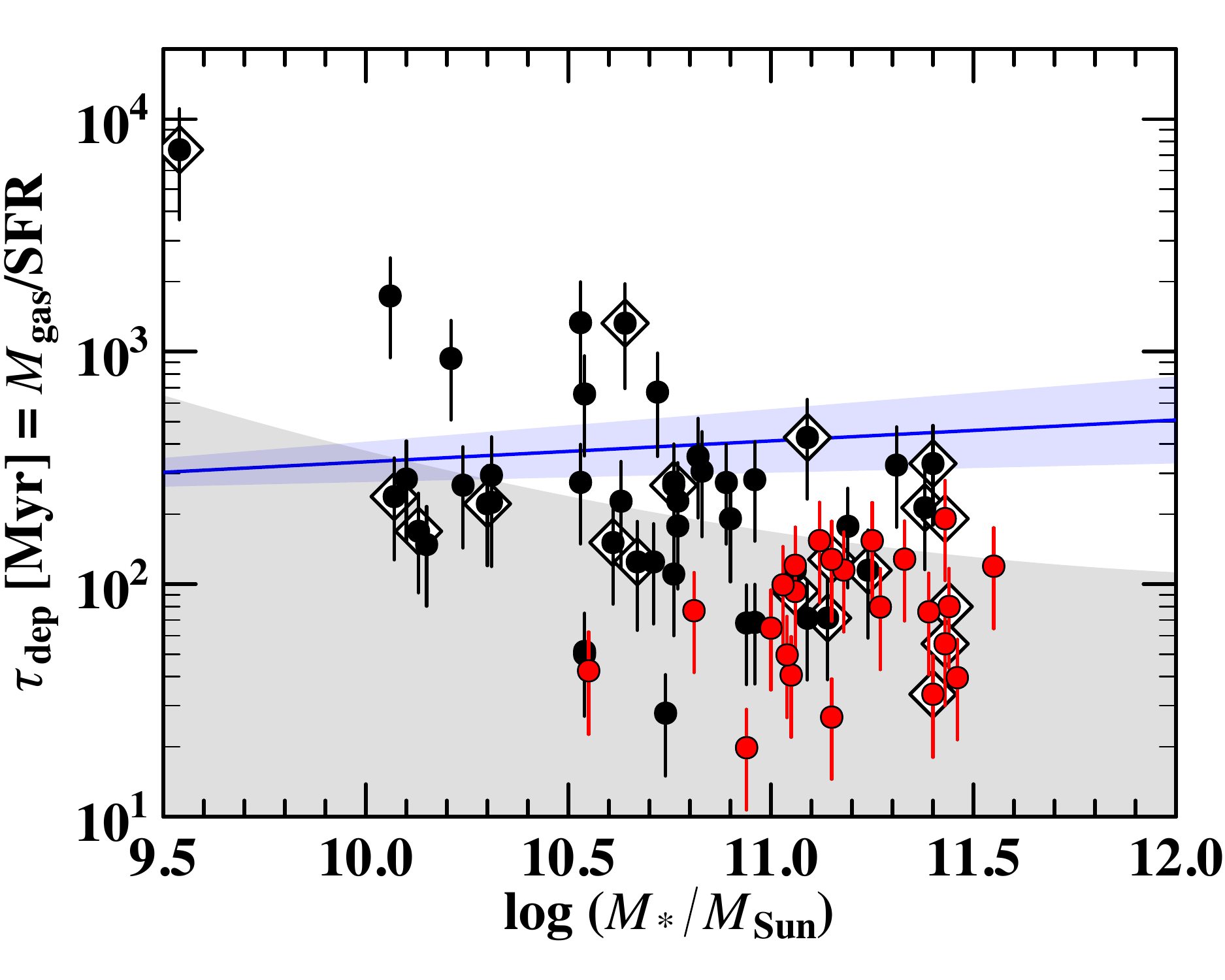}
\includegraphics[width=0.33\textwidth]{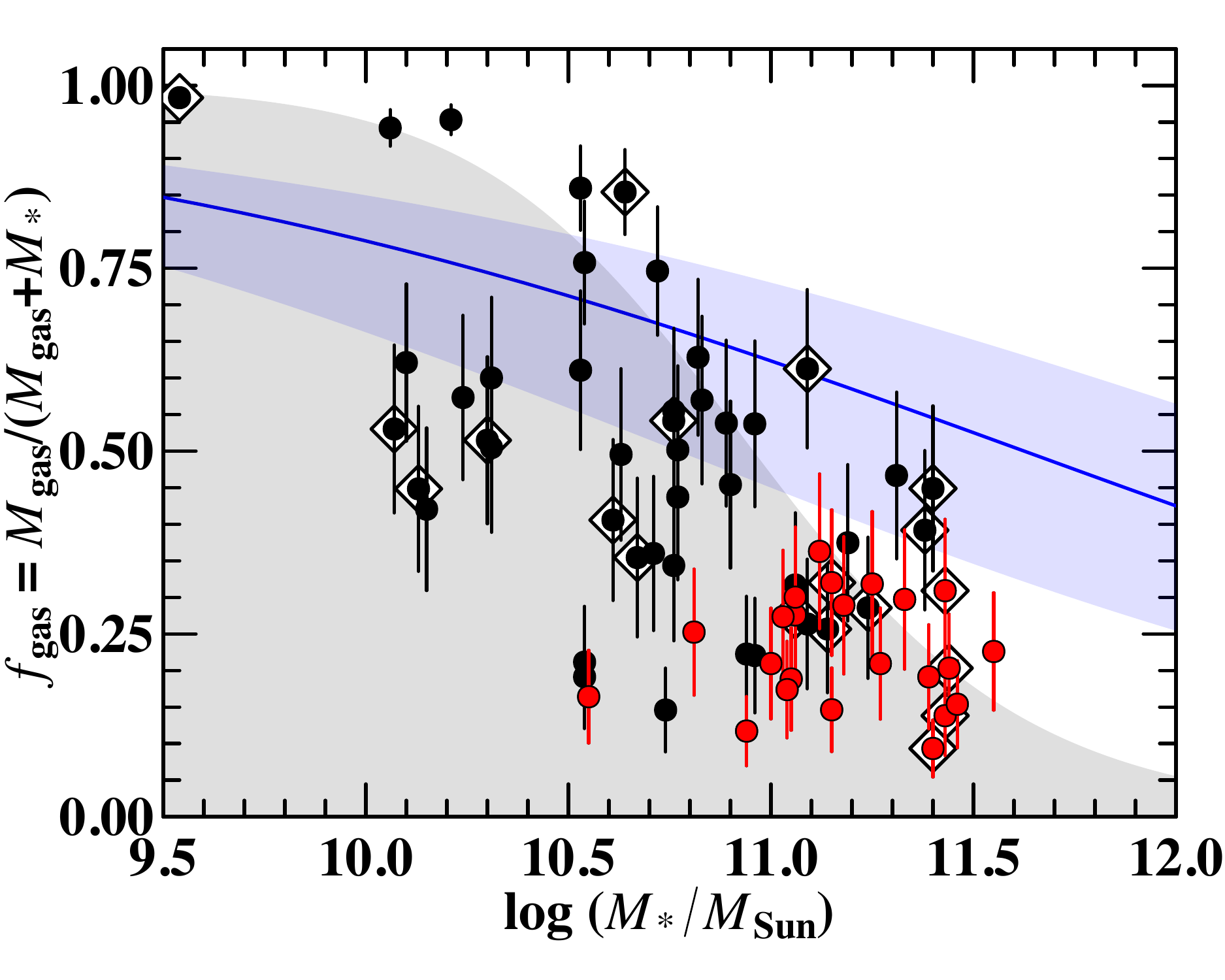}
\includegraphics[width=0.33\textwidth]{tdust_mstar.pdf}
\includegraphics[width=0.33\textwidth]{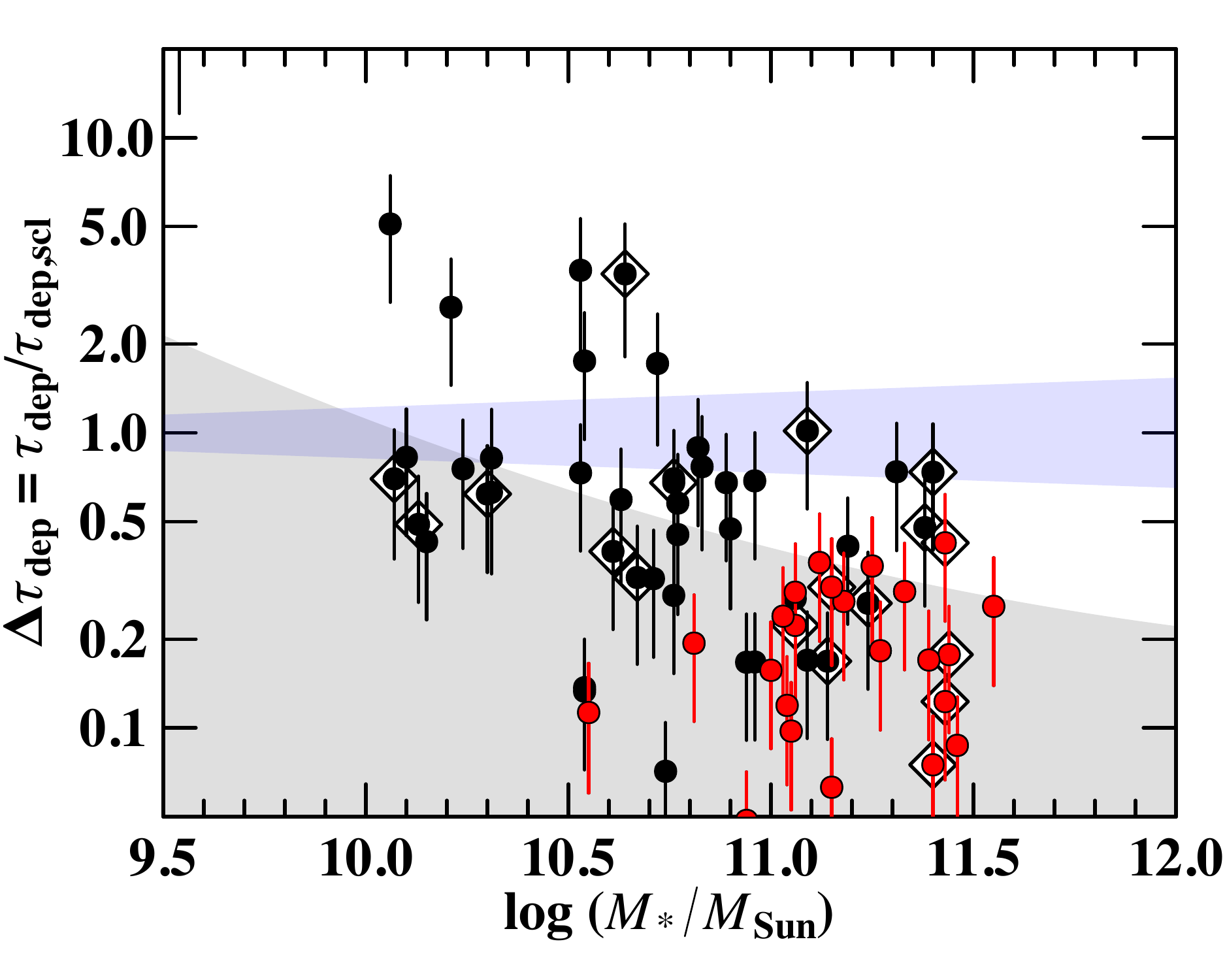}
\includegraphics[width=0.33\textwidth]{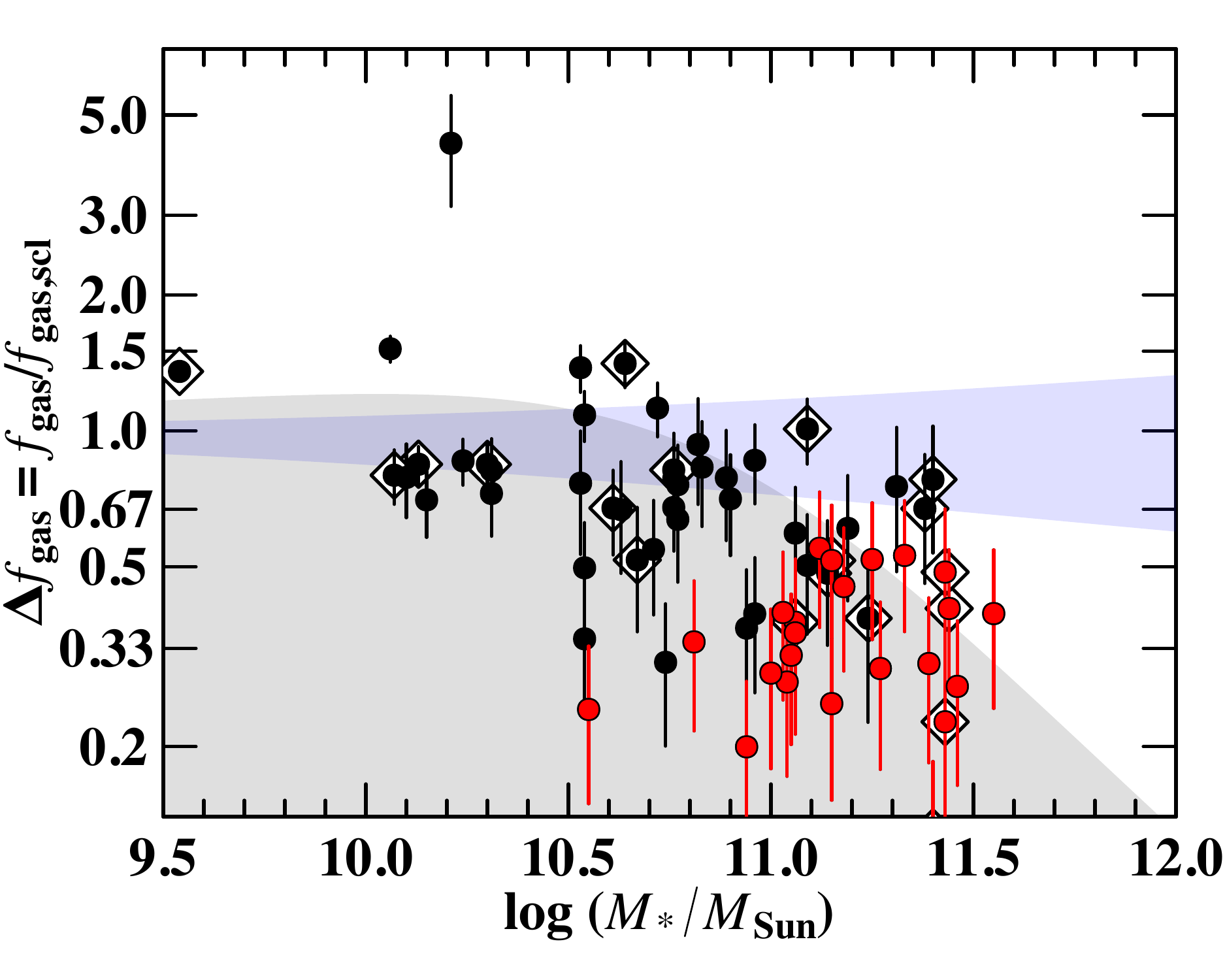}
\includegraphics[width=0.33\textwidth]{dtdust_mstar.pdf}
\caption{Same as Fig~\ref{fig:scl_rel}, but where $M_{\rm{gas}}$ estimates were obtained by using the $\delta_{\rm{GDR}}$--$Z$ with fixed solar metallicity instead of MZR. We note that as $T_{\rm{dust}}$ estimates are independent of the metallicity assumption, the third column remains unchanged. The subset of SBs in the MS highlighted in red is the same as in Fig~\ref{fig:scl_rel} selected from MZR.}
\label{fig:scl_rel_solar}
\end{center}
\end{figure*}

\begin{figure*}
\begin{center}
\includegraphics[width=0.33\textwidth]{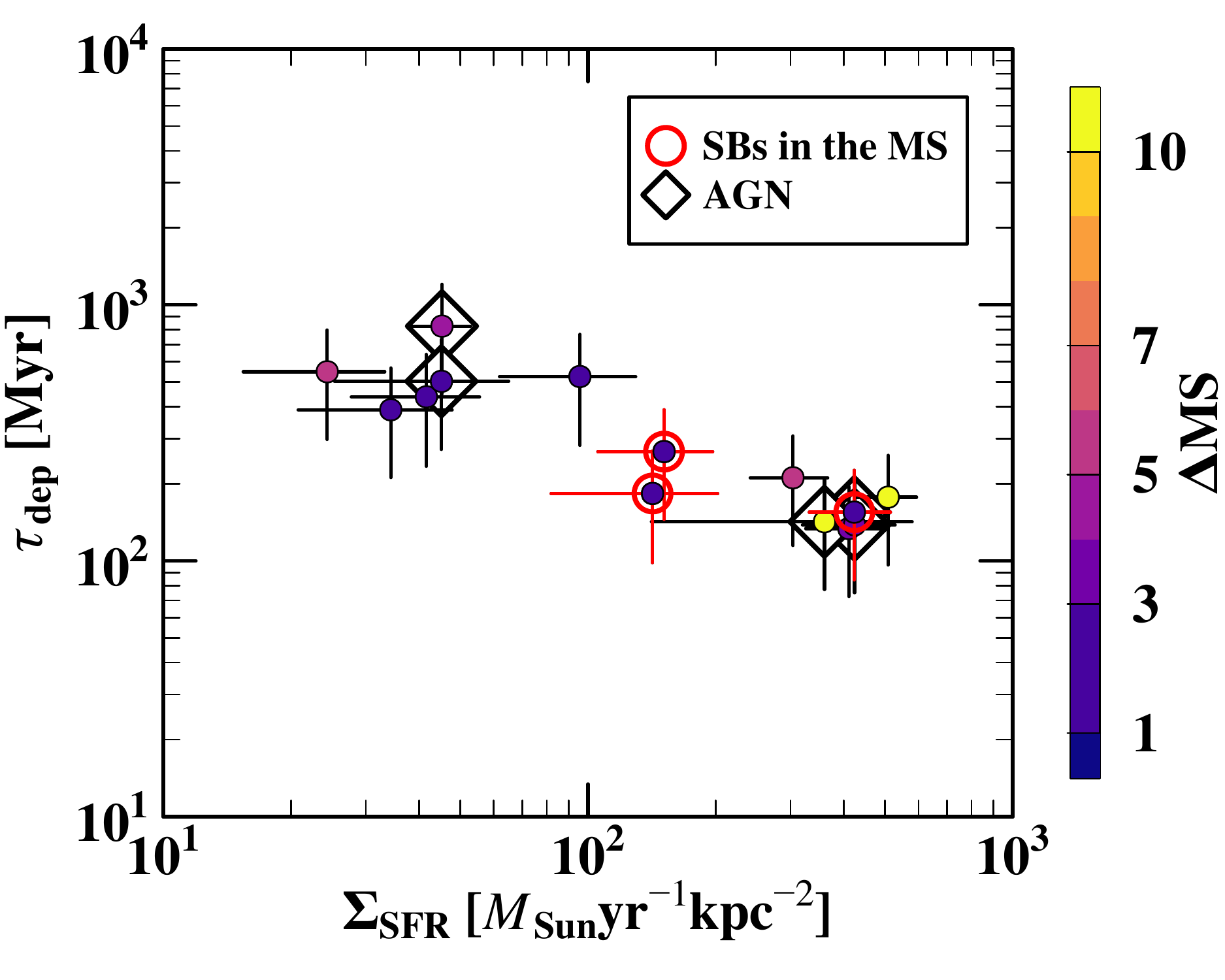}
\includegraphics[width=0.33\textwidth]{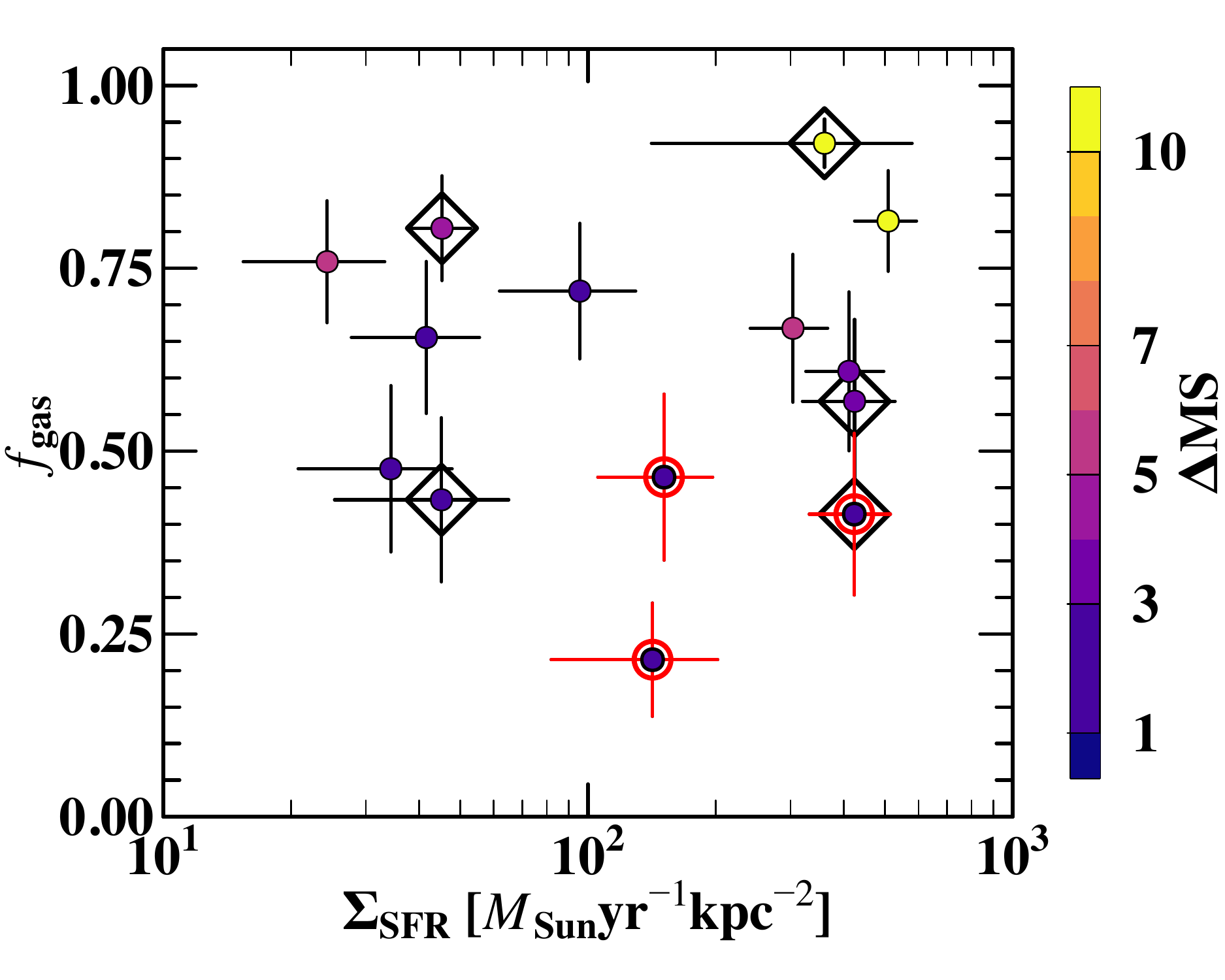}
\includegraphics[width=0.33\textwidth]{tdust_sigmasfr.pdf}
\includegraphics[width=0.33\textwidth]{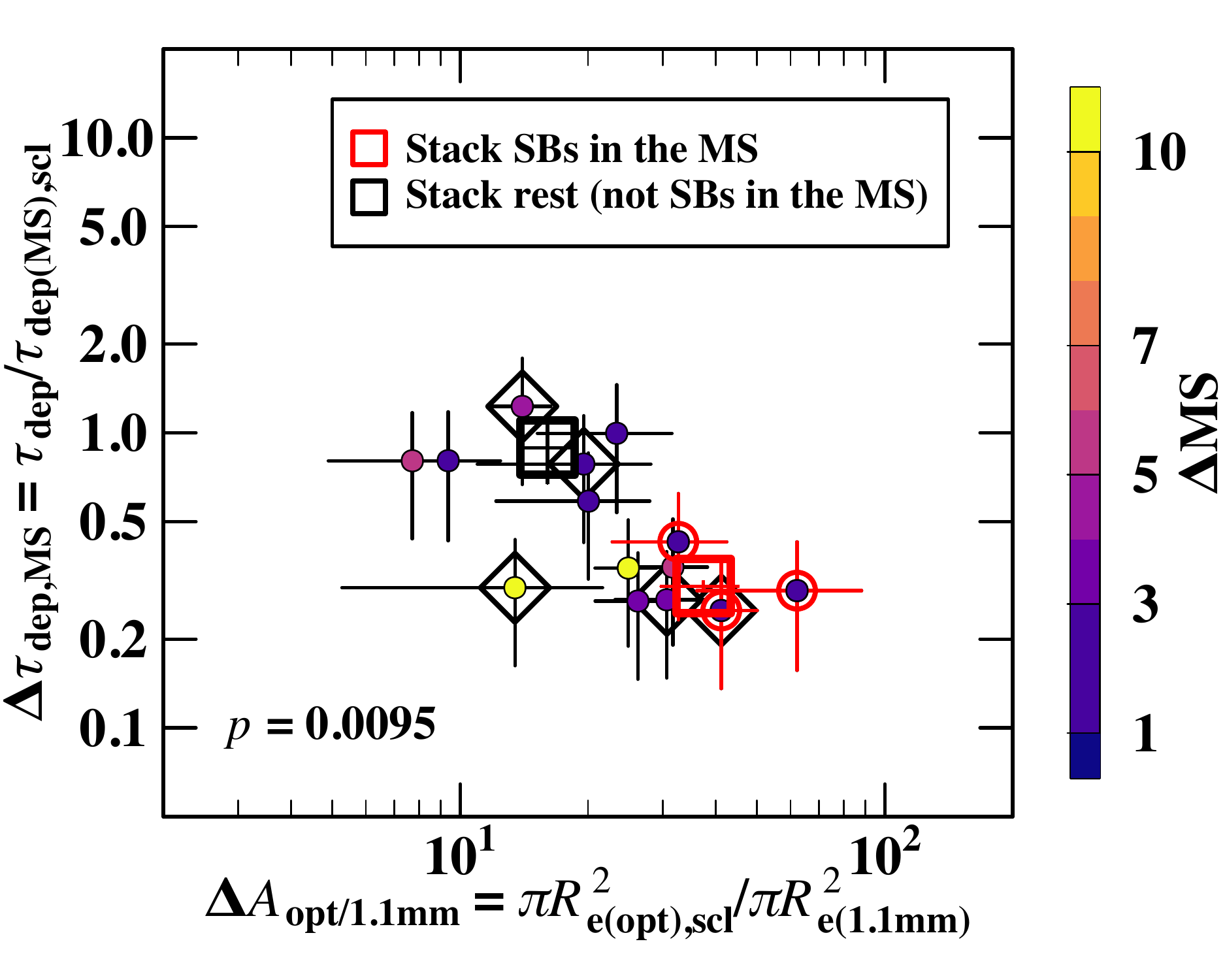}
\includegraphics[width=0.33\textwidth]{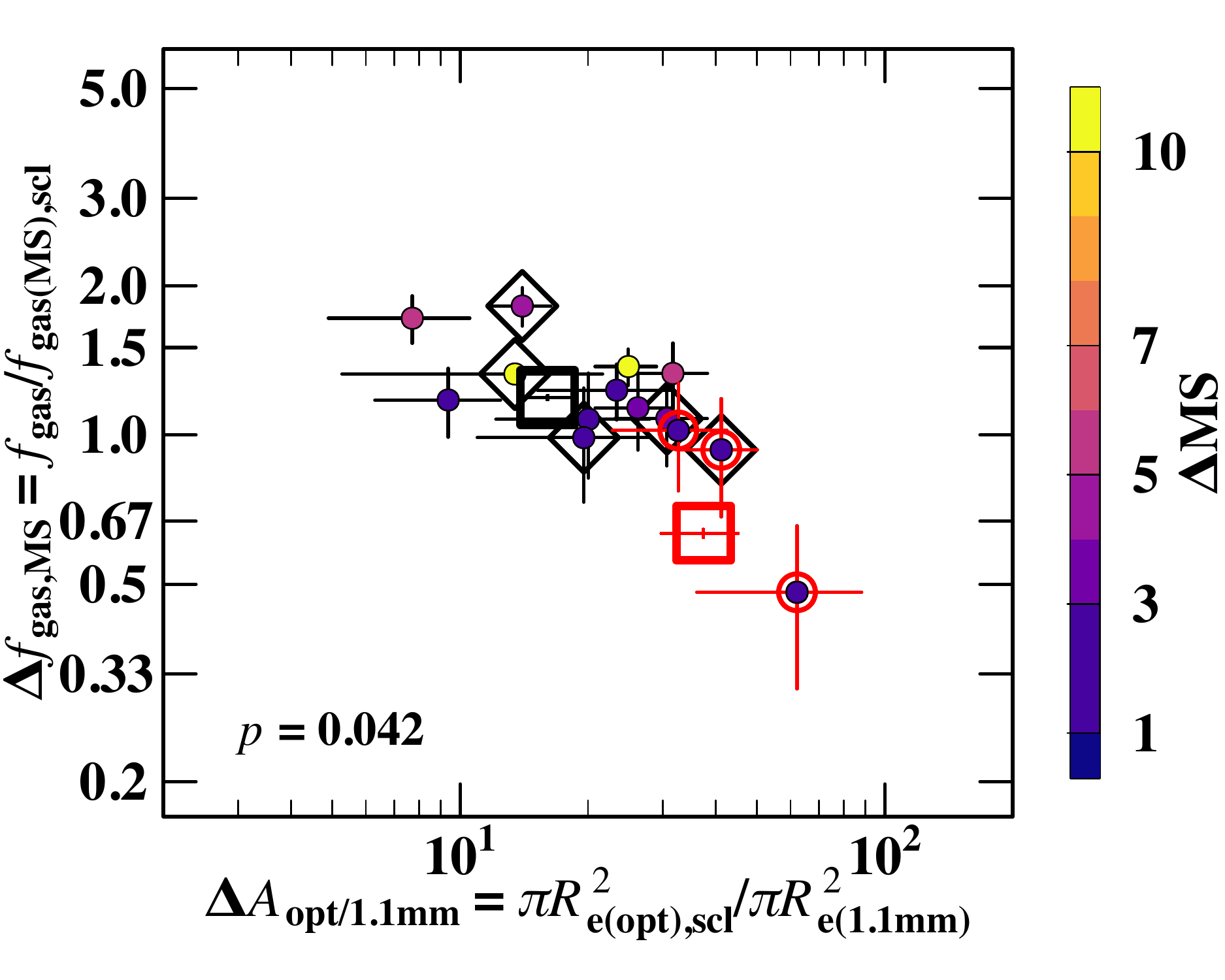}
\includegraphics[width=0.33\textwidth]{dtdust_dsize.pdf}
\caption{Same as Fig~\ref{fig:trends}, but where $M_{\rm{gas}}$ estimates were obtained by using the $\delta_{\rm{GDR}}$--$Z$ with FMR instead of MZR. We note that as $T_{\rm{dust}}$ estimates are independent of the metallicity assumption, the third column remains unchanged. The subset of SBs in the MS highlighted in red is the same as in Fig~\ref{fig:scl_rel} selected from MZR.}
\label{fig:trends_fmr}
\end{center}
\end{figure*}

\begin{figure*}
\begin{center}
\includegraphics[width=0.33\textwidth]{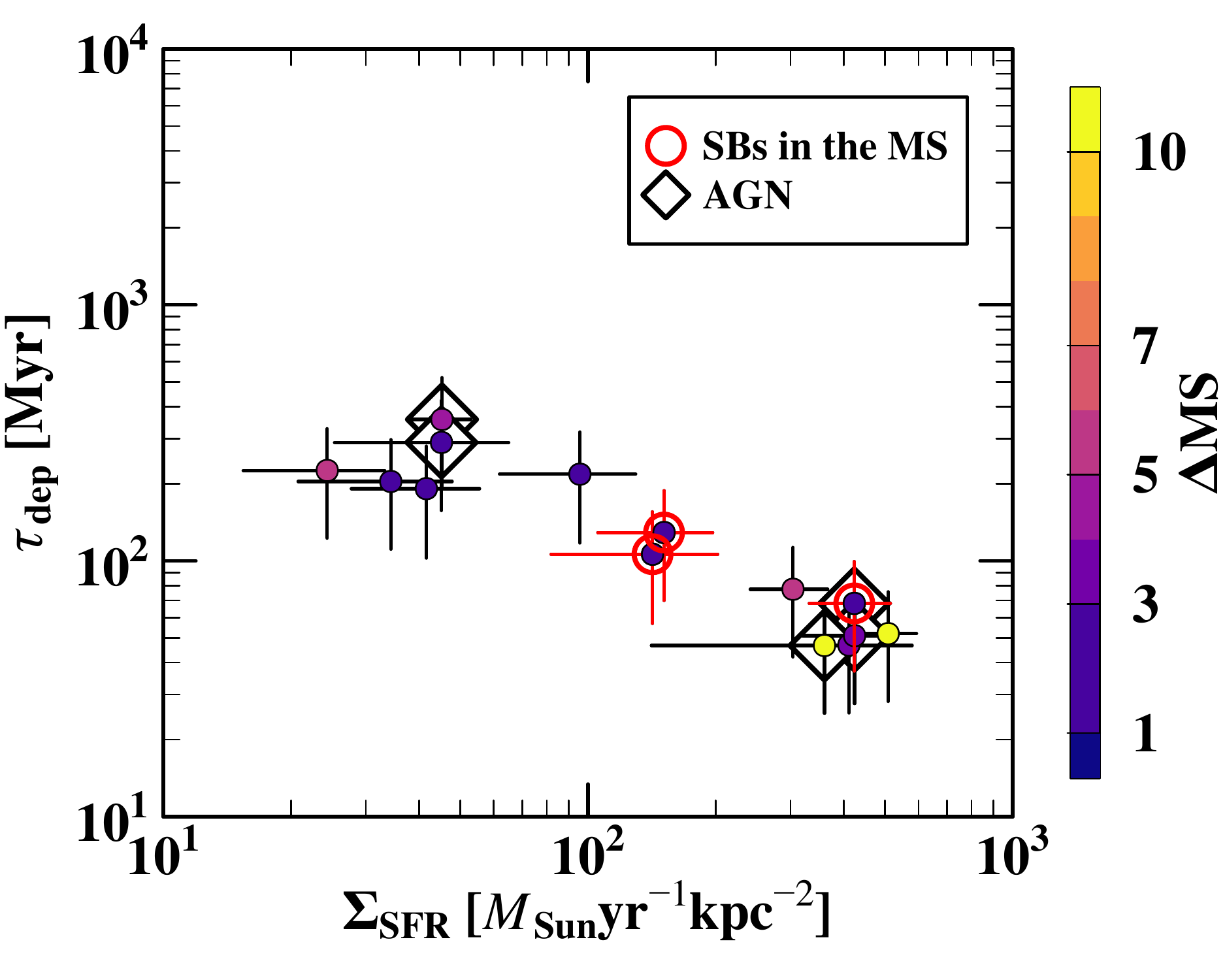}
\includegraphics[width=0.33\textwidth]{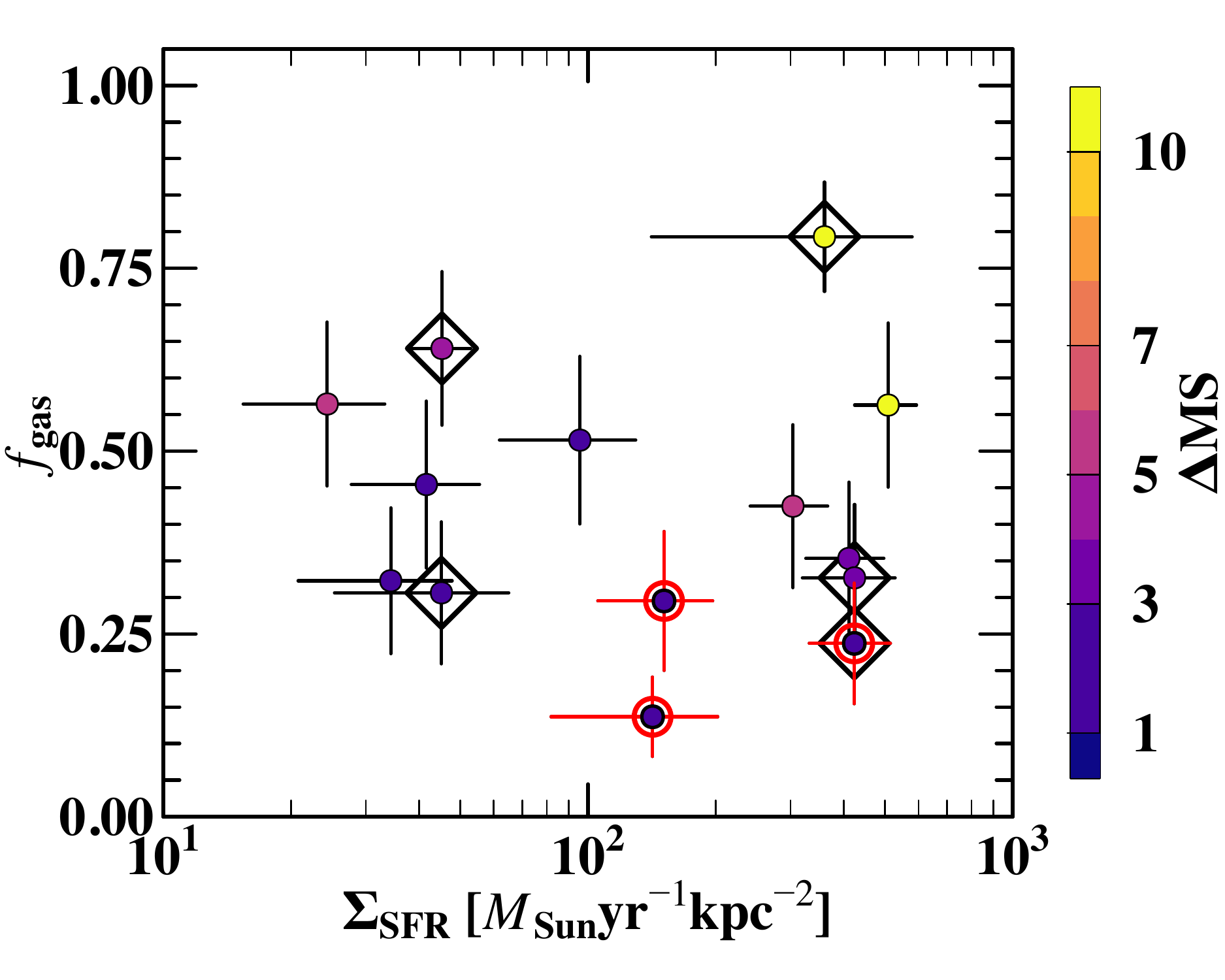}
\includegraphics[width=0.33\textwidth]{tdust_sigmasfr.pdf}
\includegraphics[width=0.33\textwidth]{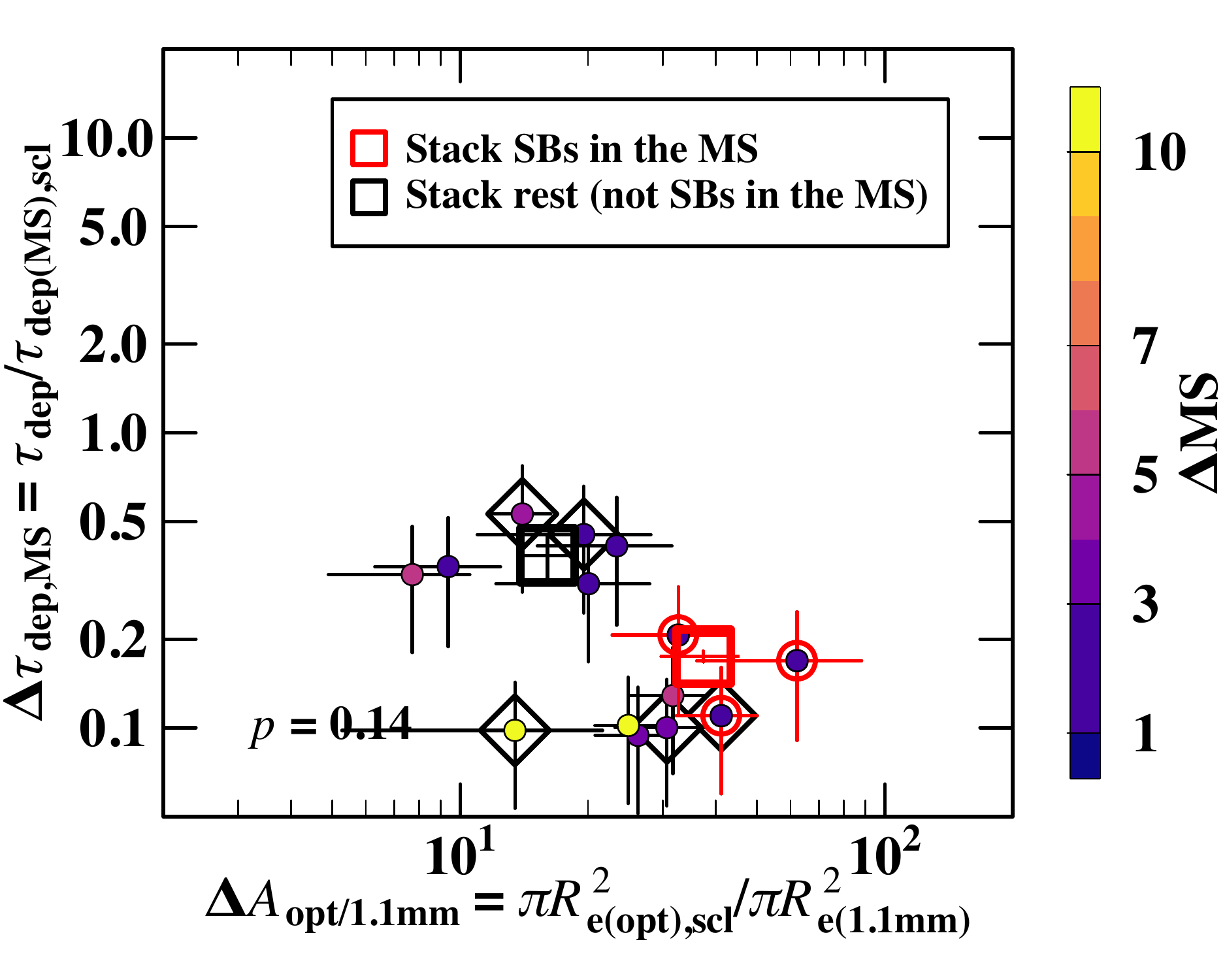}
\includegraphics[width=0.33\textwidth]{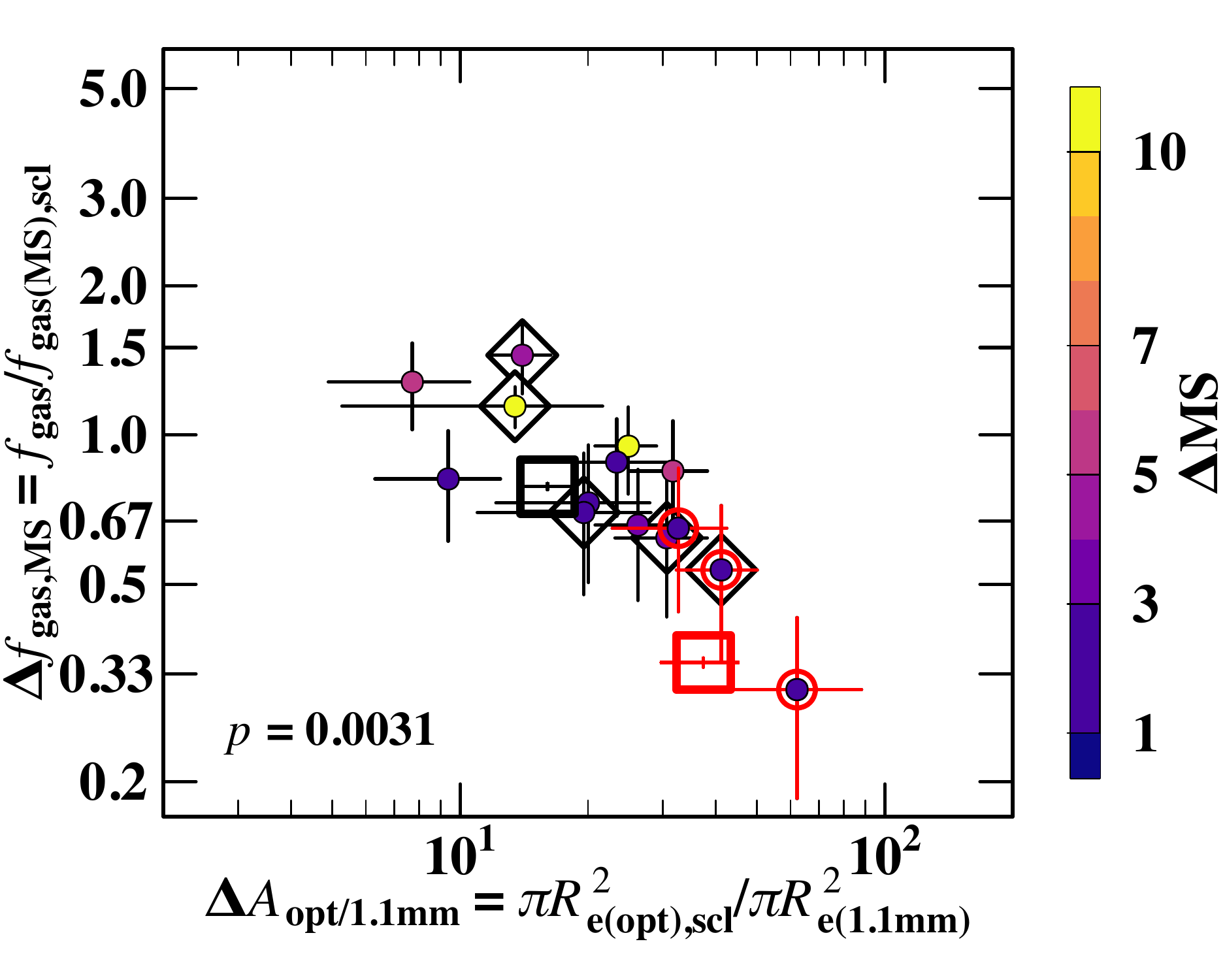}
\includegraphics[width=0.33\textwidth]{dtdust_dsize.pdf}
\caption{Same as Fig~\ref{fig:trends}, but where $M_{\rm{gas}}$ estimates were obtained by using the $\delta_{\rm{GDR}}$--$Z$ with fixed solar metallicity instead of MZR. We note that as $T_{\rm{dust}}$ estimates are independent of the metallicity assumption, the third column remains unchanged. The subset of SBs in the MS highlighted in red is the same as in Fig~\ref{fig:scl_rel} selected from MZR.}
\label{fig:trends_solar}
\end{center}
\end{figure*}

\FloatBarrier

\end{appendix}

\end{document}